\documentclass[usenatbib,useAMS]{mn2e}
\usepackage{times}
\usepackage{epsfig}

\renewcommand{\vec}[1]{\bmath{#1}}

\newcommand{\multint}[1]{\int \hspace*{-0.4em} \stackrel{({#1})}{\ldots} \hspace*{-0.4em} \int\!}

\begin{document}
\title{Stochastical distributions of lens and source properties for
observed galactic microlensing events}
\author[M. Dominik]{M. Dominik\thanks{E-mail: md35@st-andrews.ac.uk} \\
University of St Andrews, School of Physics \& Astronomy, North Haugh,
St Andrews, KY16 9SS, United Kingdom}

\maketitle

\begin{abstract}
A comprehensive new approach is presented for deriving probability densities of physical
properties characterizing the lens and source 
that constitute an observed galactic
microlensing event. While previously encountered problems are overcome, 
constraints from event anomalies and model parameter uncertainties can be 
incorporated into the estimates. Probability densities for given events need to be
carefully distinguished from the statistical distribution of the same parameters
among the underlying population from which the actual lenses and sources are drawn.
Using given model distributions of the mass spectrum, the mass density, and the velocity
distribution of Galactic disk and bulge constituents, probability densities of
lens mass, distance, and the effective lens-source velocities are derived, where
the effect on the distribution that arises from additional observations
of annual parallax or finite-source effects, or the absence of significant effects,
is shown. The presented formalism can also be
used to calculate probabilities for the lens to belong to one or another population and 
to estimate parameters that characterize anomalies. Finally, it is shown how detection
efficiency maps for binary-lens companions in the physical parameters companion mass and
orbital semi-major axis arise from values determined for the mass ratio and dimensionless projected
separation parameter, including the deprojection of the orbital motion for elliptical orbits.
Compared to the naive estimate based on 'typical values', the detection efficiency for low-mass
companions is increased by mixing in higher detection efficiencies for smaller mass ratios (i.e\
smaller masses of the primary).
\end{abstract}


\section{Introduction}
During the recent years, more than 2000 microlensing events have been observed
and corresponding model parameters have been published. However, these model
parameters in general do not coincide with the underlying physical characteristics of lens and
source star, which are their distances from the observer, $D_\rmn{L}$ and
$D_\rmn{S}$, respectively, the mass $M$ of the lens, and the relative proper motion $\mu_\rmn{LS}$
between lens and source. For 'ordinary' events, compatible with rectilinear motion
between point-like sources and lenses, the only parameter related to these characteristics
is the event time-scale $t_\rmn{E} = \theta_\rmn{E}/\mu_\rmn{LS}$, which corresponds to
the time in which the source moves by the angular Einstein radius
\begin{equation}
\theta_\rmn{E} = \sqrt{\frac{4GM}{c^2}\,\frac{D_\rmn{S}-D_\rmn{L}}{D_\rmn{L}\,
D_\rmn{S}}} = \sqrt{\frac{4GM}{c^2}\,\frac{\pi_\rmn{LS}}{1~\rmn{AU}}}
\label{eq:thetaE}
\end{equation}
relative to the lens, where
$\pi_\rmn{LS} = 1~\rmn{AU}\,(D_\rmn{L}^{-1}-D_\rmn{S}^{-1})$ denotes the 
relative lens-source parallax.

With the physical lens characteristics being statistically distributed according to
the mass density and velocity distribution of lenses and sources as well as
the mass spectrum of the lenses, the distribution of observed parameters in the ensemble
of galactic microlensing events can be used to measure these distributions.
\citet*{RJM:massmoments} have shown explicitly how statistical moments of the observed
time-scale distributions translate into moments of the underlying mass spectrum of the lenses.

A {\em different} question is posed by asking for the stochastical distribution of physical
lens and source properties given the observed model parameters for a single realized event.
In the literature, this distinction has frequently not been made strictly enough, leading to
some confusion. In particular, the probability density of the lens mass 
averaged over all observed events does {\em not} converge to the underlying mass spectrum. By quoting a
probability for the lens mass in a given event to assume a specific value, \citet{RJM:massmoments}
did not produce a meaningful result, since the probability for any random variable to assume a
specific value is zero. Further common misconceptions exist around a 
'relative probability', which is not defined, and a 'most-probable value', which does not
exist either. A finite
probability can only be attached to a finite interval of values, given as the integral over the
probability density of the considered quantity. \citet{Do98:Estimate} realized that in order to derive 
information about the lens mass and other properties for a given event, one is dealing 
with a probability density, which carries the inverse dimension of the quantity it refers to,
rather than a likelihood \citep[e.g.][]{MACHO:parallax}. In fact, likelihood functions and
probability densities are different entities, which can be seen explicitly from the
following property: If a likelihood for a 
quantity $a$ has a maximum at some value $a_0$, the likelihood for any function $f(a)$ of the
quantity $a$ has a maximum at $f(a_0)$, whereas such a property does not hold for probability densities,
i.e.\ $\left\langle f(a) \right\rangle \neq f(\left\langle a \right\rangle)$ may occur, 
where $\left\langle a \right\rangle$ denotes the expectation value
of $a$. However, like \citet{RJM:massmoments} before, \citet{Do98:Estimate} 
still failed to realize that a statistical mass spectrum of the lens population needs to
be assumed along with the space- and velocity-distributions of lenses and sources.\footnote{By
neglecting this, an implicit assumption is made.} Moreover, confusions around the 
ill assumption of a fixed-mass spectrum $\propto \delta(M-M_\rmn{fixed})$ 
led to incorrect results for power-law mass spectra, where
the power-law index would have to be shifted by unity in order to obtain correct expressions.
For determining the event characteristics of MACHO 1997-BLG-41, \citet{PLANET:M41} used a 
discretization of the statistical distributions of the basic lens and source properties
in the form of a Monte-Carlo simulation, which is a variant of the approach of \citet{Do98:Estimate}.
Some of the related ideas have been further developed into
a related formalism arguing on the basis of Bayesian statistics used in the analysis
of OGLE 2003-BLG-423 \citep{GG:Esti}, where the Galaxy model is used as the prior for the model parameters.

In this paper, a revised comprehensive framework is presented for combining
the model parameters as determined from the observations with Galaxy models in order to
estimate physical lens and source properties for a given event.
This refined approach overcomes the previously encountered problems
and allows the inclusion of model constraints from event anomalies as well as model parameter uncertainties.
Moreover, by considering different lens populations, a probability
that the observed event with its
parameters results from one or the other is obtained, which is taken into account for
deriving the probability densities of the event characteristics. 
Rather than having to rely on Monte-Carlo simulations, all results are obtained in the form of
closed expressions by means of integrals over the statistical distributions of the lens and source
properties.

In Sect.~\ref{sec:appmic}, the role of the event rate for deriving the
desired probability densities of lens and source properties is
discussed, while Sect.~\ref{sec:timescale} looks at the global properties of the ensemble of microlensing events 
such as the distribution of the event time-scale and the contribution arising from different lens populations.
The probability densities of key properties of the lens, namely its mass, distance, and relative velocity
with regard to a source at rest, that follow from a measurement of the event time-scale and the Galaxy model
are derived and discussed in Sect.~\ref{sec:probdens}, whereas Sect.~\ref{sec:anomconstr} focusses on
the implications if further constraints arise either from the measurement or from upper limits on
additional model parameters, where the two cases of annual parallax and finite source size are considered
explicitly in detail. Sect.~\ref{sec:anomdist} then discusses probability densities of further quantities
such as the parallax and finite-source parameters as well as the orbital semi-major axis and orbital period
for binary lenses, before Sect.~\ref{sec:deteffmaps} shows how the presented approach can be used to
determine the detection efficiency for companions (such as planets) to the lens star as function
of the physical properties of the system. 
Sect.~\ref{sec:summary} finally provides a summary. 
The underlying probabilistic approach is presented in 
Appendix~\ref{sec:general}, details of the adopted Galaxy model can be found in Appendix~\ref{sec:galmodel}, and Appendix~\ref{sec:orbitalproj} discusses the statistics of the orbits
of binary systems and in particular the projection factor between the actual angular separation and the
semi-major axis.

\section{Properties determining microlensing events}
\label{sec:appmic}

Microlensing relies on the chance alignment of observed source stars with
intervening massive objects acting as lenses, where the degree of
alignment is characterized by the angular Einstein radius as defined by
Eq.~(\ref{eq:thetaE}),
which depends on the lens mass $M$ as well as on the source distance 
$D_\rmn{S}$ and the lens distance $D_\rmn{L}$. With a two-dimensional angular
separation $\vec \theta$ between lens and source, the magnification of the
source star caused
by the gravitational field of the lens
in general depends only on $\vec u = \vec \theta/\theta_\rmn{E}$, 
while for a point source it even depends on
its absolute value $u = |\vec u|$ only, taking the analytical form
\citep[e.g.][]{Pac86}
\begin{equation}
A(u) = \frac{u^2+2}{u\,\sqrt{u^2+4}}\,.
\label{eq:magnification}
\end{equation}

The basic properties of point-like lenses and sources that affect the
microlensing light curve are the source magnitude $m_\rmn{S}$,
the source distance $D_\rmn{S}$, the lens mass $M$,
the lens distance $D_\rmn{L}$, the relative proper motion between
lens and source $\mu_\rmn{LS}$, taking into account the motion of the observer,
and the blend magnitude $m_\rmn{B}$.
If one considers the source distance $D_\rmn{S}$ as well-constrained
(there is no problem of including an uncertainty on this parameter as well),
and uses the fact that there is no correlation between lens properties
and the source or the blend magnitude, it is sufficient to consider
lens mass $M$, lens distance $D_\rmn{L}$ and proper motion $\mu_\rmn{LS}$ as 
the descriptive properties for a given microlensing event. 
A binary lens involves further characteristics, namely its mass ratio $q$ and 6
orbital elements that can be chosen as
the semi-major axis $a$, the eccentricity $e$, three parameters describing the
orientation of the orbit (such as the inclination, the longitude of the
ascending node, and the argument of perihelion), and finally an orbital phase
(such as the mean anomaly at a given epoch).
Distributions of the mass ratio $q$, the semi-major axis $a$, and the eccentricity $e$ are
pairwise correlated and also depend on the total mass $M$ of the system and the actual
types of stars involved, where our current knowledge on these is rather limited.

Let $v = D_\rmn{L}\,\mu_\rmn{LS}$ denote the effective velocity at the
lens distance that corresponds to the proper motion $\mu_\rmn{LS}$, while the
Einstein radius $r_\rmn{E} = D_\rmn{L}\,\theta_\rmn{E}$ 
is the physical size of the angular Einstein radius $\theta_\rmn{E}$
at this distance.
With the mass spectrum $\Phi_M(M)$ and the effective transverse 
velocity $v$ being distributed as $\Phi_v(v)$,
the contribution to the event rate by lenses in an infinitely thin sheet
at distance $D_\rmn{L}$ with masses in the range $[M, M+dM]$ and
velocities in the range $[v, v+dv]$ is given by
\begin{equation}
\rmn{d}\Gamma = w_0\,\frac{\rho(D_\rmn{L})}{M}\,\Phi_M(M)\,v\,
\Phi_v(v)\,r_\rmn{E}\,\rmn{d}M\,\rmn{d}v\,
\rmn{d}D_\rmn{L}\,,
\end{equation}
where $\rho(D_\rmn{L})$ is the volume mass density, so that the
differential area number density reads
\begin{equation}
\rmn{d}n = \frac{\rho(D_\rmn{L})}{M}\,\Phi_M(M)\,
\rmn{d}M\,\rmn{d}D_\rmn{L}\,,
\label{eq:defmassspec}
\end{equation}
and $w_0$ is a dimensionless factor representing a
characteristic width that defines the range of impact parameters for which
a microlensing event is considered to occur.
Commonly, an 'event' is defined to take place if the source happens to be
magnified by more than an adopted threshold value 
$A_\rmn{T}$, i.e.\ $A \geq A_\rmn{T}$,
where the choice $A_\rmn{T} = 3/\sqrt{5} 
\approx 1.34$ corresponds to $u_\rmn{T} = 1$, according to Eq.~(\ref{eq:magnification}),
which means that the source passes within the angular
Einstein radius of the lens, and therefore $w_0 = 2$.

Instead of $D_\rmn{L}$, let us use the dimensionless fractional
distance $x \equiv D_\rmn{L}/D_\rmn{S}$, which is distributed 
as $\Phi_x(x) = D_\rmn{S}\,\rho(x D_\rmn{S})/\Sigma$, with
$\Sigma = \int_0^{D_\rmn{S}} \rho(D_\rmn{L})\,\rmn{d}D_\rmn{L}$
being the total surface mass density. Let us further assume that
the mass spectrum $\Phi_M(M)$ is not spatially-dependent and involves
a minimal mass $M_\rmn{min}$ and a maximal mass $M_\rmn{max}$,
while the velocity distribution $\Phi_v(v,x)$ depends on the lens
(and source) distance. With these definitions and assumptions,
the event rate reads
\begin{eqnarray}
& & 
\hspace*{-2.2em}
\Gamma =
w_0\,\sqrt{\frac{4G}{c^2}}\,D_\rmn{S}^{1/2}\,\Sigma\,
\left(\int\limits_{M_\rmn{min}}^{M_\rmn{max}} \frac{\Phi_M(M)}{\sqrt{M}}\,
\rmn{d}M\right) \; \times \nonumber \\
& & \hspace*{-1.7em} \times\;\left(
\int\limits_0^1 \int\limits_0^{\infty} 
v\,\Phi_v(v,x)\,\rmn{d}v\,\sqrt{x(1-x)}\,\Phi_x(x)\,
\rmn{d}x\right)\,.
\end{eqnarray}
As discussed in detail in Appendix~\ref{sec:general}, the 
corresponding weight function
\begin{eqnarray}
\Omega(M,v,x) & = & 
w_0\,\frac{\Sigma}{M}\,v\,r_\rmn{E}(D_\rmn{S}, M, x) \nonumber \\
& = &
w_0\,\sqrt{\frac{4G}{c^2}}\,D_\rmn{S}^{1/2}\,\frac{\Sigma}{\sqrt{M}}\,
v\,\sqrt{x(1-x)}
\end{eqnarray}
for the 
basic system properties $\vec a = (M, v, x)$
provides probability densities for any lens property
that can be expressed as function of the basic properties by means
of Bayes' theorem.

By introducing a dimensionless velocity parameter
$\zeta = v/v_\rmn{c}$, where $v_\rmn{c}$ denotes a characteristic velocity,
and with
\begin{equation}
r_{\rmn{E},\sun} = \sqrt{\frac{GM_{\sun}}{c^2}\,D_\rmn{S}}
\label{eq:defrEsun}
\end{equation}
being the Einstein radius of a solar-mass lens located
half-way between observer and source ($x=0.5$), the event rate can be
written as
\begin{eqnarray}
& & 
\hspace*{-2.2em}
\Gamma = \Gamma_0\,
\left(\int\limits_{M_\rmn{min}/M_{\sun}}^{M_\rmn{max}/M_{\sun}} 
\frac{\Phi_{M/M_{\sun}}(M/M_{\sun})}{\sqrt{M/M_{\sun}}}\,
\rmn{d}(M/M_{\sun})\right) \; \times \nonumber \\
& & \hspace*{-1.7em} \times\;\left(
\int\limits_0^1 \int\limits_0^{\infty} 
\zeta\,\Phi_\zeta(\zeta,x)\,\rmn{d}\zeta\,\sqrt{x(1-x)}\,\Phi_x(x)\,
\rmn{d}x\right)\,,
\label{eq:evratenormal}
\end{eqnarray}
with $\Gamma_0 = 2\,w_0\,r_{\rmn{E},\sun}\,v_\rmn{c}\,\Sigma$ and
the dimensionless distributions $\Phi_\zeta(\zeta) = v_\rmn{c}\,\Phi_v(v_\rmn{c}\,\zeta)$ and
$\Phi_{M/M_{\sun}}(M/M_{\sun}) = M_{\sun}\,\Phi_M(M)$.
With the definition of $\Gamma_0$, the weight function takes the form
\begin{equation}
\Omega(M,\zeta,x) = \Gamma_0\,\left(M/M_{\sun}\right)^{-1/2}\,\zeta\,\sqrt{x(1-x)}\,.
\end{equation}

\section{Distribution of properties for the ensemble of events}
\label{sec:timescale}

For ordinary microlensing light curves, i.e.\ those that can be approximated
by lensing of a point-like source star by a single point-mass lens and
uniform motion of the lens relative to the line-of-sight from the observer to the
source,
the only observable that is related to the physical parameters of the system
is the time-scale
\begin{equation}
t_\rmn{E} = \theta_\rmn{E}/\mu_\rmn{LS} =
\frac{1}{v}\,\sqrt{\frac{4GM}{c^2}\,D_\rmn{S}\,x(1-x)}\,,
\end{equation}
which thus involves all the basic properties $M$, $v$, and $x$.

For an obtained best-fit estimate $t_\rmn{E}^{(0)}$, let us define 
\begin{equation}
\eta_{t_\rmn{E}}^{(0)} = \frac{t_\rmn{E}^{(0)}\,v_\rmn{c}}{r_{\rmn{E},\sun}}\,,
\label{eq:deff0}
\end{equation}
where $r_{\rmn{E},\sun}$ is given by Eq.~(\ref{eq:defrEsun}).
With $\eta_{t_\rmn{E}}^{(0)}$ depending on the
basic system properties as
$\eta_{t_\rmn{E}}^{(0)} = 2\,\sqrt{M/M_{\sun}}\,\sqrt{x(1-x)}/\zeta$,
Eq.~(\ref{eq:ratedensity}) applied to the expression
for the event rate $\Gamma$
as given by Eq.~(\ref{eq:evratenormal}) yields the corresponding event rate density as
\begin{eqnarray}
& & \hspace*{-2.2em}
\gamma_{\eta_{t_\rmn{E}}}(\eta_{t_\rmn{E}}^{(0)}) =
\frac{4\,\Gamma_0}{[\eta_{t_\rmn{E}}^{(0)}]^3}\,
\int\limits_{M_\rmn{min}/M_{\sun}}^{M_\rmn{max}/M_{\sun}} 
\sqrt{M/M_{\sun}}\; \times \nonumber \\
& & \hspace*{-1.7em} \times \;
\Phi_{M/M_{\sun}}\left(M/M_{\sun}\right)\; \times \nonumber \\
& & \hspace*{-1.7em} \times \;
\int\limits_0^1 \Phi_\zeta\left(2\,\sqrt{M/M_{\sun}}\,
\sqrt{x(1-x)}/\eta_{t_\rmn{E}}^{(0)},x\right)
\; \times \nonumber \\
& & \hspace*{-1.7em} \times \;
[x(1-x)]^{3/2}\,\Phi_x(x)\,\rmn{d}x\,\rmn{d}(M/M_{\sun})\,,
\label{eq:tEdensity}
\end{eqnarray}
while $\gamma_{t_\rmn{E}}(t_\rmn{E}) = (v_\rmn{c}/r_{\rmn{E},\sun})\,
\gamma_{\eta_{t_\rmn{E}}}[(t_{\rmn{E}}\,v_\rmn{c})/r_{\rmn{E},\sun}]$
is the corresponding density of
$t_\rmn{E}$, and ${\hat \gamma}_{t_\rmn{E}} = \gamma_{t_\rmn{E}}/\Gamma$ gives
the distribution of event time-scales arising from the lens population.\footnote{
Instead of eliminating the integration over $\rmn{d}\zeta$ by means
of the $\delta$-function, one can
alternatively eliminate the integration over $\rmn{d}x$ or
$\rmn{d}(M/M_{\sun})$, which results in
expressions that correspond to just two
of the infinitely many possibilities to transform the remaining integration variables.}

If the lens may belong to one or another population with different mass spectra, mass densities,
and velocity distributions, the event rate density $\gamma_{t_\rmn{E}}(t_\rmn{E}^{(0)})$
for the observed event time-scale $t_\rmn{E}^{(0)}$ provides a means to decide to which 
population the lens objects belongs. Namely, the probability for the lens to be drawn from
each of the populations is proportional to the corresponding event rate density. Since the
event rate density is proportional to the surface mass density along the line-of-sight, conclusions
about the latter can be drawn, e.g.\ a likelihood for a certain surface mass density can be obtained
on the assumption that the lens in the considered event with time-scale $t_\rmn{E}^{(0)}$ (or
possible additional observables) belongs to a chosen population. Such considerations are of special
interest with regard to the mass content of the Galactic halo and the still open question what fraction
of the observed microlensing events in the direction of the Magellanic Clouds is caused by lenses
in the Magellanic Clouds themselves \citep[e.g][]{SahuSahu,Gyuk:SelfLensing,Mancini:SelfLensing}.

\begin{figure}
\includegraphics[width=84mm]{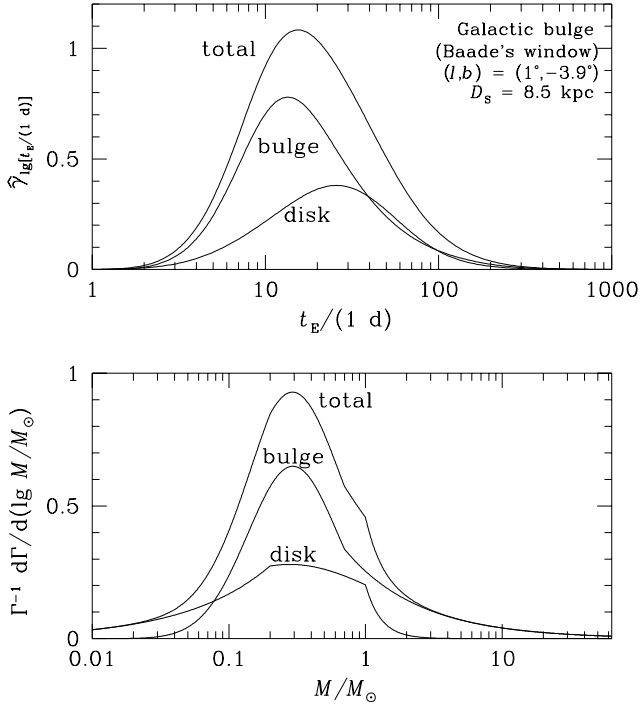}
\caption{Distribution of event time-scales $t_\rmn{E}$ and lens
masses among all created microlensing events for a source located in the 
Galactic bulge at $D_\rmn{S} = 8.5~\rmn{kpc}$ in the direction of Baade's
window $(l,b) = (1\degr,-3.9\degr)$ with the individual contributions of disk and bulge lenses.
Details of the adopted Galaxy model can be found in Appendix~\ref{sec:galmodel}.}
\label{fig:tEdist}
\end{figure}

\begin{figure}
\includegraphics[width=84mm]{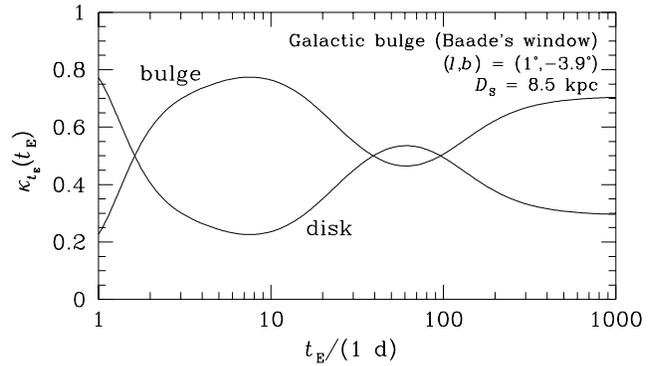}
\caption{Fractional contributions $\kappa_{t_\rmn{E}}^\rmn{disk}$ and $\kappa_{t_\rmn{E}}^\rmn{bulge}$ to the 
event rate density $\gamma_{t_\rmn{E}}(t_\rmn{E})$ as a function of the
time-scale $t_\rmn{E}$ of an observed event. For a few specific $t_\rmn{E}$,
the resulting values of $\kappa_{t_\rmn{E}}^\rmn{disk}$ and $\kappa_{t_\rmn{E}}^\rmn{bulge}$ are listed
in Table~\ref{tab:kapt}. The source is located in the Galactic bulge at $D_\rmn{S} = 8.5~\rmn{kpc}$
in the direction of Baade's window $(l,b) = (1\degr,-3.9\degr)$.}
\label{fig:kapt}
\end{figure}

For a source located in the Galactic bulge, namely in the direction of Baade's window with $(l,b) = 
(1\degr,-3.9\degr)$ at a distance of $D_\rmn{S} = 8.5~\rmn{kpc}$ and the lens residing in the
Galactic disk or bulge, Fig.~\ref{fig:tEdist} shows the distribution of time-scales and lens masses including the 
contributions of the individual lens populations, 
while Fig.~\ref{fig:kapt} shows the fractional contributions $\kappa_{t_\rmn{E}}(t_\rmn{E})$ of disk or bulge
lenses as function of the observed event time-scale $t_\rmn{E}^{(0)}$,
where 
$\kappa_{t_\rmn{E}}^{\rmn{disk}} = \gamma_{t_\rmn{E}}^\rmn{disk}/
(\gamma_{t_\rmn{E}}^\rmn{disk}+\gamma_{t_\rmn{E}}^\rmn{bulge})$, 
$\kappa_{t_\rmn{E}}^{\rmn{bulge}} = 
\gamma_{t_\rmn{E}}^{\rmn{bulge}}/
(\gamma_{t_\rmn{E}}^\rmn{disk}+\gamma_{t_\rmn{E}}^\rmn{bulge})$, and
$\gamma_{t_\rmn{E}}(t_\rmn{E})$ is given by Eq.~(\ref{eq:ratedensity}).
Table~\ref{tab:kapt} lists the fractional contributions $\kappa_{t_\rmn{E}}^{\rmn{disk}}$ and
 $\kappa_{t_\rmn{E}}^{\rmn{bulge}}$ for selected time-scales.
All details of the assumed mass spectra, mass densities, and velocity distributions
for the underlying populations can be found in Appendix~\ref{sec:galmodel}.
Among all created events, 35 per cent are caused by lenses in the Galactic disk and 65 per cent by
lenses in the Galactic bulge. Only for timescales $t_\mathrm{E} \la 2~\rmn{d}$, disk lenses provide a significantly larger contribution than
bulge lenses, whereas the latter dominate for 
$2~\rmn{d} \la t_\rmn{E} \la 40~\rmn{d}$ and for
$t_\rmn{E} \ga 100~\rmn{d}$. For $40~\rmn{d} \la t_\rmn{E}
\la 100~\rmn{d}$, both populations yield comparable contributions.
One finds a median time-scale $t_\rmn{E} \sim 18~\rmn{d}$, or
$t_\rmn{E} \sim 17~\rmn{d}$ for bulge and  
$t_\rmn{E} \sim 24~\rmn{d}$ for disk lenses.  
The distribution of ${\hat \gamma}_{t_\rmn{E}}$ does not properly 
reflect that of the time-scales observed by the experiments,
since their sensitivity for detecting an event depends on the event duration. In particular, a significant
fraction of events with short time-scales is missed with a roughly daily sampling. The median mass is $0.32~M_{\sun}$, with about 1/3 of the events 
caused by lenses more massive than $0.5~M_{\sun}$, and about 1/5 by
lenses heavier than $0.8~M_{\sun}$. 

\begin{table}
\caption{Fractional contributions $\kappa_{t_\rmn{E}}^\rmn{disk}$ and $\kappa_{t_\rmn{E}}^\rmn{bulge}$ of disk
or bulge lenses to the
event rate density $\gamma_{t_\rmn{E}}(t_\rmn{E})$ of a Galactic bulge source at 
$D_\rmn{S} = 8.5~\rmn{kpc}$ in the direction of Baade's window
by lenses in the Galactic disk or
bulge, respectively, for selected event time-scales
$t_\rmn{E}$.}
\label{tab:kapt}
\begin{footnotesize}
\vspace*{1.0ex}
\begin{tabular}{ccc}
\hline
 $t_\rmn{E}/(1~\rmn{d})$ & $\kappa_{t_\rmn{E}}^\rmn{disk}$ & $\kappa_{t_\rmn{E}}^\rmn{bulge}$ \\ \hline
5 & 0.25 & 0.75 \\
10 & 0.24 & 0.76 \\
20 & 0.35 & 0.65 \\
40 & 0.50 & 0.50 \\
80 & 0.52 & 0.48 \\
 \hline
\end{tabular}
\end{footnotesize}

\medskip
$\kappa_{t_\rmn{E}}^{\rmn{disk}} = \gamma_{t_\rmn{E}}^\rmn{disk}/
(\gamma_{t_\rmn{E}}^\rmn{disk}+\gamma_{t_\rmn{E}}^\rmn{bulge})$ and 
$\kappa_{t_\rmn{E}}^{\rmn{bulge}} = \gamma_{t_\rmn{E}}^{\rmn{bulge}}/(\gamma_{t_\rmn{E}}^\rmn{disk}+\gamma_{t_\rmn{E}}^\rmn{bulge})$,
where $\gamma_{t_\rmn{E}}$ is defined by Eq.~(\ref{eq:ratedensity}) and
can be calculated by means of Eq.~(\ref{eq:tEdensity}).
\end{table}

\section{Probability densities of system properties for ordinary events}
\label{sec:probdens}

\subsection{Lens mass}
Let us define $\mu_0 = M/M_0$, where the characteristic mass $M_0$ is assumed
for the velocity $v_\rmn{c}$ and
the lens being located half-way between observer and source ($x=0.5$). Hence,
with $\eta_{t_\rmn{E}}^{(0)}$ as defined by Eq.~(\ref{eq:deff0}),
$M_0 = [\eta_{t_\rmn{E}}^{(0)}]^2\,M_{\sun}$, while
$\mu_0^{\rmn{min}} = [\eta_{t_\rmn{E}}^{(0)}]^{-2} M_\rmn{min}/M_{\sun}$
and $\mu_0^{\rmn{max}} = [\eta_{t_\rmn{E}}^{(0)}]^{-2} M_\rmn{max}/M_{\sun}$.

With $\mu_0$ being related to the basic properties
as $\mu_0 = [\eta_{t_\rmn{E}}^{(0)}]^{-2} (M/M_{\sun})$, one easily finds
with Eqs.~(\ref{eq:pdens0}) and~(\ref{eq:tEdensity})
the probability density of $\mu_0$ for an event with measured $t_\rmn{E}^{(0)}$ 
to be
\begin{eqnarray}
& & \hspace*{-2.2em}p_{\mu_0}^{(0)}
(\mu_0; \eta_{t_\rmn{E}}^{(0)}) = \frac{4\,\Gamma_0}
{\gamma_{\eta_{t_\rmn{E}}}(\eta_{t_\rmn{E}}^{(0)})}\,
\sqrt{\mu_0}\;
\Phi_{M/M_{\sun}}\left(\mu_0 \left[\eta_{t_\rmn{E}}^{(0)}\right]^2\right)
\; \times \nonumber \\
& & \hspace*{-1.7em} \times \;
\int\limits_0^1 \Phi_\zeta\left(2\,\sqrt{\mu_0\,x(1-x)},x\right)\,
[x(1-x)]^{3/2}\,\Phi_x(x)\,\rmn{d}x\,.
\label{eq:pmu}
\end{eqnarray}

\begin{figure}
\includegraphics[width=84mm]{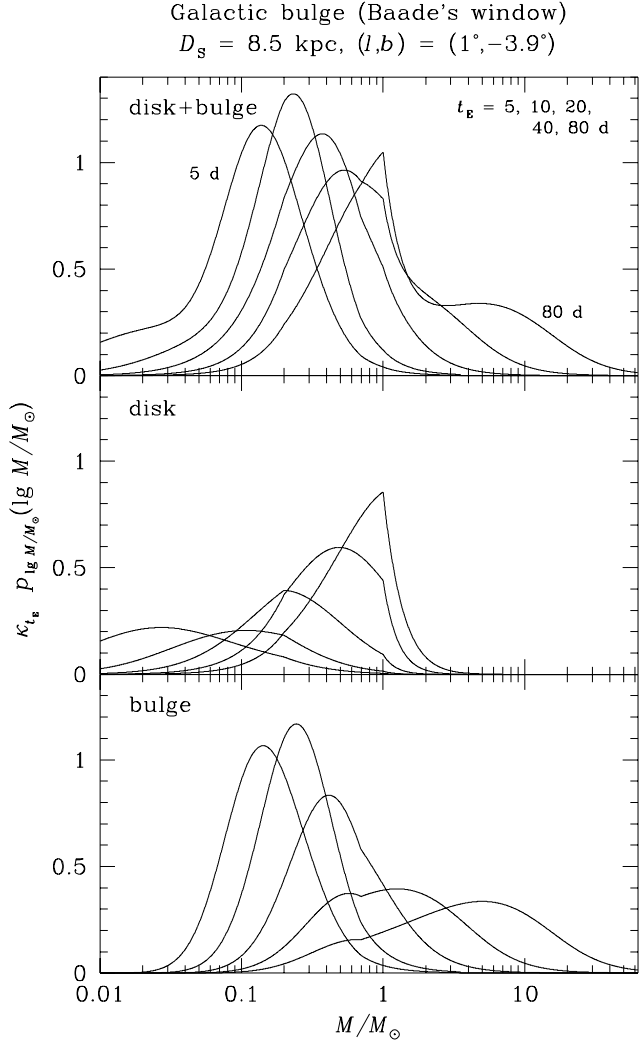}
\caption{Probability density of the lens mass $M$ for selected observed event time-scales
$t_\rmn{E}$. With the event rate density $\gamma_{t_\rmn{E}}(t_\rmn{E})$, the relative
contributions by disk or bulge lenses are $\kappa_{t_\rmn{E}}^{\rmn{disk}} = \gamma_{t_\rmn{E}}^\rmn{disk}/
(\gamma_{t_\rmn{E}}^\rmn{disk}+\gamma_{t_\rmn{E}}^\rmn{bulge})$ or
$\kappa_{t_\rmn{E}}^{\rmn{bulge}} = \gamma_{t_\rmn{E}}^{\rmn{bulge}}/(\gamma_{t_\rmn{E}}^\rmn{disk}+\gamma_{t_\rmn{E}}^\rmn{bulge})$,
respectively. The figures show the contributions $\kappa_{t_\rmn{E}}\;p_{\lg (M/M_{\sun})}$
of each population along with the total probability density 
$p_{\lg (M/M_{\sun})}$ of $\lg (M/M_{\sun})$. For the chosen values of
$t_\rmn{E}$, the corresponding fractional contributions $\kappa_{t_\rmn{E}}(t_\rmn{E})$ are listed in 
Table~\ref{tab:kapt}.}
\label{fig:pmass}
\end{figure}

By means of Eq.~(\ref{eq:pdensgen}),
the distribution for a fuzzy value of $t_\rmn{E}^{(0)}$ then follows with
$p_{\mu_0}(\mu_0; \eta_{t_\rmn{E}}^{(0)})$
being evaluated for every corresponding $\eta_{t_\rmn{E}}^{(0)}$. Frequently, it
is more adequate to represent the lens mass on a logarithmic scale.
The probability density of $\lg (M/M_{\sun})$ simply follows as
\begin{eqnarray}
& & \hspace*{-2.2em}
p_{\lg (M/M_{\sun})}(\lg (M/M_{\sun}),\eta_{t_\rmn{E}}^{(0)}) \nonumber \\
& & \hspace*{-1.7em} =\;
\left[\eta_{t_\rmn{E}}^{(0)}\right]^{-2}\,\frac{M}{M_{\sun}}\,\ln 10\;
p_{\mu_0}\left(\left[\eta_{t_\rmn{E}}^{(0)}\right]^{-2}\,\frac{M}{M_{\sun}}\right)\,.
\label{eq:plgmass}
\end{eqnarray}

For a source located at $D_\rmn{S} = 8.5~\rmn{kpc}$ in the direction of Baade's window with $(l,b) = 
(1\degr,-3.9\degr)$ as used in the previous section  
and the lens residing in the Galactic disk or bulge with the Galaxy model described
in Appendix~\ref{sec:galmodel}, Fig.~\ref{fig:pmass} shows the mass probability density 
$p_{\lg (M/M_{\sun})}$ of $\lg (M/M_{\sun})$ for selected values of the observed event
time-scale $t_\rmn{E}^{(0)}$. In addition to the resulting $p_{\lg (M/M_{\sun})}$ from
both possible lens populations, their individual contributions 
$\kappa_{t_\rmn{E}}^\rmn{disk}\;p^\rmn{disk}_{\lg (M/M_{\sun})}$ and 
$\kappa_{t_\rmn{E}}^\rmn{bulge}\;p^\rmn{bulge}_{\lg (M/M_{\sun})}$ are shown,
where the factors are determined 
by the event rate density $\gamma_{t_\rmn{E}}(t_\rmn{E})$ as
$\kappa_{t_\rmn{E}}^{\rmn{disk}} = \gamma_{t_\rmn{E}}^\rmn{disk}/
(\gamma_{t_\rmn{E}}^\rmn{disk}+\gamma_{t_\rmn{E}}^\rmn{bulge})$ and 
$\kappa_{t_\rmn{E}}^{\rmn{bulge}} = 
\gamma_{t_\rmn{E}}^{\rmn{bulge}}/(\gamma_{t_\rmn{E}}^\rmn{disk}+
\gamma_{t_\rmn{E}}^\rmn{bulge})$.
The fractional contributions $\kappa_{t_\rmn{E}}(t_\rmn{E}^{(0)})$ for the chosen values 
of $t_\rmn{E}^{(0)}$ are also listed in Table~\ref{tab:kapt}. If the uncertainty in
$t_\rmn{E}$ is less than 20 per cent, it does not have a significant effect on the probability density.

\begin{table*}
\begin{minipage}{160mm}
\caption{Expectation value and standard deviation for $\lg (M/M_{\sun})$, $x$, and
$\lg \zeta$
for selected values of the event time-scale $t_\rmn{E}$.}
\label{tab:maverage}
\vspace*{1.0ex}
\begin{tabular}{ccccccccccc}
\hline
 $t_\rmn{E}/(1~\rmn{d})$ &  $\left\langle \lg (M/M_{\sun})\right\rangle$ &
 $\sigma_{\lg (M/M_{\sun})}$ &
${\hat M}/M_{\sun}$ &  $\varsigma_{M/{\hat M}}$ & $\left\langle x\right\rangle$
 & $\sigma_x$ & 
 $\left\langle \lg \zeta\right\rangle$ &
 $\sigma_{\lg \zeta}$ &
${\hat v}/(100~\rmn{km}\,\rmn{s}^{-1})$ &  $\varsigma_{v/{\hat v}}$ \\ \hline
5 & -0.95 & 0.39 & 0.11 & 2.4 & 0.88 & 0.10 & 0.42 & 0.15 & 260 & 1.4 \\
10 & -0.70 & 0.37 & 0.20 & 2.3 & 0.83 & 0.12 & 0.33 & 0.16 & 216 & 1.4 \\
20 & -0.45 & 0.39 & 0.36 & 2.5 & 0.77 & 0.15 & 0.21 & 0.17 & 161 & 1.5 \\
40 & -0.18 & 0.46 & 0.65 & 2.9 & 0.69 & 0.20 & 0.07 & 0.21 & 117 & 1.6 \\
80 & -0.12 & 0.58 & 1.31 & 3.8 & 0.63 & 0.23 & -0.07 & 0.26 & 84 & 1.8 \\
 \hline
\end{tabular}
\end{minipage}

\medskip
In addition to the expectation value $\left\langle \lg (M/M_{\sun})\right\rangle$
and the standard deviation
$\sigma_{\lg (M/M_{\sun})}$, the corresponding exponentiated values ${\hat M}/M_{\sun} =
\exp_{10}[\left\langle \lg (M/M_{\sun})\right\rangle]$ and
$\varsigma_{M/{\hat M}} = \exp_{10}[\sigma_{\lg (M/M_{\sun})}]$ are listed. Similarly, 
${\hat v} = \exp_{10}[\left\langle \lg \zeta\right\rangle]\,v_\rmn{c}$ and 
$\varsigma_{v/{\hat v}} = \exp_{10}[\sigma_{\lg \zeta}]$ are given. The source has been assumed to
reside in the Galactic bulge at a distance $D_\rmn{S} = 8.5~\rmn{kpc}$ in the direction of
Baade's window $(l,b) = (1\degr, -3.9\degr)$ and the Galaxy model described in Appendix~\ref{sec:galmodel}
has been adopted. None of the listed values changes significantly if a 20 per cent uncertainty in
$t_\rmn{E}$ is considered, where the distributions widen by less than 2 per cent, while mass and 
velocity estimate shift by less than 0.7 per cent, and the fractional distance $x$ shifts by less 
than $0.002$.
\end{table*}

For the previously chosen selected values of $t_\rmn{E}^{(0)}$,
$\left\langle \lg (M/M_{\sun})\right\rangle$
 and  $\sigma_{\lg (M/M_{\sun})}$ and
as well as their exponentiated values are listed
in Table~\ref{tab:maverage}, while the top panel of
Fig.~\ref{fig:average} shows these values as a function
of $t_\rmn{E}$. While a mass $M \sim 0.36~M_{\sun}$ for $t_\rmn{E}^{(0)} = 20~\rmn{d}$
is in rough agreement with estimates using a 'typical' fractional 
lens distance $x$ and transverse velocity $v$, the assumed mass spectrum with a low abundance
for $M \ga 1~M_{\sun}$ forces the expected mass to be more narrowly distributed with
$t_\rmn{E}$ rather than to follow the naive $M \propto t_\rmn{E}^2$ law. In particular,
the mass ${\hat M} = \exp_{10}[\left\langle \lg (M/M_{\sun})\right\rangle]~M_{\sun}$ 
spans only 1.5 decades between $0.09~M_{\sun}$ and $3~M_{\sun}$
for time-scales $3~\rmn{d} \leq t_\rmn{E} \leq 150~\rmn{d}$,
where the inclusion of one standard deviation
extends this range to $0.03~M_{\sun} \ldots 15~M_{\sun}$.

%With a fundamental property of logarithms and the linearity of the expectation value, the expectation value
%of the logarithm of a product of quantities $\xi_i$ separates into the sum of 
%expectation values of the logarithms of the individual quantities, i.e.\
%\begin{equation}
%\left\langle \lg \prod\limits_{i=1}^{k} \xi_i^{\beta_i}\right\rangle = 
%\left\langle \sum\limits_{i=1}^{k} \beta_i\,\lg \xi_i\right\rangle =
%\sum\limits_{i=1}^{k} \beta_i\,\left\langle  \lg \xi_i\right\rangle \,.
%\end{equation}
%Similarly, one finds for the variances
%\begin{equation}
%\rmn{Var}\left(\lg \prod\limits_{i=1}^{k} \xi_i^{\beta_i}\right)
%= \sum\limits_{i=1}^{k} \sum\limits_{j=1}^{k} \beta_i \beta_j\,\rmn{Cov}\left(
%\lg \xi_i,\lg \xi_j\right)\,.
%\end{equation}
%As Eq.~\ref{eq:pmu} shows, the probability density of $\lg \mu_0 = \lg (M/M_0)$ depends on the event
%time-scale $t_\rmn{E}$ only through a shift of the mass spectrum. For events with about-average $t_\rmn{E}$,
%this is a weak dependence, so that for small offsets in $t_\rmn{E}$, $p_{\lg \mu_0}$ remains
%roughly constant along with its moments, so that
%\begin{equation}
%\rmn{Var}\left(\lg \frac{M}{M_{\sun}})\right) \approx \rmn{Var}\left(\lg \frac{M}{M_0}\right) +
%4\,\rmn{Var}\left(\lg \frac{t_\rmn{E}\,v_\rmn{c}}{r_{\rmn{E},\sun}}\right)\,.
%\end{equation}
%With $\rmn{Var}\left[\lg (M/M_0)\right] \approx \rmn{const.}$, one therefore expects an uncertainty in $t_\rmn{E}$
%of 20 per cent causing a widening in the distribution of $\lg (M/M_{\sun})$ by about 1.7 per cent,
%in agreement with the findings from the exact evaluation.

\subsection{Lens distance}
Similarly to the treatment of the lens mass, one finds the
probability density of the fractional lens distance $x$ for
an event with observed $t_\rmn{E}^{(0)}$ 
(and related $\eta_{t_\rmn{E}}^{(0)}$) with Eqs.~(\ref{eq:pdens0}) and~(\ref{eq:tEdensity})
as
\begin{eqnarray}
& & \hspace*{-2.2em}p_{x}^{(0)}
(x; \eta_{t_\rmn{E}}^{(0)}) = 
\frac{4\,\Gamma_0}
{\gamma_{\eta_{t_\rmn{E}}}(\eta_{t_\rmn{E}}^{(0)})}\,
[x(1-x)]^{3/2}\,\Phi_x(x)\; \times
\nonumber \\ 
& & \hspace*{-1.7em} \times \;
\int\limits_{\mu_0^{\rmn{min}}}^{\mu_0^{\rmn{max}}}
\sqrt{\mu_0}\,\Phi_{M/M_{\sun}}\left(
\mu_0 \left[\eta_{t_\rmn{E}}^{(0)}\right]^2\right)\; \times \nonumber \\
& & \hspace*{-1.7em} \times \;
\Phi_\zeta\left(2\,\sqrt{\mu_0\,x(1-x)},x\right)\,
\rmn{d}\mu_0
\label{eq:px1}
\end{eqnarray}
or
\begin{eqnarray}
& & \hspace*{-2.2em}p_{x}^{(0)}
(x; \eta_{t_\rmn{E}}^{(0)}) = 
\frac{4\,\Gamma_0}
{\gamma_{\eta_{t_\rmn{E}}}(\eta_{t_\rmn{E}}^{(0)})\,[\eta_{t_\rmn{E}}^{(0)}]^3}\,
[x(1-x)]^{3/2}\,\Phi_x(x)\; \times
\nonumber \\ 
& & \hspace*{-1.7em} \times \;
\int\limits_{\lg (M_\rmn{min}/M_{\sun})}^{\lg (M_\rmn{max}/M_{\sun})}
(M/M_{\sun})^{3/2}\,\Phi_{M/M_{\sun}}\left(
M/M_{\sun}\right)\; \times \nonumber \\
& & \hspace*{-1.7em} \times \;
\Phi_\zeta\left(2\,\sqrt{M/M_{\sun}}\,
\sqrt{x(1-x)}/\eta_{t_\rmn{E}}^{(0)},x\right)\,
\rmn{d}[\lg(M/M_{\sun})]\,.
\label{eq:px2}
\end{eqnarray}

\begin{figure}
\includegraphics[width=84mm]{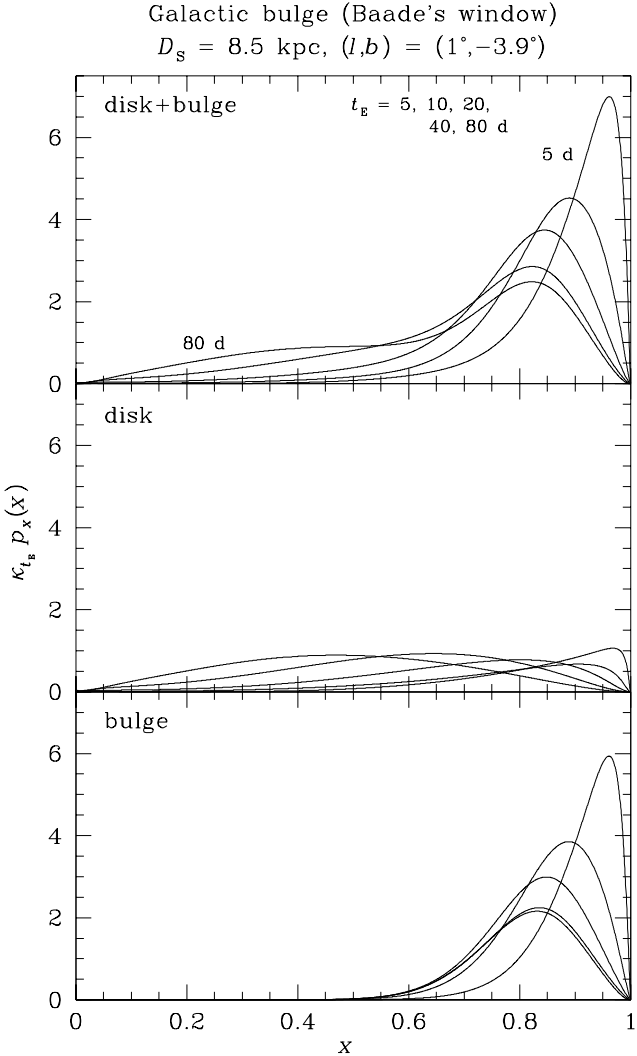}
\caption{Probability density of the fractional lens distance $x$ for selected values of the
 observed event time-scale
$t_\rmn{E}$. Similar to Fig.~(\ref{fig:pmass}), the individual contributions
$\kappa_{t_\rmn{E}}\;p_x$
 of disk and bulge lenses
are shown together with the  total probability density $p_x$, where the relative weight of 
the two lens populations is listed in Table~\ref{tab:kapt} for the chosen values of $t_\rmn{E}$.}
\label{fig:pdist}
\end{figure}

\begin{figure}
\includegraphics[width=84mm]{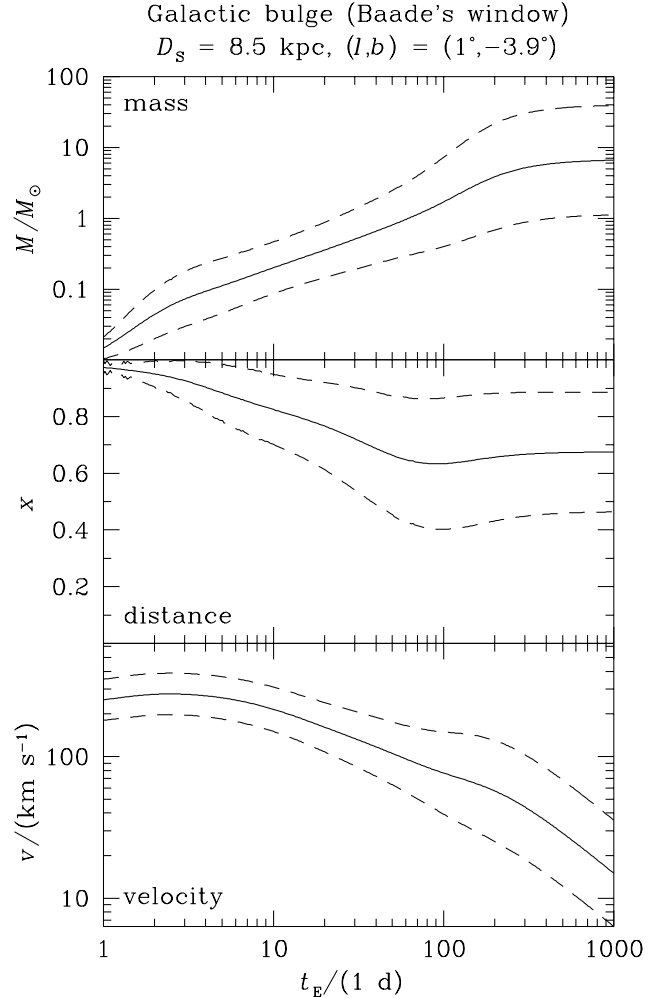}
\caption{Expectation value and standard deviation
for the logarithmic lens mass $\lg (M/M_{\sun})$, the fractional lens distance $x$ and the
logarithmic velocity $\lg \zeta$ as function of the event time-scale $t_\rmn{E}$. 
While the solid lines mark the expectation values, the dashed lines limit intervals
corresponding to the standard deviation. }
\label{fig:average}
\end{figure} 

Fig.~\ref{fig:pdist} shows the probability density $p_x$ of the fractional lens distance for
selected time-scales, while
the expectation value $\left\langle x\right\rangle$
and the standard deviation $\sigma_x$ for different 
$t_\rmn{E}^{(0)}$ are shown in the middle panel of Fig.~\ref{fig:average} as well as
in Table~\ref{tab:maverage}. As before, the source is assumed to be located in the
Galactic bulge at $D_\rmn{S} = 8.5~\rmn{kpc}$ in
the direction of Baade's window, and the Galaxy model described in Appendix~\ref{sec:galmodel} has been
adopted. Shorter time-scales favour larger fractional lens distances, while longer time-scales prefer
the lenses to be closer to the observer, in accordance with the disk population yielding the slightly
larger contribution to the event rate density for 
$40~\rmn{d} \la t_\rmn{E} \la 100~\rmn{d}$, whereas the disk dominates
for smaller time-scales unless $t_\rmn{E} \la 2~\rmn{d}$.

\subsection{Effective velocity and Einstein radius}

\begin{figure}
\includegraphics[width=84mm]{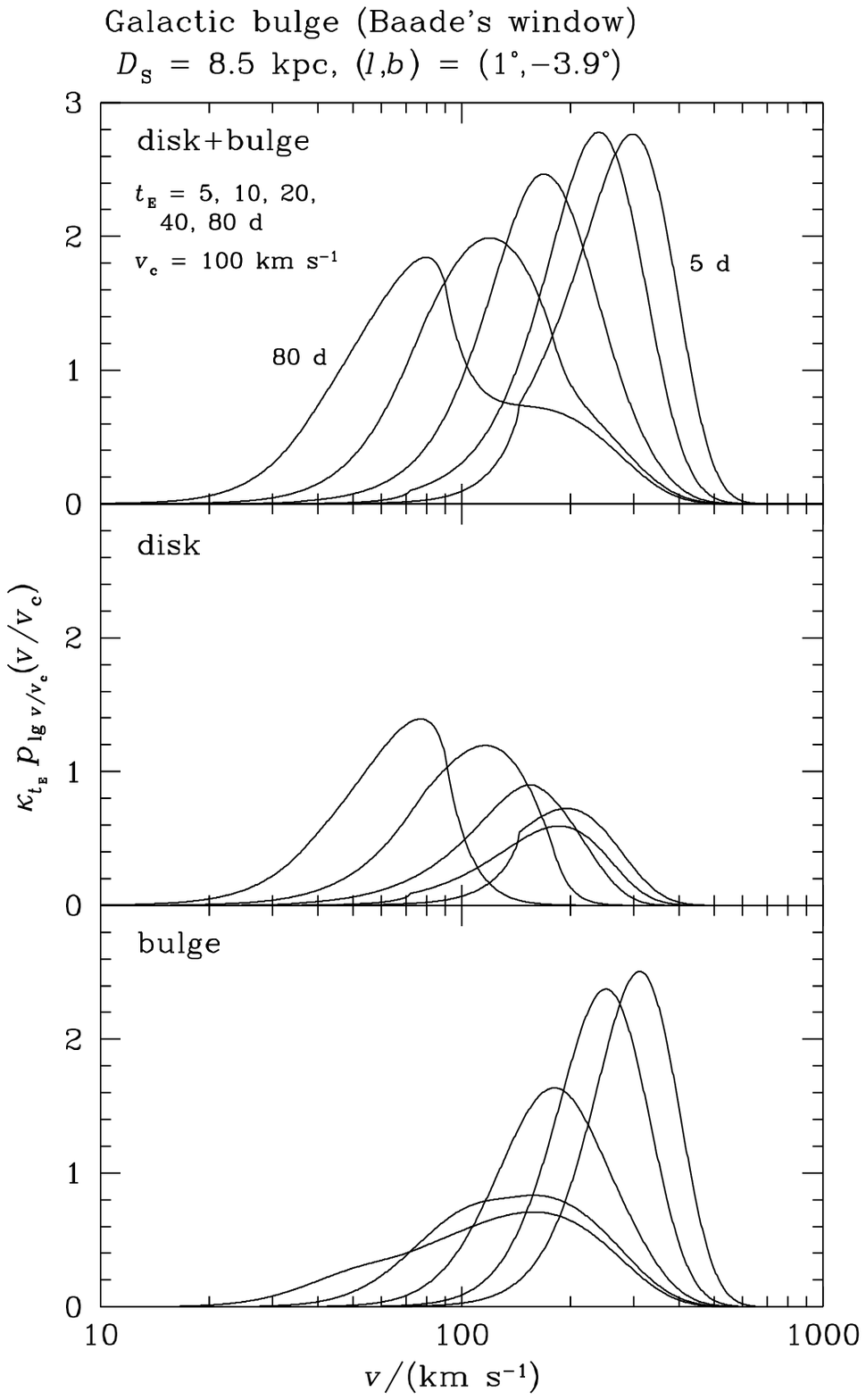}
\caption{
Probability density of the velocity for selected values of the
 observed event time-scale
$t_\rmn{E}$. As for Figs.~\ref{fig:pmass} and~\ref{fig:pdist}, the individual contributions
$\kappa_{t_\rmn{E}}\;p_{\lg \zeta}$
 of disk and bulge lenses, with the corresponding weight factors listed in Table~\ref{tab:kapt},
are shown together with the total probability density $p_{\lg \zeta}$ of $\lg \zeta$,
where $\zeta = v/v_\rmn{c}$ is the dimensionless velocity parameter and 
$v_\rmn{c} = 100~\rmn{km}\,\rmn{s}^{-1}$ has been adopted.}
\label{fig:pvel}
\end{figure}

By eliminating $x$ by means of the delta-function, Eq.~(\ref{eq:pdens0})
yields for the probability density of the velocity parameter $\zeta$ for a
fixed $\eta_{t_\rmn{E}}^{(0)}$
\begin{eqnarray}
& & \hspace*{-2.2em}p_{\zeta}^{(0)}
(\zeta; \eta_{t_\rmn{E}}^{(0)}) = 
\frac{\Gamma_0}
{4\,\gamma_{\eta_{t_\rmn{E}}}(\eta_{t_\rmn{E}}^{(0)})}\;
\zeta^4\; \times \nonumber \\
& & \hspace*{-1.7em} \times\;
\int\limits_{\mu_0^{\rmn{min}}}^{\mu_0^{\rmn{max}}}
\frac{\Theta(\mu_0-\zeta^2)}{\mu_0^2\,\sqrt{1-\zeta^2/\mu_0}}\;
\Phi_{M/M_{\sun}}\left(\mu_0 \left[\eta_{t_\rmn{E}}^{(0)}\right]^2\right)\;\times \nonumber \\
& & \hspace*{-1.7em} \times\;
\sum\limits_{\pm}
\Phi_\zeta\left(\zeta,\frac{1}{2}\left(1\pm
\sqrt{1-\zeta^2/\mu_0}\right)\right)\;\times \nonumber \\
& & \hspace*{-1.7em} \times\;
\Phi_x\left(\frac{1}{2}\left(1\pm
\sqrt{1-\zeta^2/\mu_0}\right)\right)\,\rmn{d}\mu_0\,.
\label{eq:pzeta}
\end{eqnarray}
With ${\hat x} = \sqrt{1-\zeta^2/\mu_0}$, one also finds
equivalently
\begin{eqnarray}
& & \hspace*{-2.2em}p_{\zeta}^{(0)}
(\zeta; \eta_{t_\rmn{E}}^{(0)}) = \frac{\Gamma_0}
{2\,\gamma_{\eta_{t_\rmn{E}}}(\eta_{t_\rmn{E}}^{(0)})}\;\zeta^2\,
\Theta(\mu_0^{\rmn{max}} - \zeta^2)\;\times \nonumber \\
& & \hspace*{-1.7em} \times\;
\int\limits_{{\hat x}^\rmn{min}}^{{\hat x}^\rmn{max}}
\Phi_{M/M_{\sun}}\left(\frac{\zeta^2}{1-{\hat x}^2}\,\left[\eta_{t_\rmn{E}}^{(0)}\right]^2\right)\;\times \nonumber \\
& & \hspace*{-1.7em} \times\;
\sum\limits_{\pm}
\Phi_\zeta\left(\zeta,\frac{1}{2}\left(1\pm \hat x\right)
\right)\,
\Phi_x\left(\frac{1}{2}\left(1\pm \hat x
\right)\right)\,\rmn{d}{\hat x}\,,
\label{eq:pzeta2}
\end{eqnarray}
where the integration limits are given by
\begin{eqnarray}
{\hat x}^\rmn{min} & = & \left\{\begin{array}{lcl}
\sqrt{1-\zeta^2/\mu_0^\rmn{min}} & \rmn{for} & \zeta < \mu_0^\rmn{min} \\
0 & \rmn{for} & \zeta \geq \mu_0^\rmn{min}
\end{array}\right.\,, \nonumber \\
{\hat x}^\rmn{max} & = & 
\left\{\begin{array}{lcl}
\sqrt{1-\zeta^2/\mu_0^\rmn{max}} & \rmn{for} & \zeta < \mu_0^\rmn{max} \\
0 & \rmn{for} & \zeta \geq \mu_0^\rmn{max}
\end{array}\right.\,.
\end{eqnarray}
 
Since $r_\rmn{E} = t_\rmn{E}\,v$, the distribution of the Einstein
radius $r_\rmn{E}$ follows that of the velocity $v$ for any value
of the event time-scale $t_\rmn{E}$. More precisely, if one
defines $r_{\rmn{E},0} = \eta_{t_\rmn{E}}^{(0)}\,r_{\rmn{E},\sun}$ as the Einstein
radius of the 'typical' mass $M_0$, corresponding to $v = v_\rmn{c}$ and
$x = 0.5$, one finds that 
$\rho_\rmn{E} \equiv r_\rmn{E}/r_{\rmn{E},0} = \zeta$, so
that $p_{\rho_\rmn{E}}
(\rho_\rmn{E}) = p_{\zeta}
(\zeta)$.

As for the lens mass $M$ and the fractional lens distance $x \equiv D_\rmn{L}/D_\rmn{S}$, expectation values
and standard deviations for the transverse 
velocity $v = D_\rmn{L}\,\mu = \zeta\,v_\rmn{c}$ at the lens distance
are shown in Table~\ref{tab:maverage} for selected time-scales $t_\rmn{E}$, whereas 
Fig.~\ref{fig:pvel} shows the corresponding
probability densities. As before,
the source has been assumed to be located in the direction of Baade's window,
at a distance $D_\rmn{S} = 8.5~\rmn{kpc}$. The expectation value of $\lg \zeta$ as well as its
uncertainty as function of the time-scale $t_\rmn{E}$ is also displayed in the lower panel
of Fig.~\ref{fig:average}.

\section{Constraints from parallax and finite-source effects}
\label{sec:anomconstr}

\subsection{Model parameters providing further information}
Further information about the lens mass $M$, the fractional lens distance $x$,
and the effective velocity parameter $\zeta$ exceeding that provided by the event time-scale
$t_\rmn{E}$ can be obtained from events whose light curves
are significantly affected by either the annual Earth's motion around the
Sun or the finite size of the observed source star or even both of these
effects. Either of these provides an additional relation between $(M,x,\zeta)$
from a model parameter that relates the Einstein radius $r_\rmn{E}(M,x,\zeta)$ 
to a physical scale which is either the Earth's orbital radius of $1~\rmn{AU}$
or the radius $R_\star$ of the source star. 

Moreover, for a binary lens, the total mass arises from the period $P$ and the
semi-major axis $a$, so that, as discussed by \citet{Do:Rotate},
further information about the lens properties
arises from the lens orbital motion. However, it is quite difficult to obtain
reliable measurements of the full set of orbital elements in order to
determine the period $P$ and the parameter $\rho = a/r_\rmn{E}$,
which would provide a relation between $M$, $x$, and $\zeta$.
As pointed out in the discussion of the event MACHO 1997-BLG-41 by
\citet{PLANET:M41}, the lowest-order effects can be attributed to 
the actual projected differential velocity between the components,
which restricts only a subspace with two
measured model parameters, while leaving another three undetermined.
This strongly limits the power to constrain the lens and source properties. 
A proper discussion would be quite sophisticated and needs to be tailored
to specific cases, so that it exceeds the scope of this paper.

If the light curve is significantly affected by annual parallax resulting 
from the revolution of the Earth around the Sun,
one can determine $\pi_\rmn{E} = \pi_\rmn{LS}/
\theta_\rmn{E}$ as a model parameter. In analogy to the definition of
$\eta_{t_\rmn{E}}^{(0)}$ by Eq.~(\ref{eq:deff0}), a corresponding
dimensionless parameter reads
\begin{eqnarray}
\eta_{\pi_\rmn{E}^{(0)}} & = & \frac{2\,\pi_\rmn{E}^{(0)}\,r_{\rmn{E},\sun}}{1~\rmn{AU}} 
\nonumber \\
& \approx & 2.85\,\pi_\rmn{E}^{(0)}\,\left(\frac{D_\rmn{S}}{1~\rmn{kpc}}\right)^{1/2}\,.
\label{eq:defg0}
\end{eqnarray}
Similarly, if the finite size of the source star has a significant effect on the
microlensing light curve, the time-scale
$t_\star = t_\rmn{E}\,[R_\star/(D_\rmn{S}\,\theta_\rmn{E})]$, in which the
source moves by its own angular radius relative to the lens, can be determined
as additional model parameter. Alternatively, one might use a source
size parameter $\rho_\star = t_\star/t_\rmn{E}$ instead.
A corresponding 
dimensionless parameter can be defined as
\begin{eqnarray}
\eta_{t_\star^{(0)}} & = & \frac{2\,t_\star^{(0)}\,r_{\rmn{E},\sun}}
{t_\rmn{E}^{(0)}\,R_\star} 
\nonumber \\
& \approx & 613\, {\hat \rho}_{\star,\sun}\,
\left(\frac{D_\rmn{S}}{1~\rmn{kpc}}\right)^{1/2}\,,
\label{eq:defh0}
\end{eqnarray}
with the convenient
abbreviation ${\hat \rho}_{\star,\sun} = (t_\star/t_\rmn{E})\,(R_\star/R_{\sun})^{-1}$,
where $R_{\sun} \approx 6.96\times 10^{5}~\rmn{km}$ denotes the solar radius.

\subsection{Parallax}

\begin{figure}
\includegraphics[width=84mm]{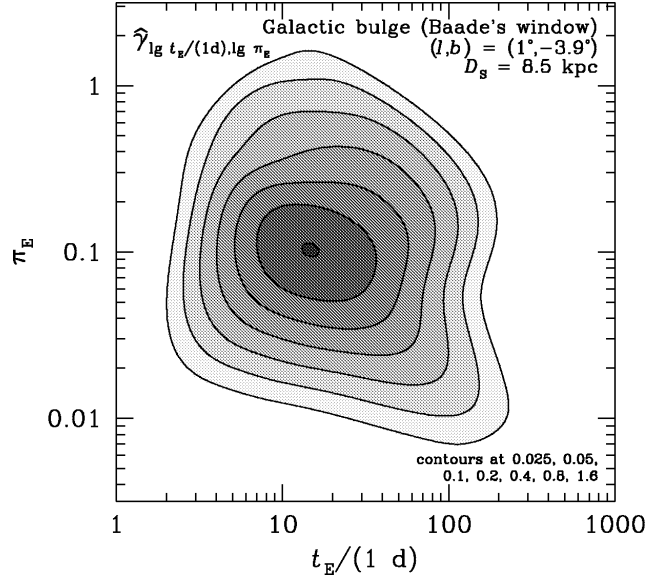}
\caption{Bivariate distribution of the event rate with the time-scale $t_\rmn{E}$ and the
parallax parameter $\pi_\rmn{E} = \pi_\rmn{LS}/\theta_\rmn{E}$. For a bulge source at
$D_\rmn{S} = 8.5~\rmn{kpc}$ in the direction of Baade's window $(l,b) = (1\degr, -3.9\degr)$
and the Galaxy model described in Appendix~\ref{sec:galmodel}, 
contours of ${\hat \gamma}_{\lg [t_\rmn{E}/(1d)],\lg \pi_\rmn{E}}$ are shown at
the levels $0.025$, $0.05$, $0.1$, $0.2$, $0.4$, $0.8$, and $1.6$. }
\label{fig:tpdens}
\end{figure}

From the expression for the event rate density for an observed $t_\rmn{E}^{(0)}$ and
the related $\eta_{t_\rmn{E}}^{(0)}$ as given by Eq.~(\ref{eq:tEdensity}), one finds 
with the additional constraint 
$\delta[\eta_{\pi_\rmn{E}}^{(0)}-\sqrt{(1-x)/x}\,(M/M_{\sun})^{-1/2}]$ 
the event rate density in both $\eta_{t_\rmn{E}}^{(0)}$
and $\eta_{\pi_\rmn{E}}^{(0)}$ to be
\begin{eqnarray}
& & \hspace*{-2.2em}\gamma_{\eta_{t_\rmn{E}},\eta_{\pi_\rmn{E}}}
(\eta_{t_\rmn{E}}^{(0)}, \eta_{\pi_\rmn{E}}^{(0)}) = 
8\,\Gamma_0\,\frac{[\eta_{\pi_\rmn{E}}^{(0)}]^4}{[\eta_{t_\rmn{E}}^{(0)}]^3}\; \times \nonumber \\
& & \hspace*{-1.7em} \times \;
\int\limits_{M_\rmn{min}/M_{\sun}}^{M_\rmn{max}/M_{\sun}} 
\frac{(M/M_{\sun})^3}
{\left\{1+
[\eta_{\pi_\rmn{E}}^{(0)}]^2\,(M/M_{\sun})\right\}^5}\; \times
\nonumber \\
& & \hspace*{-1.7em} \times \;
\Phi_{M/M_{\sun}}\left(\mu_0 \left[\eta_{t_\rmn{E}}^{(0)}\right]^2\right)
\; \times \nonumber \\
& & \hspace*{-1.7em} \times \;
\Phi_\zeta\left(
\frac{2\,(\eta_{\pi_\rmn{E}}^{(0)}/\eta_{t_\rmn{E}}^{(0)})\,
(M/M_{\sun})}{1+[\eta_{\pi_\rmn{E}}^{(0)}]^2\,(M/M_{\sun})},
\frac{1}{1+[\eta_{\pi_\rmn{E}}^{(0)}]^2\,(M/M_{\sun})}\right)\; \times
\nonumber \\
& & \hspace*{-1.7em} \times \;
\Phi_x\left(\frac{1}{1+[\eta_{\pi_\rmn{E}}^{(0)}]^2\,(M/M_{\sun})}\right)\,
\rmn{d}(M/M_{\sun})\,,
\label{eq:normfg}
\end{eqnarray}
while
\begin{eqnarray}
& & \hspace*{-2.2em}
\gamma_{t_\rmn{E},\pi_\rmn{E}}
(t_\rmn{E}^{(0)},\pi_\rmn{E}^{(0)}) \nonumber \\
& & \hspace*{-1.7em}
= \frac{2\,v_\rmn{c}}{1~\rmn{AU}}\;
\gamma_{\eta_{t_\rmn{E}},\eta_{\pi_\rmn{E}}}\left(
\frac{v_\rmn{c}}{r_{\rmn{E},\sun}}\,
t_\rmn{E}^{(0)}, 2\,\frac{r_\rmn{E},\sun}{1~\rmn{AU}}\,\pi_\rmn{E}^{(0)}\right)\,.
\label{eq:ratefg}
\end{eqnarray}
Hence, the bivariate distribution of the time-scale $t_\rmn{E}$ and the
parallax parameter $\pi_\rmn{E}$ is given by ${\hat \gamma}_{t_\rmn{E},\pi_\rmn{E}}
= \gamma_{t_\rmn{E},\pi_\rmn{E}}/\Gamma$, which is shown in Fig.~\ref{fig:tpdens}.

\begin{figure}
\includegraphics[width=84mm]{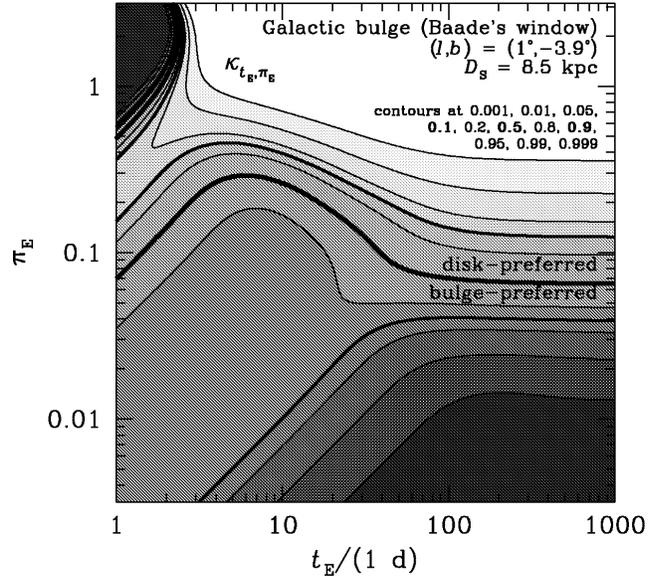}
\caption{Fractional contribution $\kappa_{t_\rmn{E},\pi_\rmn{E}}^{\rmn{disk}}$
of disk lenses and  $\kappa_{t_\rmn{E},\pi_\rmn{E}}^{\rmn{bulge}}$ of bulge
lenses to the total event rate density $\gamma_{t_\rmn{E},\pi_\rmn{E}}$ as a function 
of the time-scale $t_\rmn{E}$ and the parallax parameter $\pi_\rmn{E}$ for 
a Galactic bulge source at $D_\rmn{S} = 8.5~\rmn{kpc}$ towards Baade's window.
The set of contour levels at 0.001, 0.01, 0.05, 0.1, 0.2, 0.5, 0.8, 0.9, 0.95, 0.99, and
0.999 corresponds to both quantities, since $\kappa_{t_\rmn{E},\pi_\rmn{E}}^{\rmn{disk}}
+ \kappa_{t_\rmn{E},\pi_\rmn{E}}^{\rmn{bulge}} = 1$. The contour at 0.5 is shown in bold,
while the contours at 0.1 and 0.9 are shown in light bold. }
\label{fig:kapfg}
\end{figure}

\begin{table}
\caption{Fractional contributions 
$\kappa_{t_\rmn{E},\pi_\rmn{E}}^\rmn{disk}$ and 
$\kappa_{t_\rmn{E},\pi_\rmn{E}}^\rmn{bulge}$ of disk
or bulge lenses to the
event rate density $\gamma_{t_\rmn{E},\pi_\rmn{E}}(t_\rmn{E},\pi_\rmn{E})$ for a Galactic bulge source
by lenses in the Galactic disk or
bulge, respectively, for a typical $t_\rmn{E}$ and selected $\pi_\rmn{E}$.}
\label{tab:kaptp}
\begin{footnotesize}
\vspace*{1.0ex}
\begin{tabular}{cccc}
\hline
 $t_\rmn{E}/(1~\rmn{d})$ & $\pi_\rmn{E}$ & $\kappa_{t_\rmn{E},\pi_\rmn{E}}^\rmn{disk}$ & 
 $\kappa_{t_\rmn{E},\pi_\rmn{E}}^\rmn{bulge}$ \\ \hline
20 & --- & 0.35 & 0.65 \\
   & 0.015 & 0.05 & 0.95 \\
   & 0.06 & 0.19 & 0.81 \\
   & 0.25 & 0.80 & 0.20 \\
   & 1 & 1.0 & $3 \times 10^{-6}$\\
 \hline
\end{tabular}
\end{footnotesize}

\medskip
With the event rate density ${\gamma}_{t_\rmn{E},\pi_\rmn{E}}$ given by Eq.~(\ref{eq:ratefg}),
$\kappa_{t_\rmn{E},\pi_\rmn{E}}^{\rmn{disk}} = {\gamma}_{t_\rmn{E},\pi_\rmn{E}}^\rmn{disk}/
({\gamma}_{t_\rmn{E},\pi_\rmn{E}}^\rmn{disk}+{\gamma}_{t_\rmn{E},\pi_\rmn{E}}^\rmn{bulge})$ and 
$\kappa_{t_\rmn{E},\pi_\rmn{E}}^{\rmn{bulge}} = 
{\gamma}_{t_\rmn{E},\pi_\rmn{E}}^{\rmn{bulge}}
/({\gamma}_{t_\rmn{E},\pi_\rmn{E}}^\rmn{disk}+
{\gamma}_{t_\rmn{E},\pi_\rmn{E}}^\rmn{bulge})$.
\end{table}

The measurement of the parallax parameter $\pi_\rmn{E}^{(0)}$ in addition to the 
event time-scale $t_\rmn{E}^{(0)}$ alters the fractional contributions of the individual lens
populations to the event rate density for these values. The contributions of the Galactic
disk given by
$\kappa_{t_\rmn{E},\pi_\rmn{E}}^{\rmn{disk}} = {\gamma}_{t_\rmn{E},\pi_\rmn{E}}^\rmn{disk}/
({\gamma}_{t_\rmn{E},\pi_\rmn{E}}^\rmn{disk}+{\gamma}_{t_\rmn{E},\pi_\rmn{E}}^\rmn{bulge})$,
while bulge lenses contribute the fraction 
$\kappa_{t_\rmn{E},\pi_\rmn{E}}^{\rmn{bulge}} = 
{\gamma}_{t_\rmn{E},\pi_\rmn{E}}^{\rmn{bulge}}
/({\gamma}_{t_\rmn{E},\pi_\rmn{E}}^\rmn{disk}+
{\gamma}_{t_\rmn{E},\pi_\rmn{E}}^\rmn{bulge})$, 
are shown in Fig.~\ref{fig:kapfg}. The corresponding values for the typical time-scale 
$t_\rmn{E}^{(0)} = 20~\rmn{d}$
and a few different $\pi_\rmn{E}^{(0)}$ are also listed in Table~\ref{tab:kaptp}. Again, 
a bulge source at $D_\rmn{S} = 8.5~\rmn{kpc}$ in the direction of Baade's window has been assumed.
By considering the contours at 0.1 and 0.9 together with the distribution of $t_\rmn{E}$ and
$\pi_\rmn{E}$ as shown in Fig.~\ref{fig:tpdens}, one sees that strong preferences for one of the 
populations are not unlikely to be provided, whereas Fig.~\ref{fig:kapt} shows that from $t_\rmn{E}$
the maximal preference for any time-scale is $\sim 0.8$ in favour of the bulge, achieved for
$t_\rmn{E} \sim 8~\rmn{d}$. For $t_\rmn{E} \ga 2.5~\rmn{d}$, smaller values of
$\pi_\rmn{E}$ favour the lens to reside in the Galactic bulge, while larger $\pi_\rmn{E}$
favour the disk as lens population.
For smaller time-scales, there is an intermediate region where this order is reversed.

From the expression for the event rate density $\gamma_{\eta_{t_\rmn{E}},\eta_{\pi_\rmn{E}}}$
for an event with given
$t_\rmn{E}^{(0)}$ and $\pi_\rmn{E}^{(0)}$, given by Eq.~(\ref{eq:normfg}), one immediately
finds the corresponding probability density of $\mu_0 = M/M_0$ as
\begin{eqnarray}
& & \hspace*{-2.2em}p_{\mu_0}^{(0)}
(\mu_0; \eta_{t_\rmn{E}}^{(0)}, \eta_{\pi_\rmn{E}}^{(0)}) = 
\frac{8\,\Gamma_0}{\gamma_{\eta_{t_\rmn{E}},\eta_{\pi_\rmn{E}}}
(\eta_{t_\rmn{E}}^{(0)},\eta_{\pi_\rmn{E}}^{(0)})}\,[\eta_{t_\rmn{E}}^{(0)}]^5\,
[\eta_{\pi_\rmn{E}}^{(0)}]^4 \; \times \nonumber \\
& & \hspace*{-1.7em} \times \;
\frac{\mu_0^3}
{\left\{1+\mu_0\,
\left[\eta_{t_\rmn{E}}^{(0)}\, \eta_{\pi_\rmn{E}}^{(0)} \right]^2\right\}^5}\;
\Phi_{M/M_{\sun}}\left(\mu_0 \left[\eta_{t_\rmn{E}}^{(0)}\right]^2\right)
\; \times \nonumber \\
& & \hspace*{-1.7em} \times \;
\Phi_\zeta\left(
\frac{2\,\mu_0\,\eta_{t_\rmn{E}}^{(0)}\,\eta_{\pi_\rmn{E}}^{(0)}}{1+\mu_0\,
\left[\eta_{t_\rmn{E}}^{(0)}\,\eta_{\pi_\rmn{E}}^{(0)} \right]^2},
\frac{1}{1+\mu_0\,
\left[\eta_{t_\rmn{E}}^{(0)}\, \eta_{\pi_\rmn{E}}^{(0)} \right]^2}\right)\; \times
\nonumber \\
& & \hspace*{-1.7em} \times \;
\Phi_x\left(\frac{1}{1+\mu_0\,
\left[\eta_{t_\rmn{E}}^{(0)}\, \eta_{\pi_\rmn{E}}^{(0)} \right]^2}\right)\,.
\label{eq:pmufg}
\end{eqnarray}
The probability density of the fractional lens distance $x$ for observed
$t_\rmn{E}^{(0)}$ and $\pi_\rmn{E}^{(0)}$ can 
be obtained with Eq.~(\ref{eq:px1}) by applying the parallax constraint in
the form $\delta[\eta_{\pi_\rmn{E}}^{(0)}-\sqrt{(1-x)/(\mu_0\,x)}/\eta_{t_\rmn{E}}^{(0)}]$, yielding
\begin{eqnarray}
& & \hspace*{-2.2em}
p_{x}^{(0)}(x;\eta_{t_\rmn{E}}^{(0)}, \eta_{\pi_\rmn{E}}^{(0)})
  = 
\frac{8\,\Gamma_0}{\gamma_{\eta_{t_\rmn{E}},\eta_{\pi_\rmn{E}}}
(\eta_{t_\rmn{E}}^{(0)},\eta_{\pi_\rmn{E}}^{(0)})}\,\frac{1}{[\eta_{t_\rmn{E}}^{(0)}]^3\,
[\eta_{\pi_\rmn{E}}^{(0)}]^4} \; \times \nonumber \\  
& & \hspace*{-1.7em} \times \;
  (1-x)^3\;\Phi_{M/M_{\sun}}\left(\left[\eta_{\pi_\rmn{E}}^{(0)}\right]^{-2}\,
\frac{1-x}{x}\right)\; \times \nonumber \\
& & \hspace*{-1.7em} \times \;
\Phi_\zeta\left(\frac{2\,(1-x)}{\eta_{t_\rmn{E}}^{(0)} \eta_{\pi_\rmn{E}}^{(0)}},x\right)\,
\Phi_x(x)\,.
\end{eqnarray}
For deriving $p_{\zeta}^{(0)}(\zeta;
\eta_{t_\rmn{E}}^{(0)},\eta_{\pi_\rmn{E}}^{(0)})$,
one can first eliminate $x$ with the time-scale constraint in Eq.~(\ref{eq:pdens0}) 
in order to obtain Eq.~(\ref{eq:pzeta}) and 
then use the parallax constraint in the form $\delta[\eta_{\pi_\rmn{E}}^{(0)}-
(1\mp\sqrt{1-\zeta^2/\mu_0})/(\eta_{t_\rmn{E}}^{(0)}\,\zeta)]$,
or alternatively first eliminate $\mu_0$ and
then use $\delta[\eta_{\pi_\rmn{E}}^{(0)}-2\,(1-x)/(\zeta\,\eta_{t_\rmn{E}}^{(0)})]$
as parallax constraint. Regardless of the way of approach, one obtains
\begin{eqnarray}
& & \hspace*{-2.2em}p_{\zeta}^{(0)}
(\zeta; \eta_{t_\rmn{E}}^{(0)}, \eta_{\pi_\rmn{E}}^{(0)}) 
  = 
\frac{\Gamma_0}{2\,\gamma_{\eta_{t_\rmn{E}},\eta_{\pi_\rmn{E}}}
(\eta_{t_\rmn{E}}^{(0)},\eta_{\pi_\rmn{E}}^{(0)})}\,\eta_{t_\rmn{E}}^{(0)}\; \times \nonumber \\  
& & \hspace*{-1.7em} \times\; \zeta^3\;
\Theta\left(1-\frac{\zeta\,\eta_{t_\rmn{E}}^{(0)} \eta_{\pi_\rmn{E}}^{(0)}}{2}\right)\; \times
\nonumber \\
& & \hspace*{-1.7em} \times\;
\Phi_{M/M_{\sun}}\left(\frac{\zeta\,\eta_{t_\rmn{E}}^{(0)}}{\eta_{\pi_\rmn{E}}^{(0)}\,
\left(2-\zeta\,\eta_{t_\rmn{E}}^{(0)} \eta_{\pi_\rmn{E}}^{(0)}\right)}\right)\; \times \nonumber \\
& & \hspace*{-1.7em} \times \;
\Phi_{\zeta}\left(\zeta,1-\frac{\zeta\,\eta_{t_\rmn{E}}^{(0)} \eta_{\pi_\rmn{E}}^{(0)}}{2}\right)\,
\Phi_x\left(1-\frac{\zeta\,\eta_{t_\rmn{E}}^{(0)} \eta_{\pi_\rmn{E}}^{(0)}}{2}\right)\,.
\end{eqnarray}

\begin{figure}
\includegraphics[width=84mm]{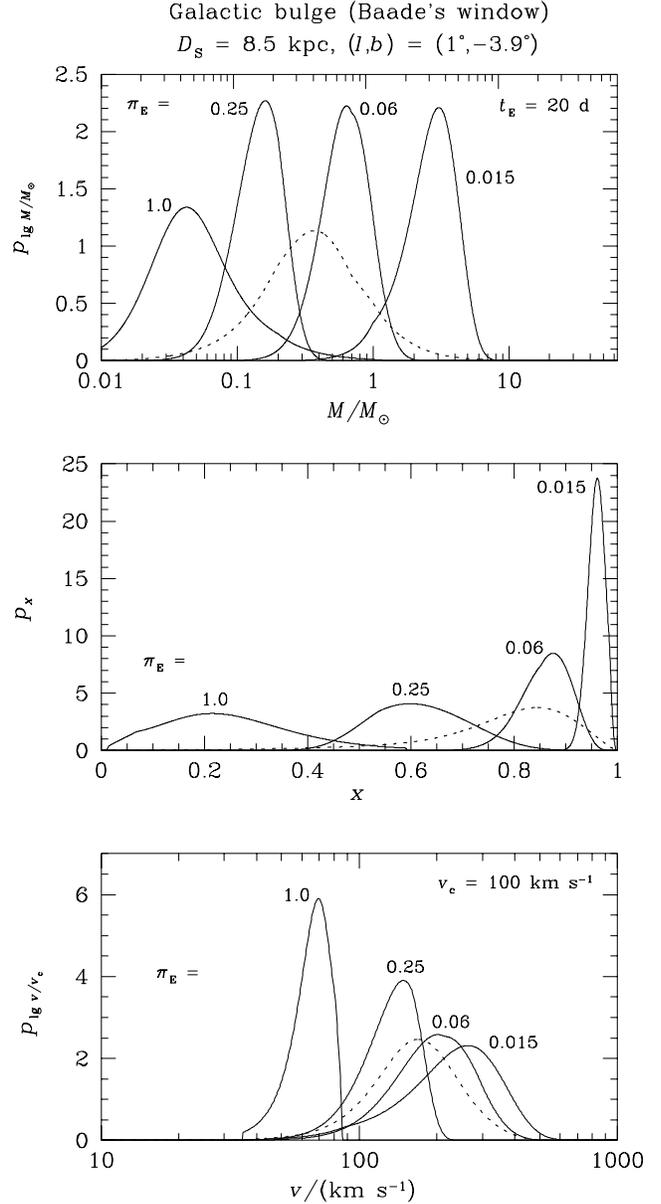}
\caption{Probability densities of $\lg (M/M_{\sun})$, the fractional lens distance $x$ and
$\lg \zeta$, where $\zeta = v/v_\rmn{c}$ and $v_\rmn{c} = 100~\rmn{km}\,\rmn{s}^{-1}$,
for $t_\rmn{E} = 20~\rmn{d}$ and 
$\pi_\rmn{E} = 0.015, 0.06, 0.25, \rmn{or}~1.0$
(solid lines) as well as that for an uncertain $\pi_\rmn{E}$, i.e.\ based
solely on the time-scale $t_\rmn{E}$ (dotted line). The source is thereby located in the Galactic bulge 
at $D_\rmn{S} = 8.5~\rmn{kpc}$ in the direction of Baade's window $(l,b) = (1\degr, -3.9\degr)$.}
\label{fig:distpar}
\end{figure}

\begin{table*}
\begin{minipage}{165mm}
\caption{Expectation value and standard deviation for the lens properties $\lg (M/M_{\sun})$, $x$, and
$\lg \zeta$ for an event with time-scale $t_\rmn{E}$ and parallax parameter $\pi_\rmn{E}$.}
\label{tab:expectpar}
\vspace*{1.0ex}
\begin{tabular}{cccccccccccc}
\hline
 $t_\rmn{E}/(1~\rmn{d})$ &  $\pi_\rmn{E}$ & $\left\langle \lg (M/M_{\sun})\right\rangle$ &
 $\sigma_{\lg (M/M_{\sun})}$ &
${\hat M}/M_{\sun}$ &  $\varsigma_{M/{\hat M}}$ & $\left\langle x\right\rangle$ & $\sigma_x$ & 
 $\left\langle \lg \zeta\right\rangle$ &
 $\sigma_{\lg \zeta}$ &
${\hat v}/(100~\rmn{km}\,\rmn{s}^{-1})$ &  $\varsigma_{v/{\hat v}}$ \\ \hline
20 & --- & -0.45 & 0.39 & 0.36 & 2.5 & 0.77 & 0.15 & 0.21 & 0.17 & 161 & 1.5 \\
  & 0.015 & 0.41 & 0.20 & 2.55 & 1.6 & 0.96 & 0.02 & 0.34 & 0.20 & 219 & 1.6 \\
  & 0.06  & -0.21 & 0.18 & 0.62 & 1.5 & 0.86 & 0.05 & 0.28 & 0.16 & 190 & 1.4 \\
  & 0.25  & -0.85 & 0.18 & 0.14 & 1.5 & 0.62 & 0.09 & 0.11 & 0.12 & 129 & 1.3 \\
  & 1 &    -1.29 & 0.34 & 0.05 & 2.2 & 0.25 & 0.01 & -0.19 & 0.08 & 64 & 1.2 \\
 \hline
\end{tabular}
\end{minipage}

\medskip
The row with $\pi_\rmn{E}$ marked '---' corresponds to an uncertain parallax parameter, i.e.\
the estimate is based solely on $t_\rmn{E}$.
Also listed are the exponentiated values ${\hat M}/M_{\sun} =
\exp_{10}[\left\langle \lg (M/M_{\sun})\right\rangle]$ and
$\varsigma_{M/{\hat M}} = \exp_{10}[\sigma_{\lg (M/M_{\sun})}]$ as well as
${\hat v} = \exp_{10}[\left\langle \lg \zeta\right\rangle]\,v_\rmn{c}$ and 
$\varsigma_{v/{\hat v}} = \exp_{10}[\sigma_{\lg \zeta}]$. The source is located 
in the Galactic bulge at  $D_\rmn{S} = 8.5~\rmn{kpc}$ in the direction of
Baade's window $(l,b) = (1\degr, -3.9\degr)$, and the Galaxy model described in Appendix~\ref{sec:galmodel}
has been adopted.
\end{table*}

For a typical time-scale $t_\rmn{E}^{(0)} = 20~\rmn{d}$ and different values for the 
parallax parameter $\pi_\rmn{E}^{(0)}$, the probability densities $p_{\lg (M/M_{\sun})}$,
$p_x$, and $p_{\lg \zeta}$ are shown in Fig.~\ref{fig:distpar}, whereas expectation values 
and standard deviations for the related quantities are listed in Table~\ref{tab:expectpar}.
For comparison, the previously obtained results for an uncertain $\pi_\rmn{E}$, i.e.\ 
based solely on the measured time-scale $t_\rmn{E}^{(0)}$ are also shown.
In most cases, the measurement of the parallax parameter results in a significant reduction of the
width of the distribution, equivalent to a reduction of the uncertainty of the considered 
lens property, where the mass estimate improves most significantly. 
If, however, the parallax constraint forces the lens
properties to fall into an a-priori disfavoured region, the expectation value is strongly
shifted and the distribution may widen. 
The uncertainty is still
dominated by the mass spectrum, mass density and velocity distributions as compared to the contribution
arising from the finite width of the time-scale distribution. 
 As a result of the finite limits on the lens mass from
the spectrum $\Phi_{M/M_{\sun}}$ and the condition $0 \leq x \leq 1$, the probability densities 
of the lens properties may face sudden cut-offs.

\subsection{Finite source size}

\begin{figure}
\includegraphics[width=84mm]{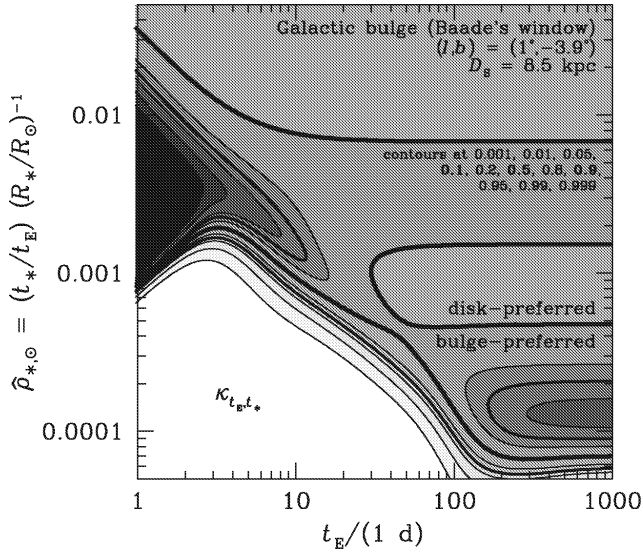}
\caption{Fractional contribution $\kappa_{t_\rmn{E},t_\star}^{\rmn{disk}}$
of disk lenses and  $\kappa_{t_\rmn{E},\star}^{\rmn{bulge}}$ of bulge
lenses to the total event rate density $\gamma_{t_\rmn{E},t_\star}$ as a function 
of the time-scale $t_\rmn{E}$ and the finite-source time-scale
$t_\star = t_\rmn{E}\,[R_\star/(D_\rmn{S}\,\theta_\rmn{E})]$ for 
a Galactic bulge source at $D_\rmn{S} = 8.5~\rmn{kpc}$ towards Baade's window.
With  $\kappa_{t_\rmn{E},\pi_\rmn{E}}^{\rmn{disk}}
+ \kappa_{t_\rmn{E},\pi_\rmn{E}}^{\rmn{bulge}} = 1$, the
set of contour levels at 0.001, 0.01, 0.05, 0.1, 0.2, 0.5, 0.8, 0.9, 0.95, 0.99, and
0.999 corresponds to both quantities. The contour at the level 0.5 is shown in bold,
while light bold has been used for the contours at 0.1 and 0.9.}
\label{fig:kapfh}
\end{figure}

For a finite-source event with observed $t_\rmn{E}^{(0)}$
and $t_\star^{(0)}$ and the related $\eta_{t_\rmn{E}}^{(0)}$ and
$\eta_{t_\star}^{(0)}$ as given by Eqs.~(\ref{eq:deff0}) and~(\ref{eq:defh0}),
the event rate density 
$\gamma_{\eta_{t_\rmn{E}},\eta_{t_\star}}(\eta_{t_\rmn{E}}^{(0)},
\eta_{t_\star}^{(0)})$ results from Eq.~(\ref{eq:tEdensity})
by applying the additional constraint
$\delta[\eta_{t_\star}^{(0)}-\sqrt{x/(1-x)}\,(M/M_{\sun})^{-1/2}]$.
If one compares this with the case of parallax effects, one finds that 
$\eta_{\pi_\rmn{E}}^{(0)}$ assumes the role of
$\eta_{t_\star}^{(0)}$ while $x$ and $1-x$ are interchanged, which is reflected in the result
\begin{eqnarray}
& & \hspace*{-2.2em}\gamma_{\eta_{t_\rmn{E}},\eta_{t_\star}}
(\eta_{t_\rmn{E}}^{(0)}, \eta_{t_\star}^{(0)}) = 
8\,\Gamma_0\,\frac{[\eta_{t_\star}^{(0)}]^4}{[\eta_{t_\rmn{E}}^{(0)}]^3}\; \times \nonumber \\
& & \hspace*{-1.7em} \times \;
\int\limits_{M_\rmn{min}/M_{\sun}}^{M_\rmn{max}/M_{\sun}} 
\frac{(M/M_{\sun})^3}
{\left\{1+
[\eta_{t_\star}^{(0)}]^2\,(M/M_{\sun})\right\}^5}\; \times
\nonumber \\
& & \hspace*{-1.7em} \times \;
\Phi_{M/M_{\sun}}\left(\mu_0 \left[\eta_{t_\rmn{E}}^{(0)}\right]^2\right)
\; \times \nonumber \\
& & \hspace*{-1.7em} \times \;
\Phi_\zeta\left(
\frac{2\,(\eta_{t_\star}^{(0)}/\eta_{t_\rmn{E}}^{(0)})\,
(M/M_{\sun})}{1+[\eta_{t_\star}^{(0)}]^2\,(M/M_{\sun})},
\frac{[\eta_{t_\star}^{(0)}]^2\,(M/M_{\sun})}{1+[\eta_{t_\star}^{(0)}]^2\,(M/M_{\sun})}\right)\; \times
\nonumber \\
& & \hspace*{-1.7em} \times \;
\Phi_x\left(\frac{[\eta_{t_\star}^{(0)}]^2\,(M/M_{\sun})}{1+[\eta_{t_\star}^{(0)}]^2\,(M/M_{\sun})}\right)\,
\rmn{d}(M/M_{\sun})\,.
\label{eq:normfh}
\end{eqnarray}
The event rate density in $(t_\rmn{E}, t_\star)$ follows directly as
\begin{eqnarray}
& & \hspace*{-2.2em}
\gamma_{t_\rmn{E},t_\star}
(t_\rmn{E}^{(0)},t_\star^{(0)}) \nonumber \\
& & \hspace*{-1.7em}
= \frac{2\,v_\rmn{c}}{R_\star\,t_\rmn{E}^{(0)}}\;
\gamma_{\eta_{t_\rmn{E}},\eta_{\pi_\rmn{E}}}\left(
\frac{v_\rmn{c}}{r_{\rmn{E},\sun}}\,
t_\rmn{E}^{(0)}, 2\,\frac{r_{\rmn{E},\sun}}{R_\star}\,\frac{t_\star^{(0)}}{t_\rmn{E}^{(0)}}\right)\,,
\label{eq:ratefh}
\end{eqnarray}
so that with ${\hat \rho}_{\star,\sun} = (t_\star/t_\rmn{E})\,(R_\star/R_{\sun})^{-1}$, one
finds $(t_\rmn{E},{\hat \rho}_{\star,\sun})$ to follow the distribution
\begin{eqnarray}
& & \hspace*{-2.2em}
{\hat \gamma}_{t_\rmn{E},{\hat \rho}_{\star,\sun}}
(t_\rmn{E}^{(0)},{\hat \rho}_{\star,\sun}^{(0)}) \nonumber \\
& & \hspace*{-1.7em}
= \frac{2}{\Gamma}\,\frac{v_\rmn{c}}{R_\star}\;
\gamma_{\eta_{t_\rmn{E}},\eta_{\pi_\rmn{E}}}\left(
\frac{v_\rmn{c}}{r_{\rmn{E},\sun}}\,
t_\rmn{E}^{(0)}, 2\,\frac{r_\rmn{E},\sun}{R_\star}\,{\hat \rho}_{\star,\sun}^{(0)}\right)\,.
\label{eq:densfh}
\end{eqnarray}
Fig.~\ref{fig:kapfh} shows the fractional contribution $\kappa_{t_\rmn{E},t_\star}$
to the event rate density 
for a given $t_\rmn{E}$ and $t_\star$ for the lens residing in either the Galactic bulge or disk.
The measurement of finite-source effects turns out to be
less powerful than that of parallax with most of the likely values not providing 
strong preference for either of the lens populations.
However, small $t_\star/t_\rmn{E}$ 
 Bulge lenses
are preferred for intermediate values $0.0015 \la (t_\star/t_\rmn{E})(R_\star/R_{\sun})^{-1} \la 0.007$,
where a strong 
preference however can only arise for $t_\rmn{E} \la 10~\rmn{d}$.
Measurements of a small $t_\star/t_\rmn{E}$ can provide a very strong
preference for the lens to reside in the disk.

\begin{figure}
\includegraphics[width=84mm]{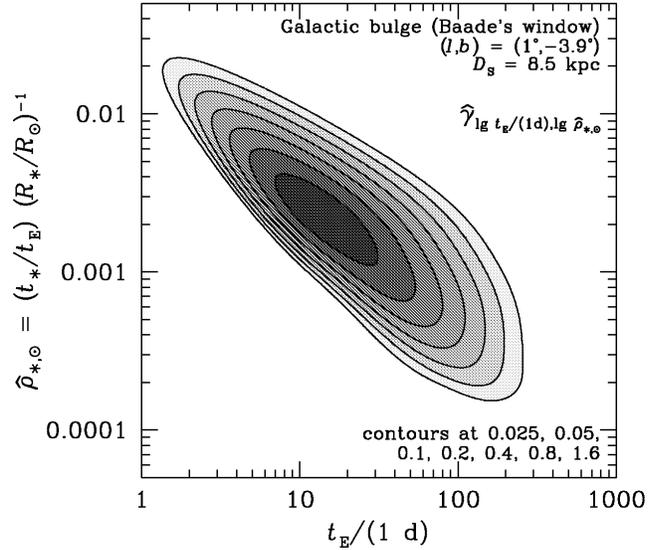}
\caption{Bivariate distribution of the event rate with the time-scale $t_\rmn{E}$ and the
finite-source parameter ${\hat \rho}_{\star,\sun} = (t_\star/t_\rmn{E})\,(R_\star/R_{\sun})^{-1}$. 
The source is located in the Galactic bulge at
$D_\rmn{S} = 8.5~\rmn{kpc}$ in the direction of Baade's window $(l,b) = (1\degr, -3.9\degr)$.
The plot shows contours of ${\hat \gamma}_{\lg [t_\rmn{E}/(1d)],\lg \pi_\rmn{E}}$ correspond to
the levels $0.025$, $0.05$, $0.1$, $0.2$, $0.4$, $0.8$, and $1.6$. }
\label{fig:tfdens}
\end{figure}

\begin{table}
\caption{Fractional contributions 
$\kappa_{t_\rmn{E},t_\star}^\rmn{disk}$ and 
$\kappa_{t_\rmn{E},t_\star}^\rmn{bulge}$ of disk
or bulge lenses to the
event rate density $\gamma_{t_\rmn{E},t_\star}(t_\rmn{E},t_\star)$ for a Galactic bulge source
by lenses in the Galactic disk or
bulge, respectively, for a typical $t_\rmn{E}$ and selected ${\hat \rho}_{\star,\sun}
= (t_\star/t_\rmn{E})\,(R_\star/R_{\sun})^{-1}$.}
\label{tab:kaptf}
\begin{footnotesize}
\vspace*{1.0ex}
\begin{tabular}{cccc}
\hline
 $t_\rmn{E}/(1~\rmn{d})$ & ${\hat \rho}_{\star,\sun}$ & 
 $\kappa_{t_\rmn{E},t_\star}^\rmn{disk}$ & 
 $\kappa_{t_\rmn{E},t_\star}^\rmn{bulge}$ \\ \hline
20 & --- & 0.35 & 0.65 \\
   & 0.0005 & 0.83 & 0.17 \\
   & 0.00125 & 0.35 & 0.65 \\
   & 0.003 & 0.34 & 0.66 \\
   & 0.0075 & 0.52 & 0.48\\
 \hline
\end{tabular}
\end{footnotesize}

\medskip
With the event rate density ${\gamma}_{t_\rmn{E},t_\star}$ given by Eq.~(\ref{eq:ratefh}),
$\kappa_{t_\rmn{E},t_\star}^{\rmn{disk}} = 
{\gamma}_{t_\rmn{E},t_\star}^\rmn{disk}/
({\gamma}_{t_\rmn{E},t_\star}^\rmn{disk}
+{\gamma}_{t_\rmn{E},t_\star}^\rmn{bulge})$ and 
$\kappa_{t_\rmn{E},t_\star}^{\rmn{bulge}} = 
{\gamma}_{t_\rmn{E},t_\star}^{\rmn{bulge}}
/({\gamma}_{t_\rmn{E},t_\star}^\rmn{disk}+
{\gamma}_{t_\rmn{E},t_\star}^\rmn{bulge})$.
\end{table}

From Eq.~(\ref{eq:normfh}), the probability density of $\mu_0$ follows as
\begin{eqnarray}
& & \hspace*{-2.2em}p_{\mu_0}^{(0)}
(\mu_0; \eta_{t_\rmn{E}}^{(0)}, \eta_{t_\star}^{(0)}) = 
\frac{8\,\Gamma_0}{\gamma_{\eta_{t_\rmn{E}},\eta_{t_\star}}
(\eta_{t_\rmn{E}}^{(0)},\eta_{t_\star}^{(0)})}\,[\eta_{t_\rmn{E}}^{(0)}]^5\,
[\eta_{t_\star}^{(0)}]^4 \; \times \nonumber \\
& & \hspace*{-1.7em} \times \;
\frac{\mu_0^3}
{\left\{1+\mu_0\,
\left[\eta_{t_\rmn{E}}^{(0)}\, \eta_{t_\star}^{(0)} \right]^2\right\}^5}\;
\Phi_{M/M_{\sun}}\left(\mu_0 \left[\eta_{t_\rmn{E}}^{(0)}\right]^2\right)
\; \times \nonumber \\
& & \hspace*{-1.7em} \times \;
\Phi_\zeta\left(
\frac{2\,\mu_0\,\eta_{t_\rmn{E}}^{(0)}\,\eta_{t_\star}^{(0)}}{1+\mu_0\,
\left[\eta_{t_\rmn{E}}^{(0)}\,\eta_{t_\star}^{(0)} \right]^2},
\frac{\mu_0\,
\left[\eta_{t_\rmn{E}}^{(0)}\, \eta_{t_\star}^{(0)} \right]^2}{1+\mu_0\,
\left[\eta_{t_\rmn{E}}^{(0)}\, \eta_{t_\star}^{(0)} \right]^2}\right)\; \times
\nonumber \\
& & \hspace*{-1.7em} \times \;
\Phi_x\left(\frac{\mu_0\,
\left[\eta_{t_\rmn{E}}^{(0)}\, \eta_{t_\star}^{(0)} \right]^2}{1+\mu_0\,
\left[\eta_{t_\rmn{E}}^{(0)}\, \eta_{t_\star}^{(0)} \right]^2}\right)\,,
\label{eq:pmufh}
\end{eqnarray}
while for the probability density of $x$, one finds
\begin{eqnarray}
& & \hspace*{-2.2em}
p_{x}^{(0)}(x;\eta_{t_\rmn{E}}^{(0)}, \eta_{t_\star}^{(0)}) =
\frac{8\,\Gamma_0}{\gamma_{\eta_{t_\rmn{E}},\eta_{t_\star}}
(\eta_{t_\rmn{E}}^{(0)},\eta_{t_\star}^{(0)})}\,\frac{1}{[\eta_{t_\rmn{E}}^{(0)}]^3\,
[\eta_{t_\star}^{(0)}]^4} \; \times \nonumber \\  
& & \hspace*{-1.7em} \times \;  
  x^3\;\Phi_{M/M_{\sun}}\left(\left[\eta_{t_\star}^{(0)}\right]^{-2}\,
\frac{x}{1-x}\right)\; \times \nonumber \\
& & \hspace*{-1.7em} \times \;
\Phi_\zeta\left(\frac{2\,x}{\eta_{t_\rmn{E}}^{(0)} \eta_{t_\star}^{(0)}},x\right)\,
\Phi_x(x)\,.
\end{eqnarray}
After elimination of $\mu_0$ using the constraint provided by $t_\rmn{E}^{(0)}$,
the finite-source constraint becomes
$\delta[\eta_{t_\star}^{(0)}-2\,x/(\zeta\,\eta_{t_\rmn{E}}^{(0)})]$,
while the elimination of $x$ yields the constraint  $\delta[\eta_{t_\star}^{(0)}-
(1\pm\sqrt{1-\zeta^2/\mu_0})/(\eta_{t_\rmn{E}}^{(0)}\,\zeta)]$, so that
either with Eq.~(\ref{eq:pzeta}) or directly from Eq.~(\ref{eq:pdens0}), one obtains
the probability density of $\zeta$ for measured $t_\rmn{E}^{(0)}$ and 
$t_\star^{(0)}$ as
\begin{eqnarray}
& & \hspace*{-2.2em}p_{\zeta}^{(0)}
(\zeta; \eta_{t_\rmn{E}}^{(0)}, \eta_{t_\star}^{(0)}) 
= \frac{\Gamma_0}{2\,\gamma_{\eta_{t_\rmn{E}},\eta_{t_\star}}
(\eta_{t_\rmn{E}}^{(0)},\eta_{t_\star}^{(0)})}\,\eta_{t_\rmn{E}}^{(0)}\; \times \nonumber \\  
& & \hspace*{-1.7em} \times\; \zeta^3\;
\Theta\left(\frac{\zeta\,\eta_{t_\rmn{E}}^{(0)} \eta_{t_\star}^{(0)}}{2}\right)
 \times\; \nonumber \\
& & \hspace*{-1.7em} \times\;
\Phi_{M/M_{\sun}}\left(\frac{\zeta\,\eta_{t_\rmn{E}}^{(0)}}{\eta_{t_\star}^{(0)}\,
\left(2-\zeta\,\eta_{t_\rmn{E}}^{(0)} \eta_{t_\star}^{(0)}\right)}\right)\; \times \nonumber \\
& & \hspace*{-1.7em} \times \;
\Phi_{\zeta}\left(\zeta,\frac{\zeta\,\eta_{t_\rmn{E}}^{(0)} \eta_{t_\star}^{(0)}}{2}\right)\,
\Phi_x\left(\frac{\zeta\,\eta_{t_\rmn{E}}^{(0)} \eta_{t_\star}^{(0)}}{2}\right)\,.
\end{eqnarray}

\begin{figure}
\includegraphics[width=84mm]{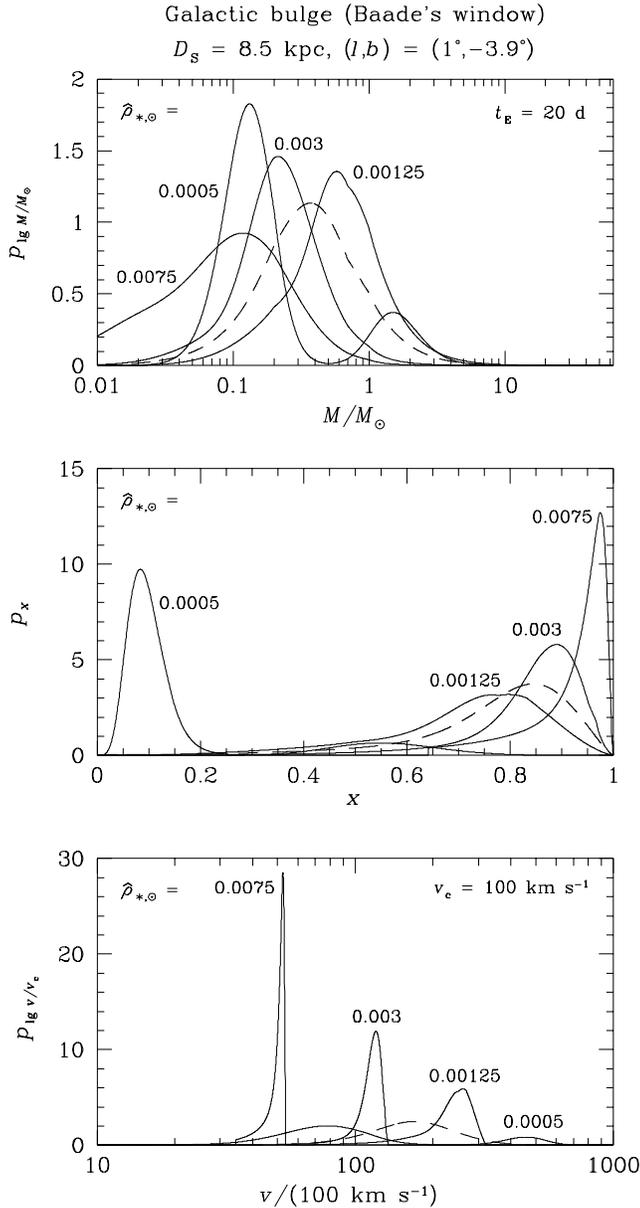}
\caption{Probability densities of $\lg (M/M_{\sun})$, the fractional lens distance $x$ and
$\lg \zeta$, where $\zeta = v/v_\rmn{c}$ and $v_\rmn{c} = 100~\rmn{km}\,\rmn{s}^{-1}$,
for $t_\rmn{E} = 20~\rmn{d}$ and 
${\hat \rho}_{\star,\sun} = (t_\star/t_\rmn{E})\,(R_\star/R_{\sun})^{-1} = 0.0005, 0.00125, 0.003,
\rmn{or}~0.0075$
(solid lines) as well as that for an uncertain $t_\star$, i.e.\ based
solely on the time-scale $t_\rmn{E}$ (dotted line). The source is located in the Galactic bulge 
at $D_\rmn{S} = 8.5~\rmn{kpc}$ in the direction of Baade's window $(l,b) = (1\degr, -3.9\degr)$.}
\label{fig:distsrc}
\end{figure}

\begin{table*}
\begin{minipage}{165mm}
\caption{Expectation value and standard deviation for the lens properties $\lg (M/M_{\sun})$, $x$, and
$\lg \zeta$ for an event with time-scale $t_\rmn{E}$ and finite-source parameter 
${\hat \rho}_{\star,\sun} = (t_\star/t_\rmn{E})\,(R_\star/R_{\sun})^{-1}$.}
\label{tab:expectsrc}
\vspace*{1.0ex}
\begin{tabular}{cccccccccccc}
\hline
 $t_\rmn{E}/(1~\rmn{d})$ &  ${\hat \rho}_{\star,\sun}$ & $\left\langle \lg (M/M_{\sun})\right\rangle$ &
 $\sigma_{\lg (M/M_{\sun})}$ &
${\hat M}/M_{\sun}$ &  $\varsigma_{M/{\hat M}}$ & $\left\langle x\right\rangle$ & $\sigma_x$ & 
 $\left\langle \lg \zeta\right\rangle$ &
 $\sigma_{\lg \zeta}$ &
${\hat v}/(100~\rmn{km}\,\rmn{s}^{-1})$ &  $\varsigma_{v/{\hat v}}$ \\ \hline
20 & --- & -0.45 & 0.39 & 0.36 & 2.5 & 0.77 & 0.15 & 0.21 & 0.17 & 161 & 1.5 \\
  & 0.0005 & -0.71 & 0.46 & 0.19 & 2.9 & 0.18 & 0.18 & 0.00 & 0.33 & 99 & 2.2 \\
  & 0.00125  & -0.26 & 0.36 & 0.56 & 2.3 & 0.71 & 0.16 & 0.35 & 0.12 & 222 & 1.3 \\
  & 0.003  &  -0.67 & 0.33 & 0.22 & 2.1 & 0.84 & 0.11 & 0.04 & 0.07 & 111 & 1.2 \\
  & 0.0075 &   -1.05 & 0.43 & 0.09 & 2.7 & 0.92 & 0.08 & -0.31 & 0.04 & 49 & 1.1 \\
 \hline
\end{tabular}
\end{minipage}

\medskip
For the row with the entry  '---' for ${\hat \rho}_{\star,\sun}$, the estimate is based 
solely on $t_\rmn{E}$, while the finite-source parameter has been considered as uncertain.
In addition to the basic estimates, the exponentiated values ${\hat M}/M_{\sun} =
\exp_{10}[\left\langle \lg (M/M_{\sun})\right\rangle]$ and
$\varsigma_{M/{\hat M}} = \exp_{10}[\sigma_{\lg (M/M_{\sun})}]$ as well as
${\hat v} = \exp_{10}[\left\langle \lg \zeta\right\rangle]\,v_\rmn{c}$ and 
$\varsigma_{v/{\hat v}} = \exp_{10}[\sigma_{\lg \zeta}]$ are listed. The source is located 
in the Galactic bulge at  $D_\rmn{S} = 8.5~\rmn{kpc}$ in the direction of
Baade's window $(l,b) = (1\degr, -3.9\degr)$, and the Galaxy model described in Appendix~\ref{sec:galmodel}
has been adopted.
\end{table*}

Fig.~\ref{fig:distsrc} shows the probability densities of $\lg (M/M_{\sun})$, the fractional
lens distance $x$, and $\lg \zeta$ for a microlensing event on a bulge source at $D_\rmn{S} = 8.5~\rmn{kpc}$
in the direction of Baade's window for which $t_\rmn{E}$ and $t_\star$ have been determined, where
a typical time-scale $t_\rmn{E}^{(0)} = 20~\rmn{d}$ has been assumed, whereas a few different
values for ${\hat \rho}_{\star,\sun}^{(0)}$ have been chosen. For the same values, Table~\ref{tab:kaptf} shows
the fractional contributions of the bulge and disk lenses to the event rate density 
$\gamma_{t_\rmn{E},t_\star}(t_\rmn{E}^{(0)},t_\star^{(0)})$.
While the parallax measurement has been found to provide the most effective reduction of uncertainty 
for the lens mass, one finds that the finite-source parameter ${\hat \rho}_{\star,\sun}$ or the
related time-scale $t_\star$ most significantly affects the uncertainty in the effective transverse
velocity $v = v_\rmn{c}\,\zeta$ or the related Einstein radius $r_\rmn{E} = t_\rmn{E}\,v$.
Some distributions show two peaks corresponding to the bulge and disk population.

\subsection{Combination of parallax and finite-source effects}
If both $\pi_\rmn{E}$ and $t_\star$ are measured, the lens
mass $M$, its fractional distance $x$, and the effective
velocity $v$ are determined, so that
\begin{equation}
p_{\mu_0}^{(0)}(\mu_0; \eta_{t_\rmn{E}}^{(0)}, \eta_{\pi_\rmn{E}}^{(0)},
\eta_{t_\star}^{(0)}) = \delta\left(\mu_0 - \frac{1}{\left[\eta_{t_\rmn{E}}^{(0)}\right]^2\,
\eta_{\pi_\rmn{E}}^{(0)}\,\eta_{t_\star}^{(0)}}\right)\,,
\end{equation}
and the uncertainty in these quantities is solely given
by the distributions of the model parameters
$t_\rmn{E}$, $\pi_\rmn{E}$, and $t_\star$ or the
related dimensionless $\eta_{t_\rmn{E}}$, $\eta_{\pi_\rmn{E}}$, and
$\eta_{t_\star}$, respectively.

As result of a fundamental property of logarithms and the linearity of the expectation value, 
the expectation value
of the logarithm of a product of arbitrary powers of quantities $\xi_i$ separates into the sum of 
multiples of the expectation values of the logarithms of the individual quantities, i.e.\
\begin{equation}
\left\langle \lg \prod\limits_{i=1}^{k} \xi_i^{\beta_i}\right\rangle = 
\left\langle \sum\limits_{i=1}^{k} \beta_i\,\lg \xi_i\right\rangle =
\sum\limits_{i=1}^{k} \beta_i\,\left\langle  \lg \xi_i\right\rangle\,.
\label{eq:logexp}
\end{equation}
Similarly, one finds for the variances
\begin{equation}
\rmn{Var}\left(\lg \prod\limits_{i=1}^{k} \xi_i^{\beta_i}\right)
= \sum\limits_{i=1}^{k} \sum\limits_{j=1}^{k} \beta_i \beta_j\,\rmn{Cov}\left(
\lg \xi_i,\lg \xi_j\right)\,,
\label{eq:logvariance}
\end{equation}
where $\rmn{Cov}(x_i,x_j) = \rmn{Cov}(x_j,x_i)$ denotes the covariance of the 
quantities $x_i$ and $x_j$, and $\rmn{Cov}(x_i,x_i) = \rmn{Var}(x_i)$.

While one naively finds the lens mass as
\begin{equation}
M = \frac{c^2}{4 G}\,\frac{1~\rmn{AU}}{\pi_\rmn{E}}\,\frac{t_\rmn{E}}{t_\star}\,\theta_\star\,,
\end{equation}
taking into account 
the finite uncertainties yields
\begin{eqnarray}
& & \hspace*{-2.2em} 
\left\langle \lg \frac{M}{M_{\sun}}\right\rangle
= \lg \left[\frac{c^2}{4 G M_{\sun}}\,(1~\rmn{AU})\right]
+\left\langle \lg \frac{t_\rmn{E}}{1~\rmn{d}}\right\rangle
- \left\langle \lg \frac{t_\star}{1~\rmn{d}}\right\rangle  \; - \nonumber \\
& & \hspace*{-1.7em} 
- \; \left\langle \lg \pi_\rmn{E}\right\rangle
+ \left\langle \lg \theta_\star\right\rangle \,, 
\end{eqnarray}
and with $\theta_\star = R_\star/D_\rmn{S}$ not being correlated with the model parameters
$t_\rmn{E}$, $t_\star$, and $\pi_\rmn{E}$, one obtains
\begin{eqnarray}
& & \hspace*{-2.2em} 
\rmn{Var}\left(\lg \frac{M}{M_{\sun}}\right) = 
\rmn{Var}\left(\lg \frac{t_\rmn{E}}{1~\rmn{d}}\right) +
\rmn{Var}\left(\lg \frac{t_\star}{1~\rmn{d}}\right)\; + \nonumber \\
& & \hspace*{-1.7em} 
+\; \rmn{Var}\left(\lg \pi_\rmn{E} \right) +
\rmn{Var}\left(\lg \frac{R_\star}{1~\rmn{d}}\right) +
\rmn{Cov}\left(\lg \frac{t_\rmn{E}}{1~\rmn{d}},\lg \frac{t_\star}{1~\rmn{d}}\right) \; + \nonumber \\
& & \hspace*{-1.7em} +\;
\rmn{Cov}\left(\lg \frac{t_\rmn{E}}{1~\rmn{d}},\lg \pi_\rmn{E} \right) +
\rmn{Cov}\left(\lg \frac{t_\star}{1~\rmn{d}},\lg \pi_\rmn{E} \right)\,.
\end{eqnarray}

\subsection{Limits arising from the absence of anomalous effects}

\begin{table}
\caption{Constraint on the fractional lens distance $x \equiv D_\rmn{L}/D_\rmn{S}$
arising from upper limits on the annual parallax or the source size.}
\label{tab:distlimits}
\begin{footnotesize}
\vspace*{1.0ex}
\begin{tabular}{cccccc}
\hline
 & & \multicolumn{4}{c}{${\tilde x}^\rmn{min}$} \\
\raisebox{1.5ex}[-1.5ex]{$\pi_\rmn{E}$} & \raisebox{1.5ex}[-1.5ex]{$\eta_{\pi_\rmn{E}}$} &
0.1~$M_{\sun}$ & 0.2~$M_{\sun}$ & 0.4~$M_{\sun}$ & 0.8~$M_{\sun}$ \\ \hline
0.015 & 0.12 & 0.998 & 0.997 & 0.994 & 0.988 \\
0.06 & 0.50 & 0.98 & 0.95 & 0.91 & 0.83 \\
0.25 & 2.1 & 0.70 & 0.54 & 0.37 & 0.22 \\
1 & 8.3 & 0.13 & 0.07 & 0.03 & 0.02 \\ 
 \hline
\end{tabular}

\bigskip

\begin{tabular}{cccccc}
\hline
 & & \multicolumn{4}{c}{${\tilde x}^\rmn{max}$} \\
\raisebox{1.5ex}[-1.5ex]{${\hat \rho}_{\star,\sun}$} & \raisebox{1.5ex}[-1.5ex]{$\eta_{t_\star}$} &
0.1~$M_{\sun}$ & 0.2~$M_{\sun}$ & 0.4~$M_{\sun}$ & 0.8~$M_{\sun}$ \\ \hline
0.0005 & 0.89 & 0.07 & 0.14 & 0.24 & 0.39 \\
0.00125 & 2.2 & 0.33 & 0.50 & 0.67 & 0.80 \\
0.003 & 5.4 & 0.74 & 0.85 & 0.92 & 0.96 \\
0.0075 & 13 & 0.95 & 0.97 & 0.986 & 0.993 \\
 \hline
\end{tabular}
\end{footnotesize}

\medskip
The source star has been assumed to be located at a distance $D_\rmn{S} = 8.5~\rmn{kpc}$
 in the direction of Baade's window. $\eta_{\pi_\rmn{E}} = 2 [r_{\rmn{E},\sun}/(1~\rmn{AU})] \pi_\rmn{E}$,
 $\eta_{t_\star} = 2 (r_{\rmn{E},\sun}/R_{\sun}) {\hat \rho}_{\star,\sun}$.
\end{table}

Frequently, anomalous effects such as those 
caused by the annual parallax or the finite source size escape detection from the photometric data
taken over the course of the microlensing event.
However, the absence of significant deviations from a lightcurve that is compatible with an
ordinary event places upper limits on the model parameters $\pi_\rmn{E}$ or $t_\star$.
Rather than "defining" a certain value by means of
$\delta$-functions, these constraints can be incorporated by including $\Theta$-functions in the respective
expressions for the probability and event rate densities. With $\eta_{\pi_\rmn{E}} = 
[(1-x)/x]^{1/2}\,(M/M_{\sun})^{-1/2}$ and $\eta_{\pi_\rmn{E}} \leq \eta_{\pi_\rmn{E}}^\rmn{max}$,
one finds a lower limit on the fractional lens distance
\begin{equation}
x \geq {\tilde x}^\rmn{min} = \frac{1}{1+\left(\eta_{\pi_\rmn{E}}^\rmn{max}\right)^2\,(M/M_{\sun})}
\end{equation}
for a given mass $M$. In analogy, for the finite source size, one finds with $\eta_{t_\star} = 
[x/(1-x)]^{1/2}\,(M/M_{\sun})^{-1/2}$ and $\eta_{t_\star} \leq \eta_{t_\star}^\rmn{max}$ an
upper limit
\begin{equation}
x \leq {\tilde x}^\rmn{max} = \frac{\left(\eta_{t_\star}^\rmn{max}\right)^2\,(M/M_{\sun})}
{1+\left(\eta_{t_\star}^\rmn{max}\right)^2\,(M/M_{\sun})}\,.
\end{equation}
Taking into account these limits yields e.g.\ the probability density of the lens 
mass $\mu_0 = M/M_0$ in generalization of Eq.~(\ref{eq:pmu}) as
\begin{eqnarray}
& & \hspace*{-2.2em}p_{\mu_0}^{(0)}
(\mu_0; \eta_{t_\rmn{E}}^{(0)}) = \frac{4\,\Gamma_0}
{\gamma_{\eta_{t_\rmn{E}}}(\eta_{t_\rmn{E}}^{(0)})}\,
\sqrt{\mu_0}\;
\Phi_{M/M_{\sun}}\left(\mu_0 \left[\eta_{t_\rmn{E}}^{(0)}\right]^2\right)
\; \times \nonumber \\
& & \hspace*{-1.7em} \times \;
\int\limits_{{\tilde x}^\rmn{min}(\mu_0)}^{{\tilde x}^\rmn{max}(\mu_0)} 
\Phi_\zeta\left(2\,\sqrt{\mu_0\,x(1-x)},x\right)\; \times \nonumber \\
& & \hspace*{-1.7em} \times \;
[x(1-x)]^{3/2}\,\Phi_x(x)\,\rmn{d}x\,.
\label{eq:pmulimited}
\end{eqnarray}

\begin{figure}
\includegraphics[width=84mm]{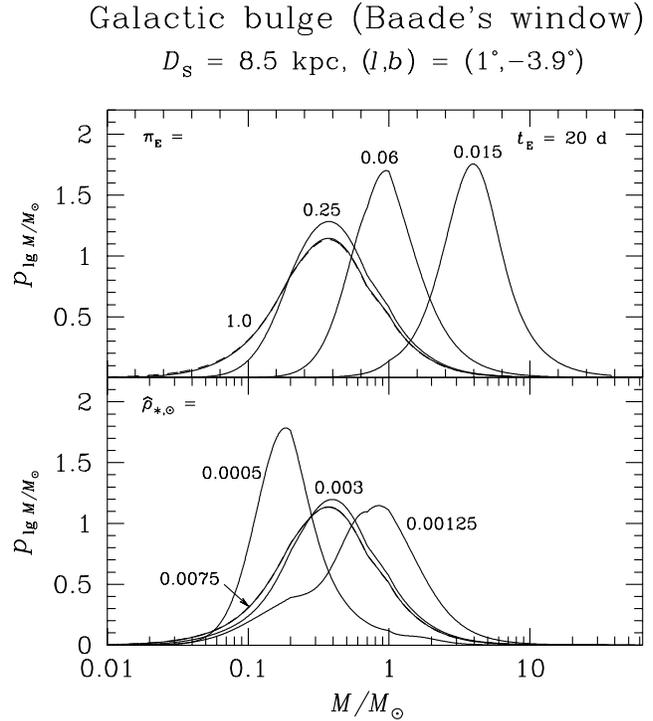}
\caption{Probability densities of $\lg (M/M_{\sun})$ for an event with time-scale $t_\rmn{E} = 20~\rmn{d}$
for a source located at $D_\rmn{S} = 8.5~\rmn{kpc}$ in the direction of Baade's window $(l,b) = (1\degr, -3.9\degr)$,
where an upper limit for the parallax $\pi_\rmn{E}$ or the source size parameter
${\hat \rho}_{\star,\sun} = (t_{\star}/t_\rmn{E})\,(R_\star/R_{\sun})^{-1}$ applies. The curves shown 
as dashed lines correspond to the limit-free case. In the lower panel,
this curve practically coincides with that for ${\hat \rho}_{\star,\sun}
= 0.0075$.}
\label{fig:distlimits}
\end{figure}

For a few selected masses, the resulting constraint on the fractional lens distance $x \equiv D_\rmn{L}/D_\rmn{S}$
that arises for selected limits for the parallax or the source size is shown in Table~\ref{tab:distlimits},
where the 'standard' source at $D_\rmn{S} = 8.5~\rmn{kpc}$ in the direction of Baade's window
has been assumed.
Both the parallax and the finite-source constraint more strongly restrict smaller lens masses, while 
larger masses remain possible at small distances with the parallax limit and at large distances with the
finite-source limit. For the same parallax and source-size limits as listed in
Table~\ref{tab:distlimits}, Fig.~\ref{fig:distlimits} shows the 
resulting probability density of the lens mass assuming an event with time-scale $t_\rmn{E} = 20~\rmn{d}$
for a source located at $D_\rmn{S} = 8.5~\rmn{kpc}$ in the direction of Baade's window
$(l,b) = (1\degr, -3.9\degr)$. Significant differences arise for $\pi_\rmn{E} \la 0.8$ or ${\hat \rho}_{\star,\sun}
\la 0.005$. 

For the annual parallax, the transition between a geocentric and a heliocentric coordinate system does
not influence the light curve and the orbital velocity is effectively absorbed into the event time-scale 
$t_\rmn{E}$ by contributing to the effective absolute perpendicular velocity. Therefore, it is the 
acceleration of the Earth's orbit that produces the lowest-order deviation \citep[e.g.][]{SMP:parallax}.
Within $t_\rmn{E}$, this acceleration induces an angular positional shift of 
$2 \upi^2 \pi_\rmn{LS}\,[t_\rmn{E}/(1~\rmn{yr})]^2$, so that 
$\kappa_\pi = 2 \upi^2 \pi_\rmn{E}\,[t_\rmn{E}/(1~\rmn{yr})]^2$ is a suitable indicator for the
prominence of parallax effects.
For an event time-scale $t_\rmn{E} = 20~\rmn{d}$, a limit $\pi_\rmn{E} \leq 0.8$ can be detected with a
sensitivity to $\kappa_\pi \sim 0.05$, whereas $\kappa_\pi = 1$ is reached for $t_\rmn{E} \sim 90~\rmn{d}$,
so that much smaller parallax limits can be obtained from such long events. If lens binarity can
be neglected, finite-source effects become apparent if the angular source size $\theta_\star$
becomes a fair fraction of the angular impact $u_0\,\theta_\rmn{E}$ between lens and source.
By requiring $u_0 \la 2\,(\theta_\star/\theta_\rmn{E}) = 2\,{\hat \rho}_{\star,\sun}\,(R_\star/R_{\sun})$,
a limit ${\hat \rho}_{\star,\sun} \leq 0.005$ for $R_\star = R_{\sun}$ is detected if $u_0 \la 0.01$,
corresponding to a peak magnification $A_0 \ga 100$, whereas an impact parameter $u_0 \la 0.1$,
corresponding to $A_0 \ga 10$,
is sufficient for $R_\star = 10~R_{\sun}$.

\section{Estimating anomaly model parameters}
\label{sec:anomdist}

In order to judge whether any anomaly is likely to have a significant effect on
the light curve, it is useful to estimate the value of parameters that quantify the
considered anomaly.
As already pointed out in Sect.~\ref{sec:anomconstr}, the size of parallax effects arising from the
orbital motion of the Earth can be modelled by the parameter $\pi_\rmn{E} =
\pi_\rmn{LS}/\theta_\rmn{E}$. With 
\begin{equation}
\pi_{\rmn{E},{\sun}} = \frac{1~\rmn{AU}}{2\,r_{\rmn{E},{\sun}}}
\end{equation}
being the value that corresponds to a solar-mass lens at $x = 0.5$, one can define a 'typical'
parallax parameter $\pi_{\rmn{E},0} = [\eta_{t_\rmn{E}}^{(0)}]^{-1}\,
\pi_{\rmn{E},{\sun}}$ for a given $t_\rmn{E}^{(0)}$ and the chosen $v_\rmn{c}$, where
$\eta_{t_\rmn{E}}^{(0)}$ is defined by Eq.~(\ref{eq:deff0}).

The corresponding ratio ${\hat \eta}_{\pi_\rmn{E}} = \pi_\rmn{E}/\pi_{\rmn{E},0}$ is related 
to the basic properties as ${\hat \eta}_{\pi_\rmn{E}} = [(1-x)/x]^{1/2}\,\mu_0^{-1/2}$, so that
Eq.~(\ref{eq:pdens0}) yields the probability density of ${\hat \eta}_{\pi_\rmn{E}}$ as
\begin{eqnarray}
& & \hspace*{-2.2em}p^{(0)}_{{\hat \eta}_{\pi_\rmn{E}}}
({\hat \eta}_{\pi_\rmn{E}} ; \eta_{t_\rmn{E}}^{(0)}) = 
\frac{8 \Gamma_0}{\gamma_{\eta_{t_\rmn{E}}}(\eta_{t_\rmn{E}}^{(0)})}\,{\hat \eta}_{\pi_\rmn{E}}^4\,
\int\limits_{\mu_0^{\rmn{min}}}^{\mu_0^{\rmn{max}}}
\frac{\mu_0^3}
{\left(1+\mu_0\,{\hat \eta}_{\pi_\rmn{E}}^2\right)^5}\; \times
\nonumber \\
& & \hspace*{-1.7em} \times \;
\Phi_{M/M_{\sun}}\left(\mu_0 \left[\eta_{t_\rmn{E}}^{(0)}\right]^2\right)
\; \times \nonumber \\
& & \hspace*{-1.7em} \times \;
\Phi_\zeta\left(
\frac{2\,\mu_0\,{\hat \eta}_{\pi_\rmn{E}}}{1+\mu_0\,
{\hat \eta}_{\pi_\rmn{E}}^2},
\frac{1}{1+\mu_0\,
{\hat \eta}_{\pi_\rmn{E}}^2}\right)\; \times
\nonumber \\
& & \hspace*{-1.7em} \times \;
\Phi_x\left(\frac{1}{1+\mu_0\,
{\hat \eta}_{\pi_\rmn{E}}^2}\right)\,\rmn{d}\mu_0\,.
\label{eq:ppiE}
\end{eqnarray}
From the respective definitions, one finds that $p_{{\hat \eta}_{\pi_\rmn{E}}}
({\hat \eta}_{\pi_\rmn{E}} ; \eta_{t_\rmn{E}}^{(0)}) = 
\gamma_{\eta_{t_\rmn{E}},\eta_{\pi_\rmn{E}}}(\eta_{t_\rmn{E}}^{(0)},
{\hat \eta}_{\pi_\rmn{E}}/\eta_{t_\rmn{E}}^{(0)})/
[\eta_{t_\rmn{E}}^{(0)}\gamma_{\eta_{t_\rmn{E}}}(\eta_{t_\rmn{E}}^{(0)})]$,
with $\gamma_{\eta_{t_\rmn{E}}}$ given by Eq.~(\ref{eq:tEdensity}) and
 $\gamma_{\eta_{t_\rmn{E}},\eta_{\pi_\rmn{E}}}$ given by Eq.~(\ref{eq:normfg}).

\begin{figure}
\includegraphics[width=84mm]{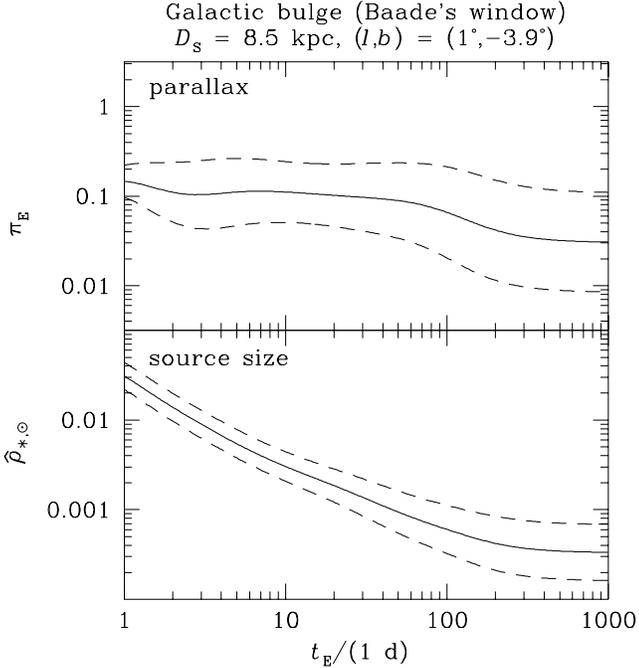}
\caption{Expectation values (solid lines) of $\lg \pi_\rmn{E}$ (top panel) and
$\lg {\hat \rho}_{\star,\sun} = \lg[(t_\star/t_\rmn{E})\,(R_\star/R_{\sun})^{-1}]$ 
as a function of the event time-scale
$t_\rmn{E}$ for a source in the direction of Baade's window located at $D_\rmn{S} = 8.5~\rmn{kpc}$.
The dashed lines indicate limits defined by the standard deviation.}
\label{fig:avtimepar}
\end{figure}

The top panel of Figure~\ref{fig:avtimepar} shows the expectation value of $\lg \pi_\rmn{E}$ 
along with its standard deviation
as a function of the event time-scale $t_\rmn{E}$.
Since the variations in the basic system properties counterbalance each other with respect to 
the parallax parameter $\pi_\rmn{E}$, its expectation value shows only a slight variation with
the event time-scale $t_\rmn{E}$, while its variance is quite substantial.
With $\kappa_\pi = 2 \upi^2 \pi_\rmn{E}\,[t_\rmn{E}/(1~\rmn{yr})]^2$ being the angular positional shift
in units of the angular Einstein radius $\theta_\rmn{E}$ induced by the acceleration of the Earth's orbit,
which is a suitable indicator for the prominence of parallax effects,
and $\pi_\rmn{E} \sim 0.1$ only weakly depending on the event time-scale,
one approximately finds $\kappa_\pi \propto t_\rmn{E}^2$, where $\kappa_\pi \sim 6\times 10^{-3}$ for
$t_\rmn{E} = 20~\rmn{d}$, while $\kappa_\pi \sim 0.1$ for 
$t_\rmn{E} \sim 80~\rmn{d}$.

Finite-source effects can be studied in analogy to the parallax case.
With the definition ${\hat \eta}_{t_\star} = \eta_{t_\star}\,\eta_{t_\rmn{E}}^{(0)}$, so that
${\hat \eta}_{\pi_\rmn{E}} = [x/(1-x)]^{1/2}\,\mu_0^{-1/2}$, 
Eq.~(\ref{eq:pdens0}) yields the corresponding probability density as
\begin{eqnarray}
& & \hspace*{-2.2em}p^{(0)}_{{\hat \eta}_{t_\star}}
({\hat \eta}_{t_\star} ; \eta_{t_\rmn{E}}^{(0)}) = 
\frac{8 \Gamma_0}{\gamma_{\eta_{t_\rmn{E}}}(\eta_{t_\rmn{E}}^{(0)})}\,
{\hat \eta}_{t_\star}^4\,
\int\limits_{\mu_0^{\rmn{min}}}^{\mu_0^{\rmn{max}}}
\frac{\mu_0^3}
{\left(1+\mu_0\,{\hat \eta}_{t_\star}^2\right)^5}\; \times
\nonumber \\
& & \hspace*{-1.7em} \times \;
\Phi_{M/M_{\sun}}\left(\mu_0 \left[\eta_{t_\rmn{E}}^{(0)}\right]^2\right)
\; \times \nonumber \\
& & \hspace*{-1.7em} \times \;
\Phi_\zeta\left(
\frac{2\,\mu_0\,{\hat \eta}_{t_\star}}{1+\mu_0\,
{\hat \eta}_{t_\star}^2},
\frac{\mu_0\,{\hat \eta}_{t_\star}^2}{1+\mu_0\,
{\hat \eta}_{t_\star}^2}\right)\; \times
\nonumber \\
& & \hspace*{-1.7em} \times \;
\Phi_x\left(\frac{\mu_0\,{\hat \eta}_{t_\star}^2}{1+\mu_0\,
{\hat \eta}_{t_\star}^2}\right)\,\rmn{d}\mu_0\,,
\label{eq:ptstar}
\end{eqnarray}
where $p_{{\hat \eta}_{t_\star}}
({\hat \eta}_{t_\star} ; \eta_{t_\rmn{E}}^{(0)}) = 
\gamma_{\eta_{t_\rmn{E}},\eta_{t_\star}}(\eta_{t_\rmn{E}}^{(0)},
{\hat \eta}_{t_\star}/\eta_{t_\rmn{E}}^{(0)})/
[\eta_{t_\rmn{E}}^{(0)}\gamma_{\eta_{t_\rmn{E}}}(\eta_{t_\rmn{E}}^{(0)})]$,
with $\gamma_{\eta_{t_\rmn{E}}}$ given by Eq.~(\ref{eq:tEdensity}) and
 $\gamma_{\eta_{t_\rmn{E}},\eta_{t_\star}}$ given by Eq.~(\ref{eq:normfh}).
As the bottom panel of Figure~\ref{fig:avtimepar} shows, events with smaller time-scales are more likely 
to show prominent finite-source effects, where for a (giant) star with $R \sim 10~R_{\sun}$, one finds
$\rho_\star = t_\star/t_\rmn{E} \sim 0.02 $ for $t_\rmn{E} = 20~\rmn{d}$, whereas 
$\rho_\star \sim 0.007$ for $t_\rmn{E} = 80~\rmn{d}$.

If the lens that caused the microlensing event is a binary object and
its orbital motion is neglected, its effects on the light curve are completely characterized 
by the mass ratio $q$, the separation parameter $d$, where
$d\,\theta_\rmn{E}$ is the angular instantaneous separation of its constituents
(considered being constant during the duration of the event), and an angle describing
the direction of the proper motion with
respect to their angular separation vector.
The probability densities of the masses of the components $M_1 = M/(1+q)$ and
$M_2 = M\,[q/(1+q)]$ simply follow from the mass ratio $q$ and the probability
density of the total mass $M$ as given by Eq.~(\ref{eq:plgmass}) or by the corresponding 
relation given in Sect.~\ref{sec:anomconstr} if parallax or finite-source effects are significant.
An estimate for the instantaneous 
projected physical separation $\hat r = d\,r_\rmn{E}$ between the lens components can be obtained
by multiplying the obtained $d$ with the corresponding statistic for the 
Einstein radius $r_\rmn{E} = D_\rmn{L}\,\theta_\rmn{E}$, so that
\begin{equation}
p_{{\hat r}/r_{\rmn{E},0}}^{(0)}({\hat r}/r_{\rmn{E},0} ;\eta_{t_\rmn{E}}^{(0)},d^{(0)}) = 
\frac{1}{d^{(0)}}\;
p_{\rho_\rmn{E}}^{(0)}\left(\frac{\hat r}{d^{(0)}\,r_{\rmn{E},0}}; \eta_{t_\rmn{E}}^{(0)}\right)
\end{equation}
with $\rho_\rmn{E} = r_\rmn{E}/r_{\rmn{E},0}$, where $r_{\rmn{E},0}$ is given by Eq.~(\ref{eq:defrEsun}),
and $p_{\rho_\rmn{E}}^{(0)} = p_{\zeta}^{(0)}$ defined by
Eq.~(\ref{eq:pzeta}). Beyond considering a fixed model parameter $d$,
$p_{\rho_\rmn{E}}^{(0)}$ can be convolved with its distribution.

With $\hat \chi = \hat r/a$ denoting the projection factor between the semi-major axis $a$ and
the actual projected separation $\hat r$, one finds 
$a = {\hat r}/{\hat \chi} = (d\,r_\rmn{E})/{\hat \chi}$. In addition to $M$, $x$, and $\zeta$, 
the value of $a$ therefore depends on the projection factor $\hat \chi$ as further property
which is distributed as $\Phi_{\hat \chi}(\hat \chi)$ as given by Eq.~(\ref{eq:phatchi}),
where $0 \leq {\hat \chi} \leq {\hat \chi}^\rmn{max}$.
For the probability density, one therefore finds
\begin{eqnarray}
& & \hspace*{-2.2em}
p_{a/r_{\rmn{E},0}}^{(0)}(a/r_{\rmn{E},0} ;\eta_{t_\rmn{E}}^{(0)},d^{(0)}) \nonumber \\
& & \hspace*{-1.7em} 
= \int\limits_0^{{\hat \chi}^\rmn{max}} \int\limits_{0}^{\infty}
p_{{\hat r}/r_{\rmn{E},0}}({\hat r}/r_{\rmn{E},0} ;\eta_{t_\rmn{E}}^{(0)},d^{(0)})
\;\times \nonumber \\
& & 
\times\;
\delta\left(\frac{a}{r_{\rmn{E},0}}-\frac{1}{\hat \chi}\,\frac{\hat r}{r_{\rmn{E},0}}\right)\,
\rmn{d}({\hat r}/r_{\rmn{E},0})\,\Phi_{\hat \chi}(\hat \chi)\,\rmn{d}{\hat \chi}
\nonumber \\
& & \hspace*{-1.7em}
= \int\limits_0^{{\hat \chi}^\rmn{max}} {\hat \chi}\;
p_{{\hat r}/r_{\rmn{E},0}}\left({\hat \chi}\,\frac{a}{r_{\rmn{E},0}};
\eta_{t_\rmn{E}}^{(0)},d^{(0)}\right)\,
\Phi_{\hat \chi}(\hat \chi)\,\rmn{d}{\hat \chi} 
\nonumber \\
& & \hspace*{-1.7em}
= \frac{1}{d^{(0)}}\,\int\limits_0^{{\hat \chi}^\rmn{max}} {\hat \chi}\;
p_{\rho_\rmn{E}}\left(\frac{\hat \chi}{d^{(0)}}\,
\frac{a}{r_{\rmn{E},0}}; \eta_{t_\rmn{E}}^{(0)}\right)\,
\Phi_{\hat \chi}(\hat \chi)\,\rmn{d}{\hat \chi}\,.
\end{eqnarray}

According to Kepler's third law, the orbital period reads $P = 2\,\upi\,[a^3/(GM)]^{1/2}$, being a function
of the total mass $M$ and the semi-major axis $a$. With 
\begin{equation}
P_0 = 2\,\upi\,\sqrt{\frac{\eta_{t_\rmn{E}}\,r_{\rmn{E},\sun}^3}{G\,M_{\sun}}}\,,
\end{equation}
${\hat \eta}_P = P/P_0$ is related to the basic system properties as 
${\hat \eta}_P = \left[2\,(d/{\hat \chi})\right]^{3/2}\,\left[x(1-x)\right]^{3/4}\,
\mu_0^{1/4}$, so that the corresponding probability density becomes
\begin{eqnarray}
& & \hspace*{-2.2em}
p_{{\hat \eta}_P}^{(0)}({\hat \eta}_P;\eta_{t_\rmn{E}}^{(0)},d^{(0)}) =
\frac{\Gamma_0}{\gamma_{\eta_{t_\rmn{E}}}(\eta_{t_\rmn{E}}^{(0)})}\,\frac{{\hat \eta}_P^5}{[d^{(0)}]^9}\; \times \nonumber \\
& & \hspace*{-1.7em} \times\;
\int\limits_0^{{\hat \chi}^\rmn{max}} 
\Theta\left(\frac{d^{(0)}\left(\mu_0^\rmn{max}\right)^{1/6}}{{\hat \eta}_P^{2/3}}-\hat \chi\right)\,
\int\limits_{{\hat x}^\rmn{min}}^{{\hat x}^\rmn{max}}
\frac{{\hat \chi}^9}{\left(1-{\hat x}^2\right)^3}\; \times \nonumber \\
& & \hspace*{-1.7em} \times\;
\Phi_{M/M_{\sun}}\left(\frac{{\hat \eta}_P^4\,{\hat \chi}^6}{[d^{(0)}]^6\,\left(1-{\hat x}^2\right)^3}\,
[\eta_{t_\rmn{E}}^{(0)}]^2\right) \; \times \nonumber \\
& & \hspace*{-1.7em} \times\;
\sum_{\pm} \Phi_{\zeta}\left(\frac{{\hat \eta}_P^2\,{\hat \chi}^3}{[d^{(0)}]^3\,\left(1-{\hat x}^2\right)},
\frac{1}{2}\left(1\pm \hat x\right)\right)\; \times \nonumber \\
& & \hspace*{-1.7em} \times\; \Phi_x\left(\frac{1}{2}\left(1\pm \hat x\right)\right)\,
\rmn{d}{\hat x}\;\Phi_{\hat \chi}(\hat \chi)\,\rmn{d}{\hat \chi}\,,
\end{eqnarray}
where
\begin{eqnarray}
{\hat x}^\rmn{min} & = & \left\{\begin{array}{l}
\sqrt{1-\frac{{\hat \eta}_P^{4/3}\,{\hat \chi}^2}{[d^{(0)}]^2\,\left(\mu_0^\rmn{min}\right)^{1/3}}} \\
\qquad \qquad \qquad \hfill \rmn{for} \quad \hat \chi < \frac{d^{(0)}\left(\mu_0^\rmn{min}\right)^{1/6}}{{\hat \eta}_P^{2/3}}\\
\rule{0cm}{3.5ex} 0 \hfill \rmn{for} \quad \hat \chi \geq 
\frac{d^{(0)}\left(\mu_0^\rmn{min}\right)^{1/6}}{{\hat \eta}_P^{2/3}}  
\end{array}\right.\,, \nonumber \\
{\hat x}^\rmn{max} & = & \left\{\begin{array}{l}
\sqrt{1-\frac{{\hat \eta}_P^{4/3}\,{\hat \chi}^2}{[d^{(0)}]^2\,\left(\mu_0^\rmn{max}\right)^{1/3}}} \\
\qquad \qquad \qquad \hfill \rmn{for} \quad \hat \chi < \frac{d^{(0)}\left(\mu_0^\rmn{max}\right)^{1/6}}{{\hat \eta}_P^{2/3}}\\
\rule{0cm}{3.5ex} 0 
\hfill \rmn{for} \quad \hat \chi \geq 
\frac{d^{(0)}\left(\mu_0^\rmn{max}\right)^{1/6}}{{\hat \eta}_P^{2/3}}  
\end{array}\right.\,.
\end{eqnarray}

Using the property for the expectation values and the variances stated by
Eqs.~(\ref{eq:logexp}) and~(\ref{eq:logvariance}), one finds that
\begin{eqnarray}
\left\langle \lg (M_1/M_0)\right\rangle  & =  & 
\left\langle \lg (M/M_0)\right\rangle  - \left\langle \lg (1+q)\right\rangle \,, \nonumber \\
\left\langle \lg (M_2/M_0)\right\rangle  & =  & 
\left\langle \lg (M/M_0)\right\rangle  + \left\langle \lg q\right\rangle  - 
\left\langle \lg (1+q)\right\rangle\,,  \nonumber \\
\left\langle \lg (\hat r/r_{\rmn{E},0})\right\rangle  & = & 
\left\langle \lg \rho_\rmn{E}\right\rangle + \left\langle \lg d\right\rangle\,,  \nonumber \\
\left\langle \lg (a/r_{\rmn{E},0})\right\rangle  & = & 
\left\langle \lg \rho_\rmn{E}\right\rangle +
\left\langle \lg d\right\rangle  - \left\langle \lg \hat \chi\right\rangle\,, \nonumber \\
\left\langle \lg (P/P_0)\right\rangle  & = & \frac{3}{2}\,\left\langle \lg (a/r_{\rmn{E},0})\right\rangle  - \frac{1}{2}\,
\left\langle \lg (M/M_0)\right\rangle\,, \nonumber \\
 & = & \frac{3}{2}\,\left[\left\langle \lg \rho_\rmn{E}\right\rangle  + \left\langle \lg d\right\rangle   
- \left\langle \lg \hat \chi\right\rangle \right] \; - \nonumber \\
& & \quad -\;\frac{1}{2}\,
\left\langle \lg (M/M_0)\right\rangle \,,
\end{eqnarray}
and
\begin{eqnarray}
\rmn{Var}\left[\lg (M_1/M_0)\right] & = & \rmn{Var}\left[\lg (M/M_0)\right] + 
\rmn{Var}\left[\lg (1+q)\right]\,,\nonumber \\
\rmn{Var}\left[\lg (M_2/M_0)\right] & = & \rmn{Var}\left[\lg (M/M_0)\right]\; + \nonumber \\
& & \quad +\;\rmn{Var}\left\{\lg [q/(1+q)]\right\}\,,\nonumber \\
\rmn{Var}\left[\lg (\hat r/r_{\rmn{E},0})\right] & = & \rmn{Var}\left(\lg \rho_\rmn{E}\right) +
\rmn{Var}\left(\lg d\right)\,,
\nonumber \\
\rmn{Var}\left[\lg (a/r_{\rmn{E},0})\right] & = & \rmn{Var}\left(\lg \rho_\rmn{E}\right) +
\rmn{Var}\left(\lg d\right) 
+  \rmn{Var}\left(\lg \hat \chi \right)\,,
\nonumber \\
\rmn{Var}\left[\lg (P/P_0)\right] & = & 
\frac{9}{4}\,[\rmn{Var}\left(\lg \rho_\rmn{E}\right) + \rmn{Var}\left(\lg d\right)\; + \nonumber \\
& & +\;\rmn{Var}\left(\lg \hat \chi \right)]  + \frac{1}{4}\,\rmn{Var}\left[\lg (M/M_0)\right] \; -
\nonumber \\
& & -\; \frac{3}{4}\,\rmn{Cov}\left[\lg \rho_\rmn{E},\lg (M/M_0)\right]\,.
\end{eqnarray}

As an example, let us consider the microlensing event OGLE 2002-BLG-099 \citep{OGLEbinaries},
where the observed light curve suggests the lens star to be a stellar
or brown-dwarf binary, while the absence of observed caustic passages
leaves the possibility that the source
is a binary instead. For the binary-lens model, the mass ratio
is $q \sim 0.25$, while $d = 1.963$ and $t_\rmn{E} = 34.4~\rmn{d}$. 
No signals of annual parallax or finite source size have been detected, 
whereas more than 60\% of the observed light at baseline arises from a source other than the lensed background star.

\begin{table}
\caption{Estimates for physical properties of the two components of the lens system that caused microlensing event
OGLE 2002-BLG-099 based on the model parameters reported by \citet{OGLEbinaries} and the Galaxy model
described in Appendix~\ref{sec:galmodel}.}
\label{tab:O71properties}
\begin{footnotesize}
\vspace*{1.0ex}
\begin{tabular}{lcc}
\hline
 & circular & elliptic \\ \hline
$M_1$ [$M_{\sun}$] & \multicolumn{2}{c}{0.38 (2.7)}  \\
$M_2$ [$M_{\sun}$] & \multicolumn{2}{c}{0.093 (2.7)} \\
$D_\rmn{L}$ [kpc]& \multicolumn{2}{c}{$5.7 \pm 1.7$} \\
$a$ [AU]  & 6.2 (1.8) & 5.8 (2.0) \\
$P$ [yr]  & 22 (2.0) & 20 (2.4) \\
 \hline
\end{tabular}
\end{footnotesize}

\medskip
The estimates are based on the
expectation values of $x$, $\lg (M/M_{\sun})$, $\lg \rho_\rmn{E}$, and $\lg \hat \chi$, where a source
distance of $D_\rmn{S} = 8.5~\rmn{kpc}$ has been adopted. Numbers in brackets denote the uncertainty factor
that corresponds to the standard deviation of the logarithm of the respective quantity. $M_1$ and $M_2$ are
the masses of the primary and secondary component of the lens binary, respectively, $D_\rmn{L}$ is their distance from the observer, 
$a$ denotes the orbital semi-major axis and $P$ denotes the orbital period. The latter quantities have been
calculated both assuming circular orbits or elliptic orbits with the eccentricity being distributed
as $\Phi_e = (4/\upi)\,\sqrt{1-e^2}$.
\end{table}

\begin{figure}
\includegraphics[width=84mm]{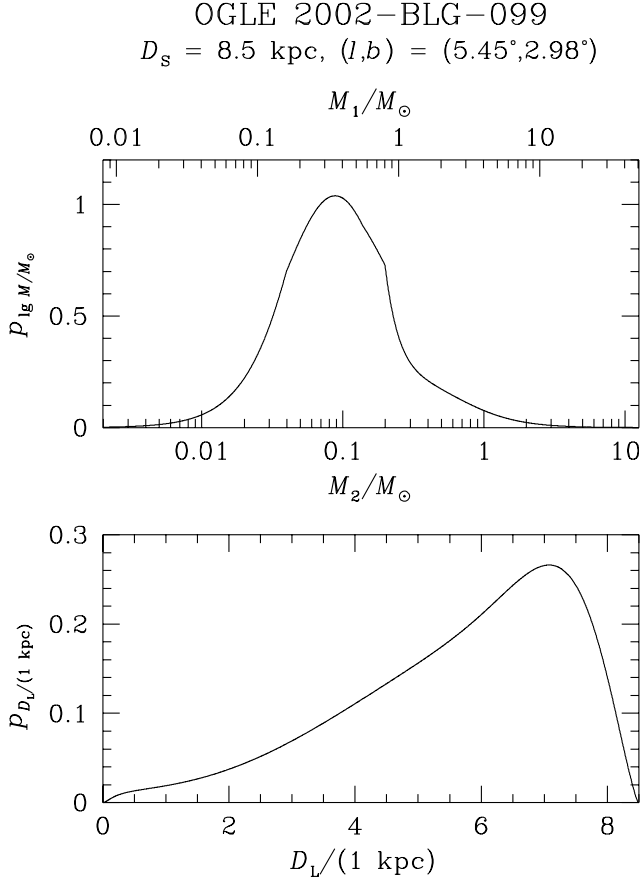}
\caption{Probability densities of the mass $M_1$ of the lens star and its planet $M_2$
as well as for their distance $D_\rmn{L}$ from the observer
for the microlensing event OGLE 2005-BLG-071 and the binary-lens
model reported by \citet{OGLEbinaries}, for which
$q \sim 0.25$, $d = 1.963$ and $t_\rmn{E} = 34.4~\rmn{d}$.}
\label{fig:O99massdist}
\end{figure}

\begin{figure}
\includegraphics[width=84mm]{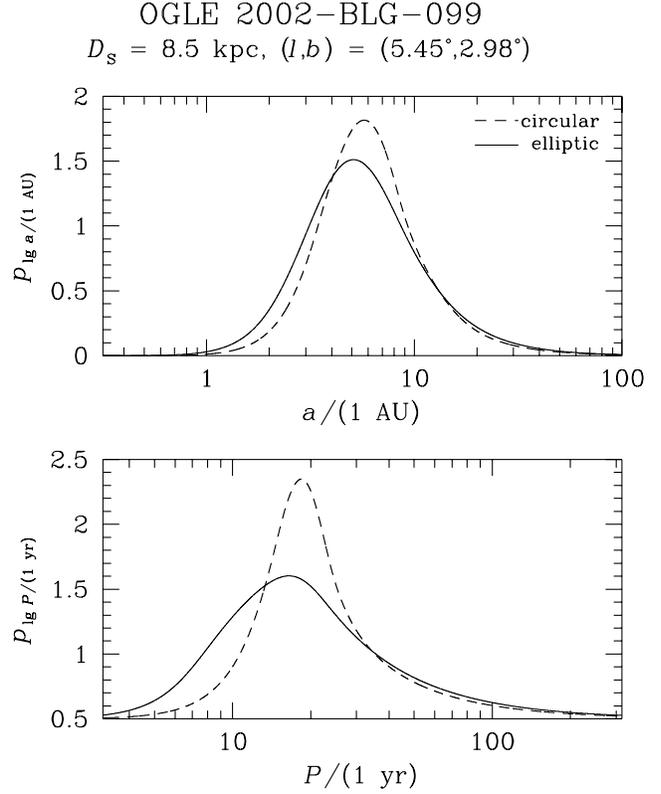}
\caption{Probability densities of the semi-major axis $a$ (top) and the orbital period $P$ (bottom) 
for the microlensing event OGLE 2002-BLG-099 \citep{OGLEbinaries}
assuming the reported binary-lens model, for which 
$q \sim 0.25$, 
$d=1.963$ and $t_\rmn{E} = 34.4~\rmn{d}$. The dashed lines correspond to the assumption of
circular orbits, while curves drawn as solid lines correspond to elliptic orbits with an eccentricity
distribution $\Phi_e = (4/\upi)\,\sqrt{1-e^2}$. }
\label{fig:O99periodaxis}
\end{figure}

With the Galaxy models
described in Appendix~\ref{sec:galmodel}, one finds the expectation values and uncertainties of the two components of the binary lens system,
the distance from the observer, the semi-major axis of the
planetary orbit, and the orbital period as listed in Table~\ref{tab:O71properties}, while the probability
densities of these quantities are shown in Figs.~\ref{fig:O99massdist} and~\ref{fig:O99periodaxis}.
For the deprojection of the orbit, either circular orbits or elliptical orbits, where the
eccentricity is distributed as $\Phi_e = (4/\upi)\,\sqrt{1-e^2}$, have been considered.
More details about
the statistics of binary orbits can be found in Appendix~\ref{sec:orbitalproj}.
For circular orbits, $\left\langle \lg \hat \chi \right\rangle = -0.133$, corresponding to a factor $\hat \chi = 0.736$, so that
$a \sim 1.36~\hat r$. The standard deviation of $\sigma_{\lg \hat \chi} = 0.183$ is equivalent 
to a factor 1.35, yielding an interval $a \sim  1.01\ldots{}1.83~\hat r$. In contrast, one finds
for elliptic orbits with the adopted distribution of eccentricities an expectation value
$\left\langle \lg \hat \chi \right\rangle = -0.104$, which yields a factor $\hat \chi = 0.787$, 
so that $a \sim 1.27~\hat r$. In this case, the standard deviation is $\sigma_{\lg \hat \chi} = 0.234$,
which corresponds to a factor 1.71, spanning an interval $a \sim  0.74\ldots{}2.18~\hat r$.
The differences between circular orbits and the elliptical orbits according to the adopted eccentricity distribution, which are 
seen in the respective probability density of the orbital period, reflect the 
distribution of the projection factor $\Phi_{\hat \chi}(\hat \chi)$ as shown in Figure~\ref{fig:orbproj}.
While for circular objects, a dominant contribution results from $\chi \la 1$, the adopted elliptical orbits
produce a small excess for large orbital periods and a significant tail towards smaller orbital periods
due to projection factors $1 < \hat \chi < 2$.

\section{Detection efficiency maps for lens companions}
\label{sec:deteffmaps}

As pointed out by \citet{MP91}, the light curve of a galactic microlensing event may reveal the existence
of companions to the lens star, which includes stellar binaries as well as planetary systems.
If the orbital motion of a binary lens does not result in a significant effect,
only two characteristics of the binary influence the light curve, namely
its mass ratio $q$ and the separation parameter $d = {\hat r}/r_\rmn{E}$, where
$\hat r$ is the actual projected separation perpendicular to the line-of-sight.
The detection efficiency $\varepsilon(d,q)$ for a companion to the lens star
with a separation parameter $d$ and a mass ratio $q$ is defined as the
probability that a detectable signal (event $S$) would arise if such a companion 
exists (event $C(d,q)$), i.e.\
\begin{equation}
\varepsilon(d,q) = P\left(S|C(d,q)\right)\,.
\end{equation}
For a given event, a detection efficiency map \citep[e.g.][]{PLANET:O14,MPSMOA:M35} can be calculated for any
combination of the parameters $(d,q)$. Let 
$F(d^-,d^+,q^-,q^+)$ denote the average number of companions
with distance parameter
$d \in [d^-,d^+]$ and 
mass ratio $q \in [q^-,q^+]$, so that
\begin{equation}
F(d^-,d^+,q^-,q^+)
= \int\limits_{d^-}^{d^+} \int\limits_{q^-}^{q^+}
f_{d,q}(d,q)\,\rmn{d}d\,\rmn{d}q\,.
\end{equation}
From a sample of $N$ events, one then expects to detect
\begin{equation}
H(d^-,d^+,q^-,q^+) = \sum_{i=1}^{N} \int\limits_{d^-}^{d^+} \int\limits_{q^-}^{q^+}
f_{d,q}(d,q)\,\varepsilon^{(i)}_{d,q}(d,q)\,\rmn{d}d\,\rmn{d}q
\end{equation}
companions, and 
the probability to detect at least one signal reads
\citep[c.f.][]{PLANET:fiveshort}
\begin{eqnarray}
& & \hspace*{-2.2em} 
{\hat \varepsilon}(d^-,d^+,q^-,q^+) \nonumber \\
& & \hspace*{-1.7em} 
= 1-\prod_{i=1}^{N} \Bigg[1 -
\int\limits_{d^-}^{d^+} \int\limits_{q^-}^{q^+}
f_{d,q}(d,q)\,
\varepsilon^{(i)}_{d,q}(d,q)\,\rmn{d}d\,\rmn{d}q  \Bigg]\,.
\end{eqnarray}
However, rather than obtaining information about the awkward $f_{d,q}(d,q)$,
one would like to investigate
the abundance of companions (such as planets) as function of the physical properties such as the
companion mass $M_2$, the semi-major axis $a$, and the orbital eccentricity $e$. While microlensing does
not provide a means to study the dependence on the orbital eccentricity, for which a distribution needs
to be assumed, the adoption of a Galaxy model allows to compare the microlensing results with 
assumed abundances $f_{a,M_2}(a,M_2)$.

For a given projected separation $\hat r$ and a companion mass $M_2$, the 
probability density of $(d,q)$ follows that of $(r_\rmn{E},M_1)$, so that the
detection efficiency in these physical lens characteristics reads
\begin{eqnarray}
& & \hspace*{-2.2em}
\varepsilon_{\hat r,M_2}({\hat r},M_2)  
=\int\limits_0^\infty \int\limits_0^\infty
 \varepsilon_{d,q}(d,q)\,p_{d,q}(d,q; \hat r, M_2, \eta_{t_\rmn{E}}^{(0)})\,
 \rmn{d}d\,\rmn{d}q \nonumber \\
& & \hspace*{-1.7em}
= \int\limits_0^\infty \int\limits_0^\infty
\varepsilon_{d,q}\left(
\frac{1}{\eta_{t_\rmn{E}}^{(0)}\,\rho_\rmn{E}}\,
\frac{\hat r}{r_{\rmn{E},\sun}},
\frac{1}{[\eta_{t_\rmn{E}}^{(0)}]^2}\,
\frac{M_2/M_{\sun}}{\mu_0}
\right)\;\times \nonumber \\
& & \qquad \qquad\times\;p_{\rho_\rmn{E},\mu_0}(\rho_\rmn{E},\mu_0;\eta_{t_\rmn{E}}^{(0)})\,
\rmn{d}\rho_\rmn{E}\,\rmn{d}\mu_0\,,
\end{eqnarray}
where $\rho_\rmn{E} = r_\rmn{E}/r_{\rmn{E},0}$ with
$r_{\rmn{E},0}$ and $M_0$ as defined in Sect.~\ref{sec:probdens}.
Rather than to the total mass $M$, the time-scale $t_\rmn{E}^{(0)}$ hereby refers to the mass
of the primary $M_1 = M/(1+q)$ and the mass spectrum is adopted as 
$\Phi_{M_1/M_{\sun}}(M_1/M_{\sun}) = \Theta(M_1-M_2)\,\Phi_{M/M_{\sun}}[(1+q)M_1/M_{\sun}]$.
Whereas for $M_1 \simeq M_2$, one needs to distinguish between close binaries,
where the best-fit single-lens time-scale refers to the total mass, and wide binaries, where it
refers to one of the constituents, for the relevant $M_2 \la 10^{-2.5} M_{\sun}$ discussed here,
$q \ll 1$ and $M_1 \approx M$ is a fair approximation. Moreover, a single $t_\rmn{E}$ for each of the events,
corresponding to the value estimated for a single lens, rather than an optimized $t_\rmn{E}$ for
each pair $(d,q)$ can be used, since as shown previously, shifts in $t_\rmn{E}$ by less than 20 per cent
can be safely neglected relative to the width of the broad distributions of lens
mass, distance, and velocity, and the uncertainties of the Galaxy models.

The distribution of $(\rho_\rmn{E}, \mu_0)$ follows from Eq.~(\ref{eq:pzeta}) as
\begin{eqnarray}
& & \hspace*{-2.2em}
p_{\rho_\rmn{E},\mu_0}^{(0)}(\rho_\rmn{E},\mu_0;
\eta_{t_\rmn{E}}^{(0)}) = 
\frac{\Gamma_0}{4\,\gamma_{\eta_{t_\rmn{E}}}(\eta_{t_\rmn{E}}^{(0)})}\; \times \nonumber \\
& & \hspace*{-1.7em} \times\;
\frac{\rho_\rmn{E}^4}{\mu_0^2}\,\frac{\Theta(\mu_0-\rho_\rmn{E}^2)}{\sqrt{1-\rho_\rmn{E}^2/\mu_0}}\;
\Phi_{M/M_{\sun}}\left(\mu_0 \left[\eta_{t_\rmn{E}}^{(0)}\right]^2\right)\;\times \nonumber \\
& & \hspace*{-1.7em} \times\;
\sum\limits_{\pm}
\Phi_\zeta\left(\rho_\rmn{E},\frac{1}{2}\left(1\pm
\sqrt{1-\rho_\rmn{E}^2/\mu_0}\right)\right)\;\times \nonumber \\
& & \hspace*{-1.7em} \times\;
\Phi_x\left(\frac{1}{2}\left(1\pm
\sqrt{1-\rho_\rmn{E}^2/\mu_0}\right)\right)\,.
\end{eqnarray}

%\begin{figure}
%\includegraphics[width=84mm]{rhomu_map.rev.eps}
%\caption{Bivariate probability density $p_{\lg r_\rmn{E}/(1 AU), \lg M/M_{\sun}}$ as function of the Einstein radius $r_\mathrm{E}$ and the
%lens mass $M$ for the event OGLE 2004-BLG-234 located at $(l,b) = (2.45\degr, -2.50\degr)$,
%for which $D_\rmn{S} = 8.5~\rmn{kpc}$ has been assumed. The bold line marks the upper limit for
%$r_\mathrm{E}$ and the mass cut-off at $M = 0.01~M_{\sun}$, which arises from the adopted mass spectrum.}
%\label{fig:rhomumap}
%\end{figure}

\begin{figure}
\includegraphics[width=84mm]{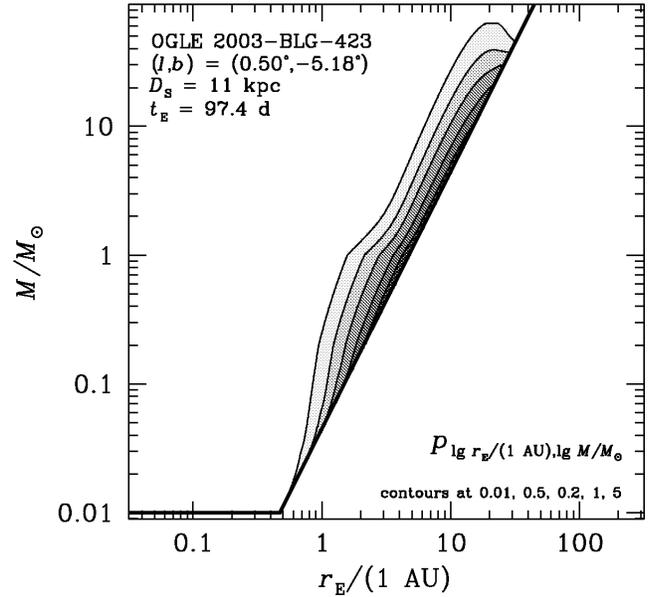}
\caption{Bivariate probability density $p_{\lg r_\rmn{E}/(1~\rmn{AU}), \lg M/M_{\sun}}$ as function of the Einstein radius $r_\mathrm{E}$ and the
lens mass $M$ for the time-scale $t_\rmn{E} = 97.4$ as determined
by \citet{GG:Esti} for OGLE 2003-BLG-423, and the corresponding location $(l,b) = (0.50\degr, -5.18\degr)$,
while a source distance
$D_\rmn{S} = 11~\rmn{kpc}$ has been assumed. The bold line marks the upper limit for
$r_\mathrm{E}$ and the mass cut-off at $M = 0.01~M_{\sun}$, which arises from the adopted mass spectrum.}
\label{fig:rhomumap}
\end{figure}

\begin{figure*}
\includegraphics[width=168mm]{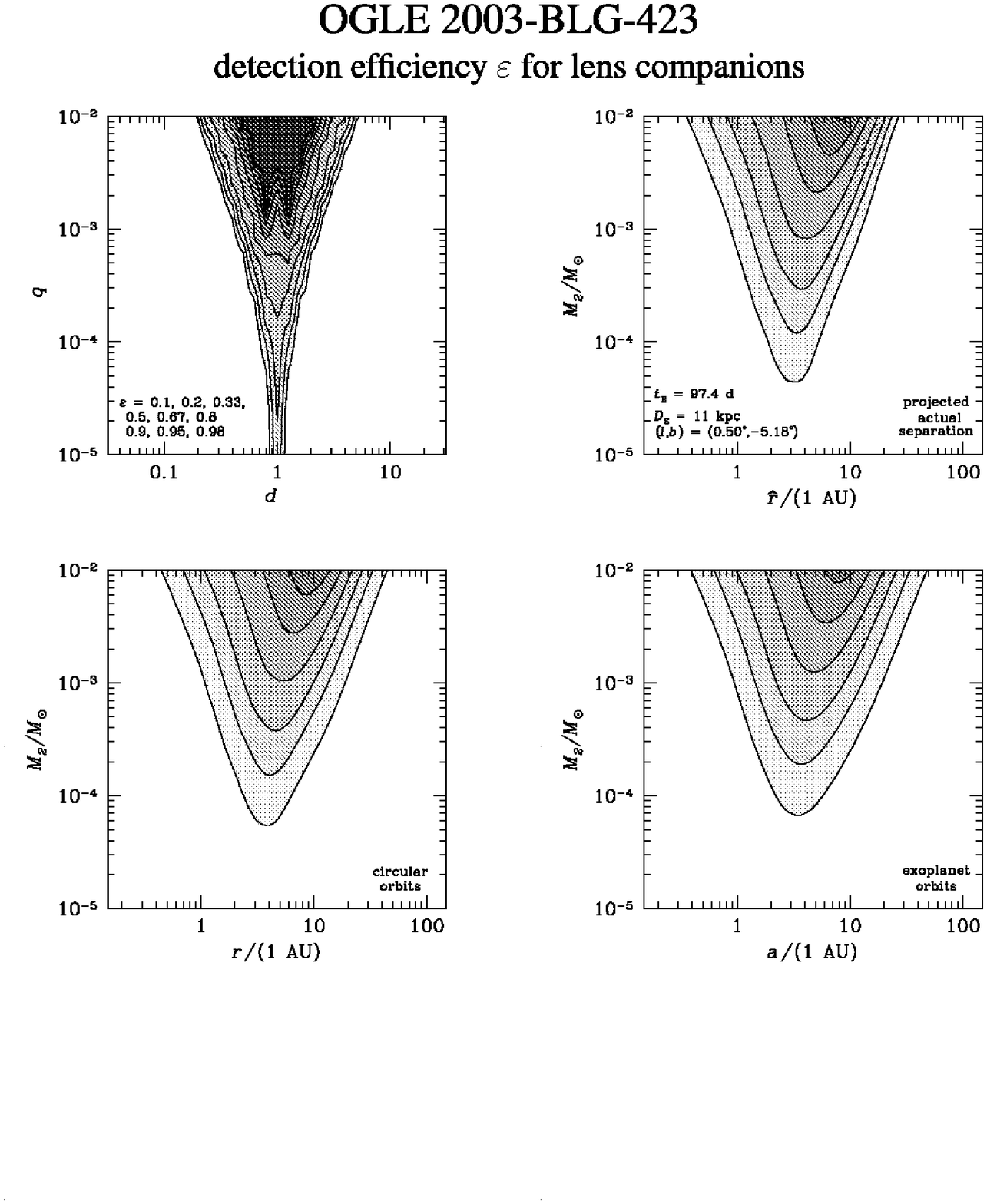}
\caption{Detection efficiency maps resulting from MicroFUN and OGLE data collected for the event
OGLE 2003-BLG-423. The top left panel shows the detection efficiency $\varepsilon_{d,q}(d,q)$
in the model parameters $(d,q)$ as calculated by \citet{GG:Esti}, where the actual angular separation between the lens constituents
is $d\,\theta_\rmn{E}$ and $q$ denotes their mass ratio. The remaining panels show the 
detection efficiency as function of the
physical lens properties derived using the Galaxy model described in 
Appendix~\ref{sec:galmodel}, using the event time-scale $t_\rmn{E} = 97.4~\rmn{d}$ resulting from 
the photometric data,
$D_\rmn{S} = 11~\rmn{kpc}$
and the Galactic coordinates $(l,b) = (0.50\degr, -5.18\degr)$. All these plots refer to the secondary mass $M_2$, but
different separations, where the top right panel shows the detection efficiency for the
actual projected distance perpendicular to the line-of-sight $\hat r = d\,r_\rmn{E}$, the bottom left
planet refers to the orbital radius $r$ assuming circular orbits, and the semi-major axis $a$ is
used in the bottom right panel assuming elliptical orbits with the eccentricity distribution
$\Phi_e = (4/\upi)\,\sqrt{1-e^2}$ (see Appendix~\ref{sec:orbitalproj}). For each of the plots, 
contour levels are shown at $\varepsilon = 0.1, 0.2, 0.33, 0.5, 0.67, 0.8, 0.9, 0.95, 0.98$. While
all these contours shown up for the plot of $\varepsilon_{d,q}(d,q)$, some contours that correspond to
larger detection efficiencies fall outside the displayed region for the other plots.}
\label{fig:efficiencymaps}
\end{figure*}

A given semi-major axis $a$ of the binary lens may result in different projected actual
separations $\hat r$ depending on the spatial orientation of the orbit, the orbital phase
and the orbital eccentricity. With $\Phi_{\hat \chi}(\hat \chi)$ denoting the probability density
of $\hat \chi = \hat r/a$ as derived in Appendix~\ref{sec:orbitalproj}, the 
detection efficiency in $(a,M_2)$ follows as
\begin{equation}
\varepsilon_{a,M_2}(a,M_2) = \int \limits_0^{{\hat \chi}^\rmn{max}}
\varepsilon_{\hat r,M_2}({\hat \chi}\,a,M_2)\,\Phi_{\hat \chi}(\hat \chi)\,\rmn{d}{\hat \chi}\,.
\label{eq:epsphys}
\end{equation}
For circular orbits, the expression for $\Phi_{\hat \chi}(\hat \chi)$ given by Eq.~(\ref{eq:phatchi0}) 
yields with 
a variable substitution in favour of $\hat w = \sqrt{1-{\hat \chi}^2}$ and ${\hat \chi}^\rmn{max} = 1$
\begin{equation}
\varepsilon_{a,M_2}(a,M_2) = \int\limits_0^1
\varepsilon_{\hat r,M_2}(\sqrt{1-{\hat w}^2}\;a,M_2)\,\rmn{d}{\hat w}\,.
\end{equation}

Following a pilot analysis of the event OGLE 1998-BUL-14 \citep{PLANET:O14}, for which the
underlying technique has been developed, the PLANET (Probing Lensing Anomalies NETwork)
collaboration has calculated 
detection efficiency maps in the parameters $(d,q)$ for its data collected from 1995 to
1999 in order to derive upper abundance limits on planetary companions to the
lens stars \citep{PLANET:fiveshort,PLANET:fivelong}. 
Detection efficiency maps have also been derived
by the MPS (Microlensing Planet Search) and MOA (Microlensing Observations in Astrophysics)
collaborations for
the event MACHO 1998-BLG-35 \citep{MPSMOA:M35}, and several other groups for suitable events
\citep{Tsapras:StAndSearch, MOA:planets, GG:Esti, Dong},
while \citet{Tsapras:OGLElimits} and \citet{Snodgrass:OGLElimits} have determined 
planetary abundance limits from OGLE (Optical Gravitational Lens Experiment) data.
The largest sensitivity to planets so far was achieved for the event MOA~2003-BLG-32/OGLE~2003-BLG-219
 \citep{Abe:bestlimits} that showed an extreme peak magnification in excess of 500.
PLANET is in the progress of carrying out a new comprehensive analysis including
the more recently observed events \citep{PLANET:newdeteff}, where, based on the results presented in
this section, a planetary abundance $f_{a,M_2}(a,M_2)$ rather than $f_{d,q}(d,q)$ is considered.

%For the parameters of the event OGLE 2004-BLG-234, where the source star is located towards
%$(l,b) = (2.45\degr, -2.50\degr)$ and the best-fitting event time-scale assuming a point lens 
%is $t_\rmn{E} = 11.8~\rmn{d}$,
%Figure~\ref{fig:rhomumap} shows the bivariate probability density $p_{\lg r_\rmn{E}/(1 AU), \lg M/M_{\sun}}$
%as function of the Einstein radius $r_\mathrm{E}$ and the
%lens mass $M$, where
%$D_\rmn{S} = 8.5~\rmn{kpc}$ has been assumed. 

For the parameters of the event OGLE 2003-BLG-423, where the source star is located towards
$(l,b) = (0.50\degr, -5.18\degr)$ and the best-fitting event time-scale assuming a point lens 
is $t_\rmn{E} = 97.4~\rmn{d}$,
Figure~\ref{fig:rhomumap} shows the bivariate probability density $p_{\lg r_\rmn{E}/(1~\rmn{AU}), \lg M/M_{\sun}}$
as function of the Einstein radius $r_\mathrm{E}$ and the
lens mass $M$, where
$D_\rmn{S} = 11~\rmn{kpc}$ has been assumed.
The bold line marks the upper limit for
$r_\mathrm{E}$, which corresponds to the lens being half-way between source and observer ($x = 0.5$),
and the mass cut-off at $M = 0.01~M_{\sun}$, resulting from the adopted mass spectrum.
As the Einstein radius $r_\rmn{E}$ approaches its maximal value, $p_{\lg r_\rmn{E}/(1~\rmn{AU}), \lg M/M_{\sun}}$
diverges.

%The detection efficiency as function of the model parameters $(d,q)$
%that has been calculated by \citet{PLANET:newdeteff} for this event is shown in the top left
%panel of Fig.~\ref{fig:efficiencymaps}.
The detection efficiency as function of the model parameters $(d,q)$
that has been calculated by \citet{GG:Esti} based on data collected by MicroFUN (Microlensing
Follow-Up Network) and OGLE for this event is shown in the top left
panel of Fig.~\ref{fig:efficiencymaps}. Data from the OGLE survey are made available 
on-line\footnote{{\tt http://www.astrouw.edu.pl/\~{}ogle/ogle3/ews/ews.html}} 
as the events are progressing \citep{OGLE:general3}, 
which significantly eases the assessment of their parameters and thereby allows the optimization
of follow-up observations.
Using the expressions presented in this section and adopting the
Galaxy model described in Appendix~\ref{sec:galmodel}, corresponding detection efficiency maps in
physical quantities have been determined, where the single time-scale 
%$t_\rmn{E} = 11.8~\rmn{d}$
$t_\rmn{E} = 97.4~\rmn{d}$
has
been used rather than the more exact best-fitting value for each $(d,q)$ that refers to the primary mass,
and a source distance of 
%$D_\rmn{S} = 8.5~\rmn{kpc}$
$D_\rmn{S} = 11~\rmn{kpc}$
has been assumed.
These maps are shown in the remaining panels of Fig.~\ref{fig:efficiencymaps}, where the
detection efficiency refers to the secondary mass $M_2$ and the projected actual separation $\hat r$,
the orbital radius $r$ for circular orbits, or the semi-major axis $a$ for elliptical orbits.
The distribution of the eccentricity $e$ for elliptical orbits has been chosen to be $\Phi_e(e) 
= (4/\upi) \sqrt{1-e^2}$, which provides a rough model of
the eccentricities for planetary orbits found by radial velocity searches
(see Appendix~\ref{sec:orbitalproj}).

All detection efficiency maps show the similar pattern of a maximum efficiency for
a characteristic separation 
decreasing both towards smaller and larger separations and a decrease of detection efficiency towards
smaller masses. As compared to the detection efficiency in the model parameters $(d,q)$, the other 
panels show the detection efficiency being stretched over a broader range of parameter space, so that
peak detection efficiencies are reduced, while smaller values occupy wider regions. 
The main broadening occurs on the transition from $(d,q)$ to $(\hat r,M_2)$, so that the width of the
distributions of the lens mass, distance, and velocity yield the dominant contribution rather
than the orbital projection, which has a more moderate but highly significant effect. While the map
in the projected actual separation $\hat r$ reflects the upper limit of the Einstein radius $r_\rmn{E}$,
this is smeared out by the distribution of the projection factor when considering the semi-major axis
instead. The average orbital radius $r$ for circular orbits exceeds the average semi-major axis
for the considered elliptic orbits, and for elliptic orbits, the detection efficiency is also
stretched towards smaller
$a$, with $1 < \hat\chi = \hat r/a < 2$ being possible. A substantial detection efficiency for small 
planetary masses results from the large abundance of parent stars with small masses,
whereas stars much heavier than the Sun are rare,
and the fact that large detection efficiencies from larger mass ratios provide substantial
contributions with the finite
width of the mass distribution for a given event time-scale.

\section{Summary and outlook}
\label{sec:summary}

This paper provides a comprehensive 
theoretical framework for the estimation of lens and source properties on the basis of 
the related model parameters that are estimated from the observational data.
This formalism can be used to answer a large variety of questions about the nature of 
individual microlensing events.
With the adopted Galaxy model and a source star residing in the direction of Baade's window
$(l,b) = (1\degr, -3.9\degr)$ at $D_\rmn{S} = 8.5~\rmn{kpc}$, 35 per cent of all ongoing events 
(not identical with the monitored or detected ones)
are caused by lenses in the Galactic disk and 65 per cent by
lenses in the Galactic bulge. While the bulge lenses clearly dominate the total event rate only for very small
time-scales $t_\rmn{E} \la 2~\rmn{d}$, 
the disk lenses yield the larger contribution for time-scales
$2~\rmn{d} \la t_\rmn{E} \la 40~\rmn{d}$
and $t_\rmn{E} \ga 100~\rmn{d}$, where the latter however yield only a small contribution to the total rate. For $2~\rmn{d} \la t_\rmn{E} \la 40~\rmn{d}$, bulge and disk lenses yield comparable contributions.
The provision of probability densities of the underlying characteristics of the 
lens and source system such as the lens mass $M$, the distance $D_\rmn{L}$ and
the effective transverse absolute velocity $v$ under the assumption of mass spectra,
mass densities and velocity distributions yields the largest amount of information that
can be extracted from the observations, i.e.\ much more than by a finite number of moments.
While a mass $M \sim 0.36~M_{\sun}$ for $t_\rmn{E} = 20~\rmn{d}$
is in rough agreement with estimates using a 'typical' fractional 
lens distance $x$ and transverse velocity $v$, the assumed mass spectrum with a low abundance
for $M \ga 1~M_{\sun}$ forces the expected mass to be more narrowly distributed with
$t_\rmn{E}$ rather than to follow the naive $M \propto t_\rmn{E}^2$ law. In particular,
there are only 1.5 decades $0.09~M_{\sun} \leq M \leq 3~M_{\sun}$ for the expected mass if 
time-scales $3~\rmn{d} \leq t_\rmn{E} \leq 150~\rmn{d}$ are considered
where the inclusion of one standard deviation
extends this range to $0.03~M_{\sun} \leq M \leq 15~M_{\sun}$.

Additional constraints such as those resulting from a measurement of the relative proper motion
between lens and source from observed finite-source effects or the relative lens-source parallax
as well as upper limits on these quantities resulting from the absence of related effects can be
incorporated. Explicitly one sees how uncertainties in $M$, $D_\rmn{L}$, and
$v$ are reduced,
although the respective probability densities can also widen if the additional constraint forces
the lens to assume values that fall into regions disfavoured by the given time-scale.
For any set of observables,
one also obtains a probability for the lens to reside in any of
the potential lens populations. Unless there are sufficient constraints to yield a sharp value for the
lens mass, distance, and velocity for a given set of model parameters, the uncertainties of the latter
can be neglected against the broad distributions of the relevant characteristics of the lens populations
and the Galaxy model uncertainties. 
With significant effects by annual parallax on the light curve starting at $\pi_\rmn{E} \la 0.8$,
such a limit can be detected in an event with time-scale $t_\rmn{E} = 20~\rmn{d}$ with a 
sensitivity to an angular positional shift within $t_\rmn{E}$ of $\kappa_\pi\,\theta_\rmn{E}
\sim 0.05~\theta_\rmn{E}$, whereas $\kappa_\pi = 1$ is reached for $t_\rmn{E} \sim 90~\rmn{d}$.
Similarly, finite-source effects become apparent if the angular source size $\theta_\star$
becomes a fair fraction of the angular impact $u_0\,\theta_\rmn{E}$ between lens and source.
By requiring $u_0 \la 2\,(\theta_\star/\theta_\rmn{E}) = 2\,{\hat \rho}_{\star,\sun}\,(R_\star/R_{\sun})$,
a limit ${\hat \rho}_{\star,\sun} \leq 0.005$ for $R_\star = R_{\sun}$ is detected if $u_0 \la 0.01$,
corresponding to a peak magnification $A_0 \ga 100$, whereas an impact parameter $u_0 \la 0.1$,
corresponding to $A_0 \ga 10$,
is sufficient for $R_\star = 10~R_{\sun}$.

In addition to the basic quantities, probability densities of the orbital semi-major axis
and the orbital period for binary lenses, as well as of any quantity that depends on the
basic characteristics, can be 
obtained. The bivariate probability density of the Einstein radius $r_\rmn{E}$ and the lens mass $M$
together with statistics of binary orbits yields detection efficiency maps for planetary companions to the 
lens star as function of the planet mass $M_2$ and its orbital semi-major axis $a$ rather than of the
model parameters $d$ and $q$, where $d\,\theta_\rmn{E}$ is the actual angular separation from its parent
star and $q$ is the planet-to-star mass ratio. The presented formalism has been applied to
some first examples and will be used for discussing the implications of many further events.
This paper explicitly shows the distributions of event properties
for the binary-lens model of microlensing event OGLE 2002-BLG-099 
\citep{OGLEbinaries}, namely of the masses of the lens components and
their distance from the observer, as well as of 
the orbital semi-major axis and period. Moreover, it shows the detection efficiency
map in $(a,M_2)$ resulting from 
%PLANET and OGLE data \citep{PLANET:newdeteff} for the event OGLE 2005-BLG-234.
MicroFUN and OGLE data \citep{GG:Esti} for the event OGLE 2003-BLG-423.
As a function of $(a,M_2)$, the detection efficiency stretches
over a much broader range of parameter space than 
for the $(d,q)$-map. In particular, this results in a larger detection efficiency for low-mass planets
than one would expect from typical values.

\section*{Acknowledgments}
This work has been made possible by postdoctoral support 
on the PPARC rolling grant
PPA/G/O/2001/00475. The basic ideas that are presented here
have developed steadily over the last few years, where work
has been supported by research grant Do~629/1-1 from
Deutsche Forschungsgemeinschaft (DFG), Marie Curie fellowship
ERBFMBICT972457, and award GBE 614-21-009 from Nederlandse Organisatie
voor Wetenschappelijk Onderzoek (NWO).
During this time, some discussions with A.~C.~Hirshfeld, 
K.~C.~Sahu, P.~D.~Sackett, P.~Jetzer, K.~Horne, D.~Bennett, and H.~Zhao added some valuable insight.
Careful reading of the manuscript by P.~Fouqu{\'e} and J.~Caldwell
helped eliminating some mistakes.
I would like to thank the MicroFUN and OGLE collaborations, in particular J.~Yoo, for
the provision of a detection efficiency map as function of the binary-lens
model parameters for event OGLE 2003-BLG-423 and the permission to show
a corresponding figure in this paper, and the PLANET collaboration, in particular
A.~Cassan and D.~Kubas, for providing me
with detection efficiency maps of several events, on which I could test my routines.
Last but not least, the success
of microlensing observations crucially depends on the 
provision of on-line alerts, as offered by past and present
survey collaborations such as EROS, MACHO, OGLE, and MOA.

\bibliographystyle{mn2e}
\bibliography{estimate_mn_rev}

\appendix

\section{General probabilistic approach}
\label{sec:general}

Let us consider a system characterized by $k$ properties $a_i$ ($i = 1
\ldots k$) that are distributed statistically, where $\Phi_{a_i}(a_1 \ldots a_k)\,
\rmn{d}a_i$ gives the probability to find the property $a_i$ in the
interval $[a_i, a_i+\rmn{d}a_i]$ which might depend on all
$a_1 \ldots a_k$. Further consider any realization of 
these system properties yielding a specific contribution to observed events
described by a weight function $\Omega(a_1 \ldots a_k)$ which may be chosen
appropriately to include selection effects caused by the experiment, so that the
total event rate according to their statistical representation is given by
\begin{eqnarray}
& & \hspace*{-2.2em} 
\Gamma(\Phi_{a_i} \ldots \Phi_{a_k}; \Omega)  \nonumber \\
& & \hspace*{-1.7em} = \;
\multint{k} \Omega(a_1 \ldots a_k) \left\{\prod_{i=1}^{k} \Phi_{a_i}(a_1 \ldots a_k)\right\}
\left\{\prod_{j=1}^{k} \rmn{d}a_j\right\},
\end{eqnarray}
where the notation refers to a $k$-dimensional integral.

Hence, the probability density of the
properties $a_1 \ldots a_k$ among all observed events is proportional to
\begin{eqnarray}
& & \hspace*{-2.2em}
\widetilde{p}_{a_1 \ldots a_k}(a_1 \ldots a_k; \Phi_{a_1} \ldots
\Phi_{a_k}, \Omega) \nonumber \\
& & \qquad \quad = \;
\Omega(a_1 \ldots a_k) \prod\limits_{i=1}^{k} \Phi_{a_i}(a_1 \ldots a_k)\,,
\end{eqnarray}
so that an appropriately normalized probability density is given by
\begin{eqnarray}
& & \hspace*{-2.2em}
p_{a_1 \ldots a_k}(a_1 \ldots a_k; \Phi_{a_1} \ldots
\Phi_{a_k}, \Omega) \nonumber \\
& & \hspace*{-1.7em} = \; \widetilde{p}_{a_1 \ldots a_k}(a_1 \ldots a_k; \Phi_{a_1} \ldots
\Phi_{a_k}, \Omega) \; \Bigg/ \nonumber\\
& & \hspace*{-1.7em}
\Bigg/ \multint{k}
\widetilde{p}_{a_1 \ldots a_k}(a_1 \ldots a_k; \Phi_{a_1} \ldots
\Phi_{a_k}, \Omega)\,\prod\limits_{i=1}^{k} \rmn{d}a_i\,,
\label{eq:normalize}
\end{eqnarray}
which does not depend on any constant factors in $\Omega$.
This means that $p_{a_1 \ldots a_k}(a_1 \ldots a_k; \Phi_{a_1}\ldots
\Phi_{a_k}, \Omega)$ is obtained by
weighting the intrinsic 
statistical distribution $\prod_{i=1}^{k} \Phi_{a_i}(a_i)$ 
by $\Omega(a_1 \ldots a_k)$ and normalizing the resulting product, so that
Eq.~(\ref{eq:normalize})
corresponds to Bayes' theorem. 

A specific event involves a set of $n$ observed 
parameters $f_l(a_1 \ldots a_k)$, where $l = 1 \ldots{}n$, which in general depend on the
$k$ basic underlying properties $a_i$, but are not necessarily identical
to these. With specific realizations $f_l^{(0)}$ for an observed event, the
event rate can be written as integral over these realizations 
\begin{eqnarray}
& & \hspace*{-2.2em}
\Gamma(\Phi_{a_i} \ldots \Phi_{a_k}, \Omega) \nonumber \\
& & \hspace*{-1.7em} = \;
\multint{n}
\gamma_{f_1 \ldots f_n}(f_1^{(0)} \ldots f_n^{(0)};
\Phi_{a_i} \ldots \Phi_{a_k}, \Omega) \; \times \nonumber \\
& & \hspace*{-1.7em} \times \;
\prod\limits_{l=1}^{n}\,\rmn{d}f_l^{(0)}
\end{eqnarray}
with the event rate density
\begin{eqnarray}
& & \hspace*{-2.2em}
\gamma_{f_1 \ldots f_n}(f_1^{(0)} \ldots f_n^{(0)};
\Phi_{a_i} \ldots \Phi_{a_k}, \Omega) \nonumber \\
& & \hspace*{-1.7em} = \;
\multint{k} \Omega(a_1 \ldots a_k) 
\left\{\prod\limits_{l=1}^{n} \delta\left(f_l(a_1 \ldots a_k)-f_l^{(0)}\right)
\right\} \; \times \nonumber \\
& & \hspace*{-1.7em} \times \;
\left\{
\prod\limits_{i=1}^{k} \Phi_{a_i}(a_i)\right\}\,
\left\{
\prod\limits_{j=1}^{k} \rmn{d}a_j\right\}\,,
\label{eq:ratedensity}
\end{eqnarray}
so that
the corresponding probability density of the basic properties
$a_1 \ldots a_k$ is given by
\begin{eqnarray}
& & \hspace*{-2.2em}
p^{(0)}_{a_1 \ldots a_k} (a_1 \ldots a_k, f_1^{(0)} \ldots f_n^{(0)}; 
\Phi_{a_1} \ldots \Phi_{a_k}, \Omega)  \nonumber \\
& & \hspace*{-1.7em} 
= [\gamma_{f_1 \ldots f_n}(f_1^{(0)} \ldots f_n^{(0)};
\Phi_{a_i} \ldots \Phi_{a_k}, \Omega)]^{-1}\; \times \nonumber \\
& & \hspace*{-1.7em} \times \;
\Omega(a_1 \ldots a_k) 
\bigg\{\prod\limits_{l=1}^{n} \delta\left(f_l(a_1 \ldots a_k)-f_l^{(0)}\right)\bigg\}
\; \times \nonumber \\
 & & \hspace*{-1.7em} \times \;
\bigg\{\prod\limits_{i=1}^{k} \Phi_{a_i}(a_i)\bigg\}\,,
\label{eq:p0}
\end{eqnarray}
while the probability density of a single property $a_r$ reads
\begin{eqnarray}
& & \hspace*{-2.2em}
p^{(0)}_{a_r}(a_r; f_1^{(0)} \ldots f_n^{(0)}; 
\Phi_{a_1} \ldots \Phi_{a_k}, \Omega)  \nonumber \\
& & \hspace*{-1.7em} 
= [\gamma_{f_1 \ldots f_n}(f_1^{(0)} \ldots f_n^{(0)};
\Phi_{a_i} \ldots \Phi_{a_k}, \Omega)]^{-1}\; \times \nonumber \\
& & \hspace*{-1.7em} \times \;
\multint{k-1} \Omega(a_1 \ldots a_k) 
\bigg\{\prod\limits_{l=1}^{n} \delta\left(f_l(a_1 \ldots a_k)-f_l^{(0)}\right)\bigg\}
\; \times \nonumber \\
& & \hspace*{-1.7em} \times \;
\bigg\{\prod\limits_{i=1}^{k} \Phi_{a_i}(a_i)\,\bigg\}\,
\bigg\{\prod\limits_{j=1 \atop j \neq r}^{k}\rmn{d}a_j\bigg\}
\,.
\end{eqnarray}

If the observables $f_1 \ldots f_n$ for one or more events follow
a distribution $\Psi_{f_1 \ldots f_n}(f_1 \ldots f_n)$, the probability density of the
basic properties $a_1 \ldots a_k$ arises from an integral over the
probability density $p^{(0)}$ for the fixed values $f_1^{(0)} \ldots
f_n^{(0)}$, given by Eq.~(\ref{eq:p0}), as
\begin{eqnarray}
& & \hspace*{-2.2em}
p_{a_1 \ldots a_k}^{\psi} (a_1 \ldots a_k; 
\Phi_{a_1} \ldots \Phi_{a_k}, \Omega) \nonumber \\
& & \hspace*{-1.7em} = \;
\multint{n} p^{(0)}_{a_1 \ldots a_k} (a_1 \ldots a_k,
f_1 \ldots f_n; 
\Phi_{a_1} \ldots \Phi_{a_k}, \Omega)\; \times \nonumber \\
& & \hspace*{-1.7em} \times \;
\Psi_{f_1 \ldots f_n}(f_1 \ldots f_n)\,
\prod_{l=1}^{n} \rmn{d}f_l\,.
\label{eq:ppsi}
\end{eqnarray}
If the observables are statistically independent, their distribution
factorizes as $\Psi_{f_1 \ldots f_n}(f_1 \ldots f_n) =
\prod_{l=1}^{n} \Psi_{f_l}(f_l)$.
While fixed values of $f_l$ correspond to the distribution
$\Psi_{f_l}(f_l) = \delta(f_l-f_l^{(0)})$, distributions around a
central value $f_l^{(0)}$ with a standard deviation $\sigma_{f_l}$ 
can be approximated by the Gaussian distribution 
\begin{equation}
\Psi_{f_l}^\rmn{Gauss}(f_l) = \frac{1}{
\sqrt{2\upi}\sigma_{f_l}}\,\exp\left\{-\frac{\left(f_l-f_l^{(0)}\right)^2}{2\sigma_{f_l}^2}\right\}\,.
\end{equation}
In case the $n$ observables ${\vec f} = (f_1,\ldots,f_n)$ are correlated, 
$\Psi_{f_1 \ldots f_n}(f_1 \ldots f_n)$ can be modelled as a multivariate Gaussian distribution
\begin{eqnarray}
& & \hspace*{-2.2em}
\Psi_{f_1 \ldots f_n}(f_1 \ldots f_n) = \frac{1}{(2\upi)^{n/2}\,|{\cal C}|^{1/2}}\; \times \nonumber \\
& & \hspace*{-1.7em} \times \;
\exp\left\{-\frac{1}{2}\,\left(\vec f - {\vec f}^{(0)}\right)^{\rmn{T}} {\cal C}^{-1}
 \left(\vec f - {\vec f}^{(0)}\right)\right\}\,, 
\end{eqnarray}
where ${\cal C}^{-1}$ is the inverse and $|\cal C|$ is the determinant of the 
covariance matrix $\cal C$, and
${\vec f}^{(0)} = (f_1^{(0)},\ldots,f_n^{(0)})$.

The moments of any property $g(a_1 \ldots a_k)$ for fixed values of
the observables $f_1^{(0)} \ldots f_n^{(0)}$ follow from the expectation
values
\begin{eqnarray}
& & \hspace*{-2.2em}
{\left\langle g^\beta(a_1 \ldots a_k) \right\rangle}^{(0)} = 
\multint{k}
g^\beta(a_1 \ldots a_k) \; \times \nonumber \\
& & \hspace*{-1.7em}
\times \; p^{(0)}_{a_1 \ldots a_k} (a_1 \ldots a_k, f_1^{(0)} \ldots f_n^{(0)}; 
\Phi_{a_1} \ldots \Phi_{a_k}, \Omega) \; \times \nonumber \\
& & \hspace*{-1.7em} \times \;
\prod\limits_{i=1}^{k} \rmn{d}a_i\,,
\label{eq:expval0}
\end{eqnarray}
and for a distribution $\Psi_{f_1 \ldots f_n}$, one finds in
analogy to Eq.~(\ref{eq:ppsi})
\begin{eqnarray}
& & \hspace*{-2.2em}
{\left\langle g^\beta(a_1 \ldots a_k)\right\rangle}^{\Psi}
= 
\multint{n}\
{\left\langle g^\beta(a_1 \ldots a_k)\right\rangle}^{(0)}\;\times \nonumber \\
 & & \hspace*{-1.7em}
 \times \;
 \Psi_{f_1 \ldots f_n}(f_1 \ldots f_n)\,
\prod_{l=1}^{n} \rmn{d}f_l\,,
\end{eqnarray}
where interchanging the integrations over $\rmn{d}a_1 \ldots \rmn{d}a_k$ and
over $\rmn{d}f_1 \ldots \rmn{d}f_n$ yields the equivalent expression
\begin{eqnarray}
& & \hspace*{-2.2em}
{\left\langle g^\beta(a_1 \ldots a_k)\right\rangle}^{\Psi} = 
\multint{k}
g^\beta(a_1 \ldots a_k) \; \times \nonumber \\
& & \hspace*{-1.7em}
\times \; p^{\Psi}_{a_1 \ldots a_k} (a_1 \ldots a_k, f_1^{(0)} \ldots f_n^{(0)}; 
\Phi_{a_1} \ldots \Phi_{a_k}, \Omega) \; \times \nonumber \\
& & \hspace*{-1.7em} \times \;
\prod\limits_{i=1}^{k} \rmn{d}a_i\,.
\label{eq:expvalpsi}
\end{eqnarray}
In particular, the standard deviation is given by
\begin{equation}
\sigma_g = \sqrt{\left\langle g^2(a_1 \ldots a_k) \right\rangle
- \left\langle g(a_1 \ldots a_k) \right\rangle^2}\,.
\end{equation}

Beyond the moments, one finds the complete probability density of
a general property $g(a_1 \ldots a_k)$ for fixed values of
the observables $f_1^{(0)} \ldots f_n^{(0)}$ to be
\begin{eqnarray}
& & \hspace*{-2.2em}
p^{(0)}_{g}(g; f_1^{(0)} \ldots f_n^{(0)}; 
\Phi_{a_1} \ldots \Phi_{a_k}, \Omega) \nonumber \\
& & \hspace*{-1.7em} 
= [\gamma_{f_1 \ldots f_n}(f_1^{(0)} \ldots f_n^{(0)};
\Phi_{a_i} \ldots \Phi_{a_k}, \Omega)]^{-1}\; \times \nonumber \\
& & \hspace*{-1.7em} \times \;
\multint{k} \Omega(a_1 \ldots a_k)\,
\delta(g-g(a_1 \ldots a_k)) \; \times \nonumber \\
& & \hspace*{-1.7em} \times \;
\bigg\{\prod\limits_{l=1}^{n} \delta\left(f_l(a_1 \ldots a_k)-f_l^{(0)}\right)\bigg\}
\bigg\{\prod\limits_{i=1}^{k} \Phi_{a_i}(a_i)\,\rmn{d}a_i\bigg\}\,,
\label{eq:pdens0}
\end{eqnarray}
while for a distribution $\Psi_{f_1 \ldots f_n}$, one obtains
\begin{eqnarray}
& & \hspace*{-2.2em}
p_{g}^\Psi(g;
\Phi_{a_1} \ldots \Phi_{a_k}, \Omega) \nonumber \\
& & \hspace*{-1.7em} = \; 
\multint{n} p^{(0)}_{g} (g,
f_1 \ldots f_n; 
\Phi_{a_1} \ldots \Phi_{a_k}, \Omega) \; \times  \nonumber \\
& & \hspace*{-1.7em} \times \;
 \Psi_{f_1 \ldots f_n}(f_1 \ldots f_n)\,
\prod_{l=1}^{n} \rmn{d}f_l\,.
\label{eq:pdensgen}
\end{eqnarray}

It is important to distinguish carefully
the different quantities that have been defined in this section.
The system properties $a_1 \ldots a_k$ are distributed {\em statistically}
among the population according to $\Phi_{a_1} \ldots \Phi_{a_k}$.
With $\Omega(a_1 \ldots a_k)$ being the weight of any realization to the
number of produced events, one expects $a_1 \ldots a_k$ being distributed
as $p_{a_1 \ldots a_k}(a_1 \ldots a_k; \Phi_{a_1} \ldots \Phi_{a_k}, 
\Omega)$ among all events. For a given event, with a set of 
observables $f_1(a_1 \ldots a_k) \ldots f_n(a_1 \ldots a_k)$ being 
realized as $f_1^{(0)} \ldots f_n^{(0)}$, one infers a {\em stochastical}
probability density $p^{(0)}_{a_1 \ldots a_k}(a_1 \ldots a_k,
f_1^{(0)} \ldots f_n^{(0)}; \Phi_{a_1} \ldots \Phi_{a_k}, 
\Omega)$ of $a_1 \ldots a_k$ or $p_{g}^{(0)}(g,
f_1^{(0)} \ldots f_n^{(0)}; \Phi_{a_1} \ldots \Phi_{a_k}, 
\Omega)$ of any specific property $g(a_1 \ldots a_k)$, which does not need 
to be an observable $f_l$ or a basic property $a_j$.
Finally, one can consider the observables
$f_1 \ldots f_n$ to follow a {\em stochastical} distribution for a single 
event and/or a {\em statistical} distribution from several events, namely
$\Psi_{f_1 \ldots f_n}$ yielding
the probability densities $p_{a_1 \ldots a_k}^{\Psi}(a_1 \ldots a_k;
\Phi_{a_1} \ldots \Phi_{a_k}, 
\Omega)$  or $p_{g}^{\Psi}(g, \Phi_{a_1} \ldots \Phi_{a_k}, 
\Omega)$.

\section{Model of the Galaxy}
\label{sec:galmodel}

\subsection{Mass spectrum}

\begin{table*}
\begin{minipage}{126mm}
\caption{Coefficients for the mass laws for different parts of the 
mass spectrum (following \citet{Chabrier:massfunc})}
\label{tab:masslaws}
\begin{footnotesize}
\vspace*{1.0ex}
\begin{tabular}{lccccc}
& $\lg (M_\rmn{min}/M_{\sun})$ & $\lg (M_\rmn{max}/M_{\sun})$ &
$\alpha$ & $\lg (M_\rmn{c}/M_{\sun})$  & $\sigma_\rmn{c}$ \\ \hline
{\bf disk} 
& $-2.0$ & $-0.7$ & $-0.2$ & --- & --- \\
 & $-0.7$ & $0.0$ & --- & $-1.102$ & $0.69$ \\
 & $0.0$ & $0.54$ & $4.37$ & --- & --- \\
 & $0.54$ & $1.26$ & $3.53$ & --- & --- \\
 & $1.26$ & $1.8$ &  $2.11$ & --- & --- \\
 \hline
{\bf bulge} & -2.0 & -0.155 & --- & -0.658 & 0.33 \\
	& -0.155 & 1.8 & 1.3 & --- & --- \\
\end{tabular}
\end{footnotesize}

\medskip
For $M_\rmn{min} \leq M \leq M_\rmn{max}$, either a 
power-law mass function $\hat \Phi_{\lg (M/M_{\sun})}[\lg (M/M_{\sun})]
\propto (M/M_{\sun})^{-\alpha}$ or a Gaussian distribution 
$\hat \Phi_{\lg (M/M_{\sun})}[\lg (M/M_{\sun})] \propto  
\exp\{-0.5\,[\lg (M/M_{\sun}) - \lg (M_\rmn{c}/M_{\sun})]^2
/(\sigma_\rmn{c})^2\}$ is adopted.
\end{minipage}
\end{table*}

\begin{figure}
\includegraphics[width=84mm]{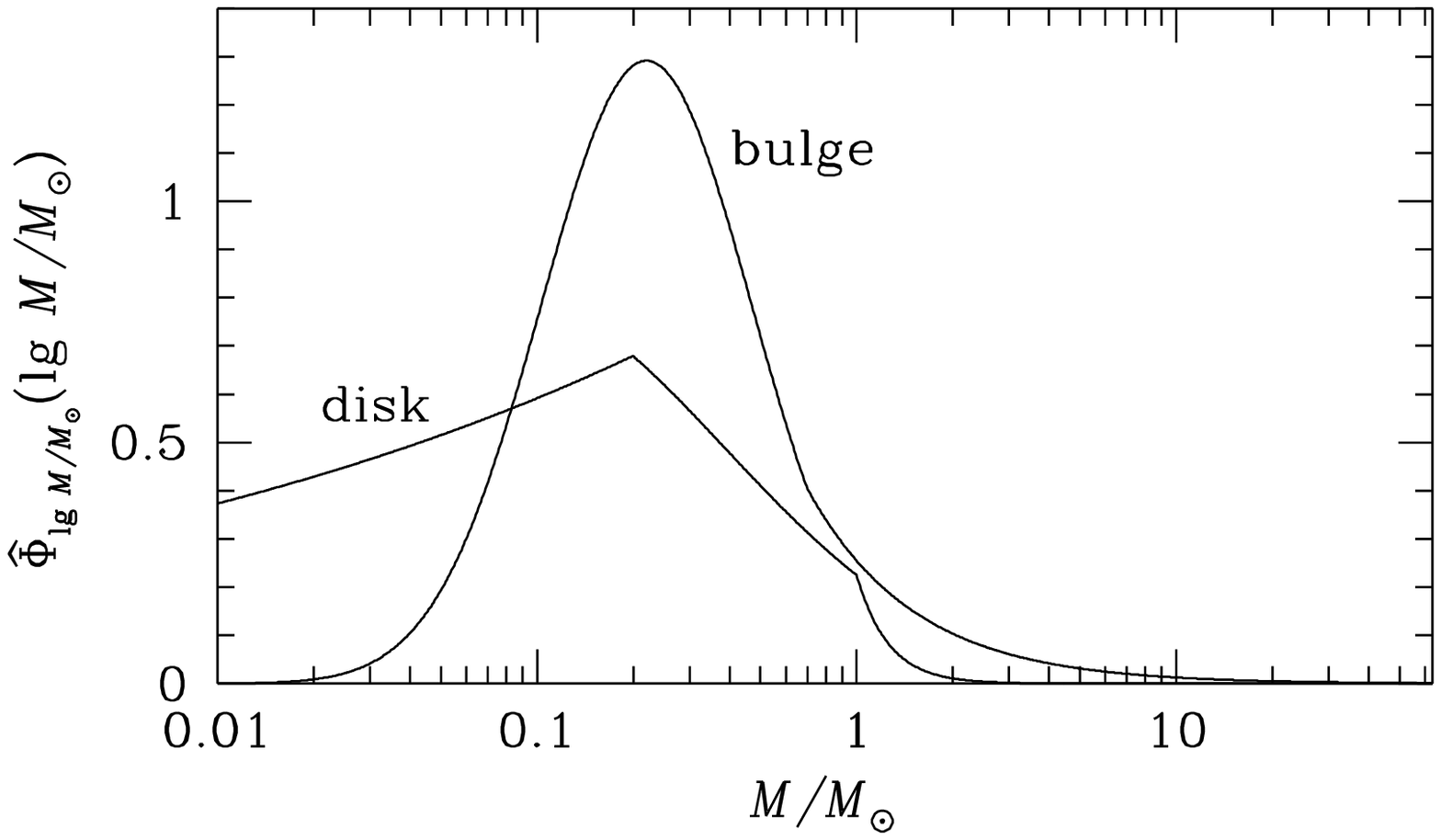}
\caption{Adopted mass function $\hat \Phi_{\lg M/M_{\sun}} \propto M^{-1}\,\Phi_{\lg M/M_{\sun}}$ for lens objects in the Galactic disk or bulge whose
parameters are listed in Table~\ref{tab:masslaws}.}
\label{fig:origmass}
\end{figure} 

While the mass spectrum $\Phi_M(M)$ as defined by 
Eq.~(\ref{eq:defmassspec})
gives the decomposition of the mass density into objects 
with mass in the range $[M, M+\rmn{d}M]$, the decomposition of the 
number density is given by the mass function
$\hat \Phi_M(M) \propto M^{-1}\,\Phi_M(M)$.
As pointed out e.g.\ in the review of \citet{Chabrier:massfunc}, the mass
function $\hat \Phi_{\lg (M/M_{\sun})}$ can be fairly well approximated piecewise
by different kinds of analytic functions. In particular,
for selected ranges
$M_\rmn{min}^{(i)} \leq M \leq M_\rmn{max}^{(i)}$, one may
consider a power-law  mass
spectrum in $M$, i.e.
\begin{equation}
\hat \Phi^{(i)}_{\lg (M/M_{\sun})}[\lg (M/M_{\sun})]
\propto (M/M_{\sun})^{-\alpha^{(i)}}
\end{equation}
 with the
power index $\alpha^{(i)}$, or a 
Gaussian distribution in $\lg (M/M_{\sun})$, i.e.
\begin{eqnarray}
& & \hspace*{-2.2em}
\hat \Phi^{(i)}_{\lg (M/M_{\sun})}[\lg (M/M_{\sun})]\; \propto  \nonumber \\
& & \hspace*{-1.7em} \propto \;
\exp\left\{-0.5\,[\lg (M/M_{\sun}) - \lg (M_\rmn{c}^{(i)}/M_{\sun})]^2
/(\sigma_\rmn{c}^{(i)})^2\right\}
\end{eqnarray}
with the characteristic mass $M_\rmn{c}^{(i)}$ and the 
width of the distribution $\sigma_\rmn{c}^{(i)}$.
The proportionality factors thereby have to be chosen so that
the mass spectrum
$\Phi_{\lg (M/M_{\sun})}$ is continuous at all $M_\rmn{max}^{(i)}
= M_\rmn{min}^{(i+1)}$, and its integral over all $\lg (M/M_{\sun})$
becomes unity. The choice for the parameters corresponding to different selected mass
ranges for
disk or bulge lenses following \citet{Chabrier:massfunc}, that is adopted
for this paper, is shown in Table~\ref{tab:masslaws}, and the corresponding 
mass function $\hat \Phi_{\lg (M/M_{\sun})}(\lg (M/M_{\sun}))$ is shown
in Fig.~\ref{fig:origmass}.

\subsection{Mass density}

The view from the observer to the source is reflected by a 
coordinate system with a basis 
$(\vec e_x, \vec e_y, \vec e_z)$
where $\vec e_x$ points from the observer to the source, while
$\vec e_y$ and $\vec e_z$ 
span a plane perpendicular to the line-of-sight. 
For describing properties of the Galaxy, however, it is more appropriate
to refer to the galactic coordinates $(D_\rmn{L},l,b)$ which are the
spherical coordinates that refer to the basis
$({\vec E}_X,{\vec E}_Y, 
{\vec E}_Z)$,
where ${\vec E}_X$ points to the
Galactic centre $(l,b) = (0\degr,0\degr)$,
${\vec E}_Y$ towards the direction of local circular motion 
$(l,b) = (90\degr,0\degr)$, and ${\vec E}_Z$ towards
 Galactic North  $b =90\degr$, so that 
 ${\vec X} = D_\rmn{L}\,(\cos l \cos b,
 \sin l \cos b,\sin b)$. The basis $(\vec e_x, \vec e_y, \vec e_z)$ arises
 from $({\vec E}_X,{\vec E}_Y, {\vec E}_Z)$
by a rotation ${\cal R}({\vec E}_Z; l)$  around ${\vec E}_Z$
by the angle $l$ and a subsequent rotation
${\cal R}(-{\cal R}({\vec E}_Z; l){\vec E}_Y; b)$ around
$-{\cal R}({\vec E}_Z; l) {\vec E}_Y$  by the angle $b$, i.e.\ 
${\vec e}_i = \sum_j {\cal T}_{ij}(l,b) {\vec E}_j$, where
\begin{eqnarray}
{\cal T}(l,b) & = &
\left(\begin{array}{ccc} \cos b & 0 & \sin b \\
0 & 1 & 0 \\ -\sin b & 0 & \cos b \end{array}\right)\,
\left(\begin{array}{ccc} \cos l & \sin l & 0 \\
- \sin l & \cos l & 0 \\ 0 & 0 & 1 \end{array}\right)  \nonumber \\
& = &
\left(\begin{array}{ccc}
\cos l \cos b & \sin l \cos b & \sin b \\
-\sin l & \cos l & 0 \\
-\cos l \sin b & -\sin l \sin b & \cos b \end{array}\right)\,,
\label{eq:galtrafo}
\end{eqnarray}
so that vector components transform as $\vec x =
\left({\cal T}^{-1}\right)^{\rmn{T}}(l,b)\,{\vec X} = {\cal T}(l,b)\,{\vec X}$ 
or ${\vec X} = {\cal T}^{-1}(l,b)\,{\vec x} = 
{\cal T}^{\rmn{T}}(l,b)\,{\vec x}$.

The density of matter in the Galaxy is more easily expressed in coordinate
frames whose origins are at the Galactic centre rather than at the 
position of the Sun. Just by subtracting the corresponding difference
location vector, one finds for such a system with
the fractional lens distance $x \equiv D_\rmn{L}/D_\rmn{S}$
and the galactic coordinates $(l,b)$ of the source star:
\begin{eqnarray}
\hat X & = & X - R_0 \;=\; x\,D_\rmn{S}\,\cos l\,\cos b - R_0\,,\nonumber \\
\hat Y & = & Y \;=\; x\,D_\rmn{S}\,\sin l\,\cos b \nonumber\,, \\
\hat Z & = & Z \;=\; x\,D_\rmn{S}\,\sin b\,.
\end{eqnarray}
Hence, the cylindrical distance from the Galactic centre is
\begin{eqnarray}
R & = & \sqrt{{\hat X}^2+{\hat Y}^2} \nonumber \\
& = & 
R_0\,\sqrt{[\cos l - x\,(D_\rmn{S}/R_0)\,\cos b]^2 +
\sin^2 l}\,,
\label{eq:R2D}
\end{eqnarray}
while the spherical distance reads
\begin{eqnarray}
r & = & \sqrt{{\hat X}^2+{\hat Y}^2+{\hat Z}^2} \nonumber \\
& = & \sqrt{R_0^2 + x^2 D_\rmn{S}^2 - 2\,x D_\rmn{S} R_0 \cos l \cos b}\,.
\end{eqnarray}

\begin{figure}
\includegraphics[width=84mm]{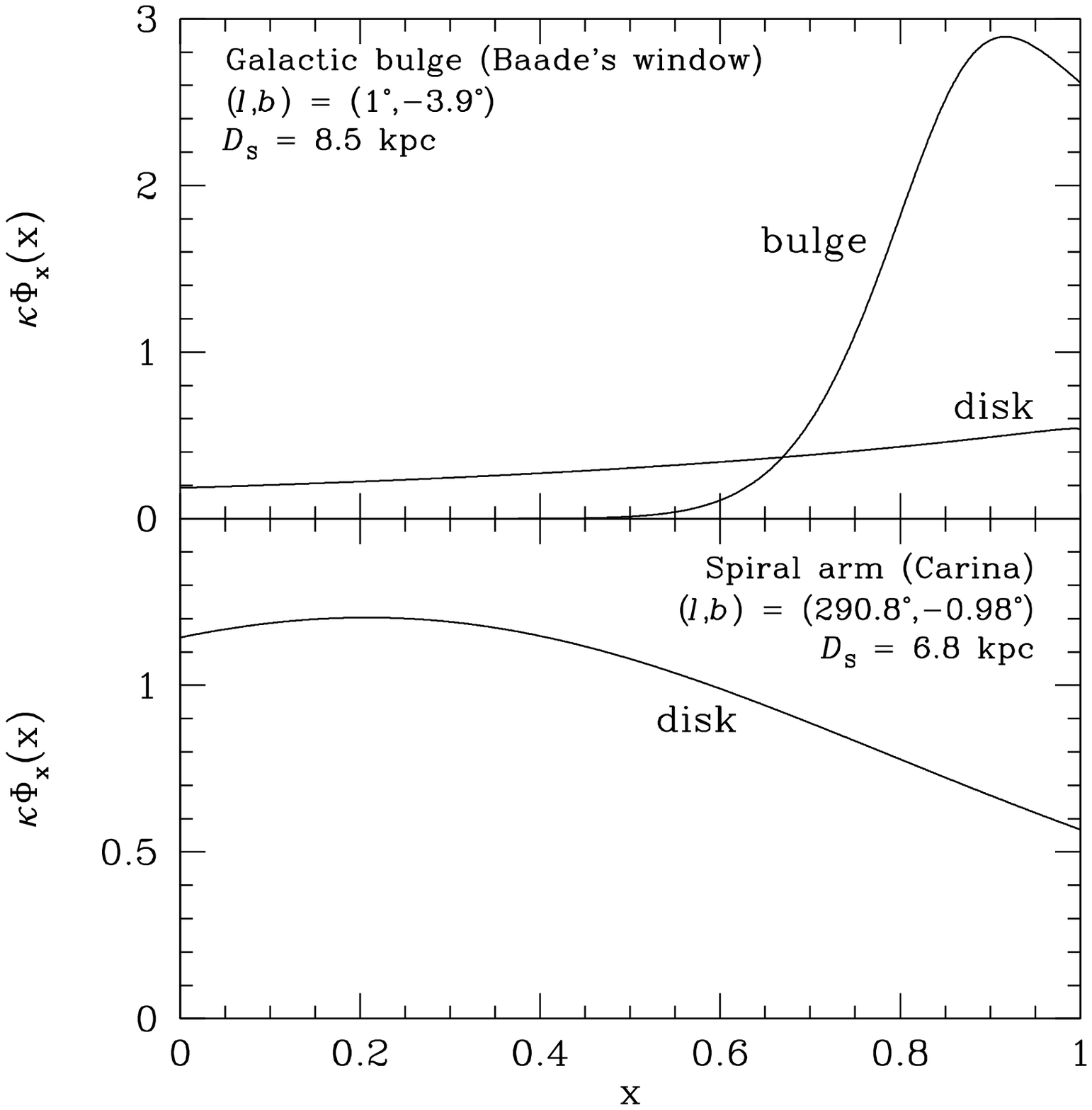}
\caption{Weighted probability density $\kappa\,\Phi_x(x)$ of the fractional lens
distance $x = D_\rmn{L}/D_\rmn{S}$ for two different positions of the source, and
lenses in the Galactic disk or bulge. The weight factors are given by
$\kappa^\rmn{disk} = \Sigma^\rmn{disk}/(\Sigma^\rmn{disk}+\Sigma^\rmn{bulge})$ and
$\kappa^\rmn{bulge} = \Sigma^\rmn{bulge}/(\Sigma^\rmn{disk}+\Sigma^\rmn{bulge})$, so
that $\kappa\,\Phi_x(x)$ reflects the mass density, where
$\int_0^1 [\kappa^\rmn{disk}\Phi_x^{\rmn{disk}} + \kappa^\rmn{bulge}\Phi_x^{\rmn{bulge}}]
\,\rmn{d}x = 1$, i.e.\ $\kappa^\rmn{disk} + \kappa^\rmn{bulge} = 1$. While
$\kappa^\rmn{disk} \sim 2\times 10^{-5}$ for the source in the spiral arm, one finds
$\kappa^\rmn{disk} = 0.33$ and $\kappa^\rmn{bulge} = 0.67$ for the bulge source.}
\label{fig:origdist}
\end{figure} 

As chosen by \citet{Jetzer:gal1}, the mass density of the disk is modelled
by two double-exponential disks, following \citet*{Bahcall:gal} and
\citet*{Gilmore:gal}, where
\begin{eqnarray}
& & \hspace*{-2.2em}
\rho^{\rmn{disk}}(R,Z) =
0.5\,\exp\left(-\frac{R-R_0}{h}\right)\; \times \nonumber \\
& & \hspace*{-1.7em} \times \;
\left[\frac{\Sigma_\rmn{thin}}{H_\rmn{thin}}
\exp\left(-\frac{|Z|}{H_\rmn{thin}}\right)+
\frac{\Sigma_\rmn{thick}}{H_\rmn{thick}}
\exp\left(-\frac{|Z|}{H_\rmn{thick}}\right)\right]
\end{eqnarray}
with $h = 3.5~\rmn{kpc}$ being the scale length in the galactic plane,
while $H_\rmn{thin} = 0.3~\rmn{kpc}$ and $H_\rmn{thick} =
1.0~\rmn{kpc}$ are the scale lengths of a thin and a thick disk perpendicular
to the galactic plane, and the corresponding column mass densities are $\Sigma_\rmn{thin} =
25~M_{\sun}\,\rmn{pc}^{-2}$ and $\Sigma_\rmn{thick} = 
35~M_{\sun}\,\rmn{pc}^{-2}$.

Similar to the discussion of \citet{HanGould:gal1,HanGould:gal2}
and \citet{Jetzer:gal1}, let us adopt a 
model of a barred bulge based on the COBE data \citep{Dwek:COBEbar},
which is tilted by an angle $\theta = 20\degr$ with respect to the
direction of the Galactic centre, so that coordinates along its main
axes with origin at the
Galactic centre are given by
\begin{eqnarray}
\hat X' & = & \hat X \cos \theta + \hat Y \sin \theta\,, \nonumber \\
\hat Y' & = & -\hat X \sin \theta + \hat Y \cos \theta\,, \nonumber \\
\hat Z' & = & \hat Z\,.
\end{eqnarray}
In these coordinates, the mass density of the Galactic bulge can be expressed by
\begin{equation}
\rho^{\rmn{bulge}}(\hat X', \hat Y', \hat Z') 
= \rho^{\rmn{bulge}}_0\,\exp\left\{-s^2/2\right\}\,,
\end{equation}
where 
\begin{equation}
s^2 = \sqrt{\left[\left(\hat X'/a\right)^2 + 
\left(\hat Y'/b\right)^2\right]^2 + \left(\hat Z'/c\right)^4}
\end{equation}
with $a = 1.58~\rmn{kpc}$, $b= 0.62~\rmn{kpc}$, and $c = 0.43~\rmn{kpc}$.
A total mass $M_\rmn{bulge} = 1.8 \times 10^{10}~M_{\sun}$ implies
$\rho^{\rmn{bulge}}_0 = M_\rmn{bulge}/(6.57\,\upi\,abc) = 
2.1 \times 10^{9}~M_{\sun}\,\rmn{kpc}^{-3}$.

For a source in the Galactic bulge towards Baade's window at $(l,b) = (1\degr,-3.9\degr)$ at 
a distance $D_\rmn{S} = 8.5~\rmn{kpc}$ as well as for a source in the Carina spiral
arm at $(l,b) = (290.8\degr,-0.98\degr)$ and a distance $D_\rmn{S} = 6.8~\rmn{kpc}$ as example
for an off-bulge target, 
the weighted probability densities $\kappa\,\Phi_x(x)$ of the fractional lens distance 
$x \equiv D_\rmn{L}/D_\rmn{S}$ are shown in Fig.~\ref{fig:origdist}. The weight factors
$\kappa^\rmn{disk} = \Sigma^\rmn{disk}/(\Sigma^\rmn{disk}+\Sigma^\rmn{bulge})$ and
$\kappa^\rmn{bulge} = \Sigma^\rmn{bulge}/(\Sigma^\rmn{disk}+\Sigma^\rmn{bulge})$ have
been chosen so
that $\int_0^1 [\kappa^\rmn{disk}\Phi_x^{\rmn{disk}} + \kappa^\rmn{bulge}\Phi_x^{\rmn{bulge}}]
\,\rmn{d}x = 1$ and $\kappa^\rmn{disk} + \kappa^\rmn{bulge} = 1$.
Not surprisingly, the contribution of bulge lenses is negligible for a source 
in the spiral arm, where for the chosen parameters, $\kappa^{\rmn{bulge}} = 2\times 10^{-5}$.
In contrast, for 
the bulge source, one finds contributions of comparable order, where $\kappa^{\rmn{disk}} = 0.33$  
and $\kappa^{\rmn{bulge}} = 0.67$.
While the lens mass density for the source in the spiral arm shows a broad distribution favouring
smaller lens distances, one finds
a moderate increase with distance for disk lenses and a source in the Galactic bulge, while bulge lenses yields 
significant contributions only for $x \ga 0.6$.

\subsection{Effective transverse velocity}

The effective transverse velocity in a plane at the lens distance 
$D_\rmn{L} = x\,D_\rmn{S}$ perpendicular to the line-of-sight is
given by 
\begin{equation}
\vec v(x) = \vec v_{\rmn{L}}
- x\, \vec v_{\rmn{S}} - (1-x)\, \vec v_{\rmn{O}}\,,
\end{equation}
where
$\vec v_{\rmn{L}}$, $\vec v_{\rmn{S}}$, and 
$\vec v_{\rmn{O}}$ denote the perpendicular velocities 
of the lens, source, or observer, respectively. 
Let us consider an expectation value ${\vec v}^0 = \left\langle \vec v\right\rangle$,
and the source and lens velocities follow Gaussian distributions, where
isotropic velocity dispersions are assumed for both the Galactic disk
and bulge. While the introduction of anisotropies heavily complicates
both the discussion and the calculation, the results are only 
marginally affected, and the arising differences
do not exceed those resulting from uncertainties in the velocity dispersions
themselves. Discussions of anisotropic 
velocity dispersions must not miss the non-diagonal elements of the 
velocity dispersion tensor for directions that do not coincide with
the main axes of the velocity dispersion ellipsoid.

In this paper, values of $\sigma^{\rmn{disk}} = 30~\rmn{km}\,\rmn{s}^{-1}$
for the Galactic disk and $\sigma^{\rmn{bulge}} = 100~\rmn{km}\,\rmn{s}^{-1}$
for the Galactic bulge have been adopted. Therefore, for bulge sources, the
total velocity dispersion is 
$\sigma(x) = \sqrt{1+x^2}\,\sigma^{\rmn{bulge}}$ for bulge lenses or
$\sigma(x) = \sqrt{x^2\,\left(\sigma^{\rmn{bulge}}\right)^2 
+ \left(\sigma^{\rmn{disk}}\right)^2}$ for disk lenses, while for disk
sources, where the lens also resides in the Galactic disk,
$\sigma(x) = \sqrt{1+x^2}\,\sigma^{\rmn{disk}}$.

The probability density $\Phi_v(v)$ of the absolute effective
velocity therefore takes the form
\begin{eqnarray}
& & 
\hspace*{-2.2em}\Phi_v(v,x) \nonumber \\
& & \hspace*{-1.7em}
= \frac{v}{2\upi\,[\sigma(x)]^2}
\int\limits_0^{2\upi}\,
\exp\left\{-\frac{1}{2\,[\sigma(x)]^2}\,\left(\vec v - {\vec v}^0(x)\right)^2\right\}\,
\rmn{d}\varphi
\nonumber \\
& & \hspace*{-1.7em}
= \frac{v}{2\upi\,[\sigma(x)]^2}\,
\exp\left\{-\frac{v^2+[v^0(x)]^2}{2\,[\sigma(x)]^2}\right\}\;\times \nonumber \\
& & \hspace*{-1.7em} \qquad \times\;
\int\limits_0^{2\upi}
\exp\left\{\frac{v\,v^0(x)}{[\sigma(x)]^2}\,\cos\varphi\right\}\,
\rmn{d}\varphi
\nonumber \\
& & \hspace*{-1.7em}
= \frac{v}{[\sigma(x)]^2}\,
\exp\left\{-\frac{v^2+[v^0(x)]^2}{2\,[\sigma(x)]^2}\right\}\,
I_0\left(\frac{v\,v^0(x)}{[\sigma(x)]^2}\right)
\end{eqnarray}
with $\varphi$ being the angle between $\vec v$ and $\vec v^0(x)$,
$v = |\vec v|$, $v^0(x) = |\vec v^0(x)|$, and $I_0(\eta)$ denoting the modified
Bessel function of the first kind to the order zero.

Hence, with dimensionless $\zeta = v/v_\rmn{c}$, $\zeta^0(x) = v^0(x)/v_\rmn{c}$,
and $\hat \sigma(x) = \sigma(x)/v_\rmn{c}$, one finds
\begin{eqnarray}
& & 
\hspace*{-2.2em}\Phi_\zeta(\zeta,x) \nonumber \\
& & \hspace*{-1.7em}
= \frac{\zeta}{[\hat\sigma(x)]^2}\,
\exp\left\{-\frac{\zeta^2+[\zeta^0(x)]^2}{2\,[\hat\sigma(x)]^2}\right\}\,
I_0\left(\frac{\zeta\,\zeta^0(x)}{[\hat\sigma(x)]^2}\right)\,.
\end{eqnarray}

While the direction of the velocity vector for Bulge objects is purely
random, disk lenses as well as the Sun perform a systematic rotation
around the Galactic
centre with a velocity $v_\rmn{circ}(R)$ depending on the cylindrical distance
$R$. With $R$ given by Eq.~(\ref{eq:R2D}),
the systematic lens motion reads
\begin{eqnarray}
v_{\rmn{L},y}^0(x) & = & v_\rmn{circ}(R)\;
\frac{R_0\,\cos l - x\,D_\rmn{S}\,\cos b}{R}\,, \nonumber  \\
v_{\rmn{L},z}^0(x) & = & 
v_\rmn{circ}(R)\;
\frac{R_0\,\sin l \sin b}{R}\,
\,.
\end{eqnarray}
The rotation velocity can be effectively described by
the model introduced by \citet*{NFW:galvel},
where the mass density is given by
\begin{equation}
\rho(x) \propto \frac{1}{r(r+r_\rmn{S})^2}\,,
\end{equation}
so that with
\begin{equation}
M(r) = 4\upi \int\limits_0^{r} \rho(\tilde r)\,{\tilde r}^2\,\rmn{d}
{\tilde r} 
\end{equation}
and $v_\rmn{circ}(r) = [G\,M(r)/r]^{1/2}$, one finds
\begin{eqnarray}
& & \hspace*{-2.2em}
v_\rmn{circ}(R) = v_\rmn{circ}(R_0)\,f^0_\rmn{NFW}\; \times \nonumber \\
& & \hspace*{-1.7em} \times\; 
\sqrt{\frac{R_0}{R}}\,
\sqrt{\ln\left[1+R/r_\rmn{S}\right]
- \frac{R}{R+r_\rmn{S}}}\,,
\end{eqnarray}
where the choices $v_{\rmn{circ}}(R_0) = 220\,\rmn{km}\,\rmn{s}^{-1}$
for the reference distance $R_0 = 8.5~\rmn{kpc}$ 
and $r_\rmn{S} = 20\,\rmn{kpc}$ yield $f^0_\rmn{NFW} = 4.23$.

\begin{figure}
\includegraphics[width=84mm]{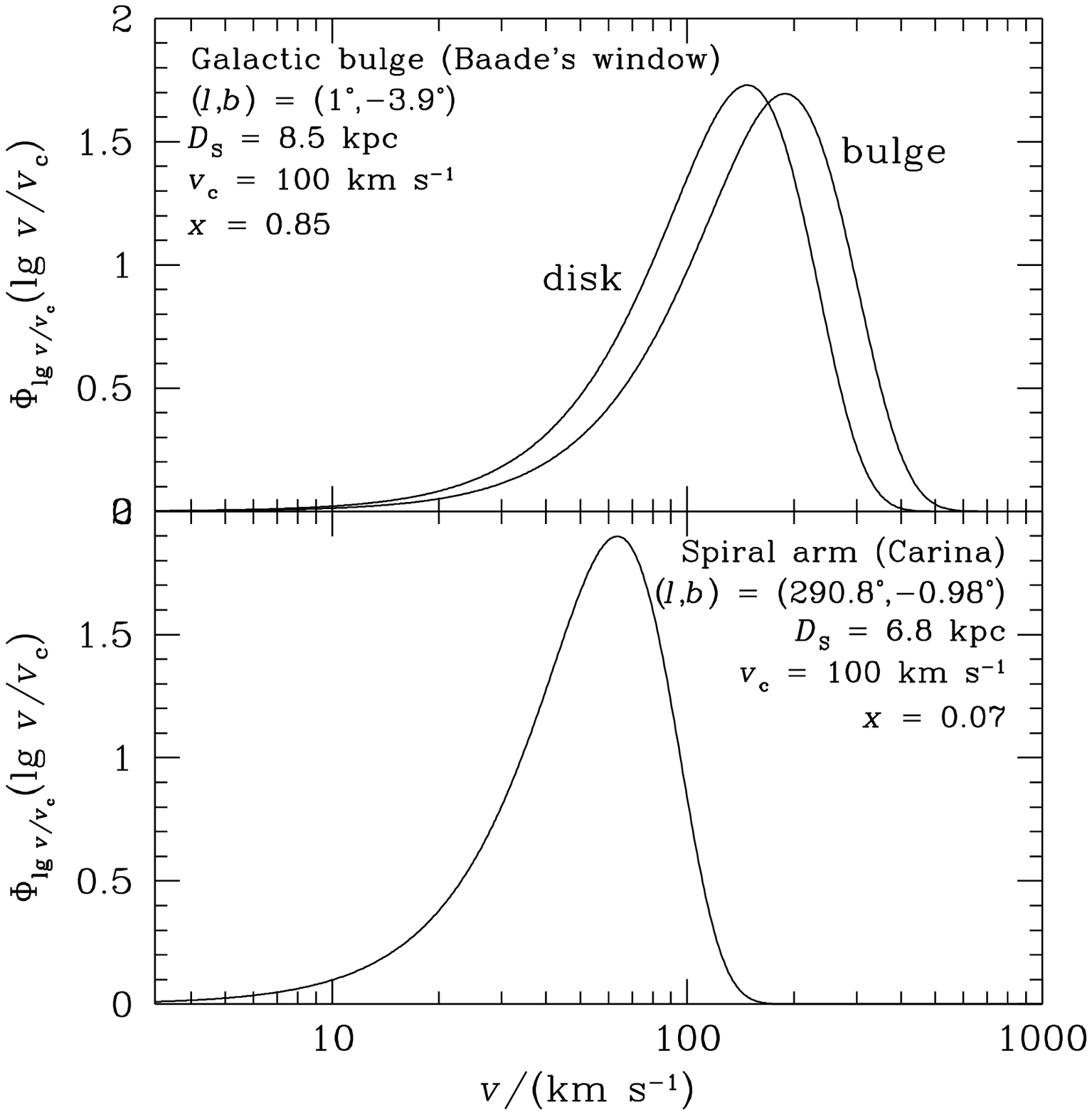}
\caption{Probability density $\Phi_{\lg \zeta}(\lg \zeta) = \zeta\,\ln 10\;
\Phi_{\zeta}(\zeta)$, where $\zeta = v/v_\rmn{c}$.
The two panels correspond to different positions of the source,
where the lens is either located in the Galactic disk or bulge
at a fractional lens distance $x \equiv D_\rmn{L}/D_\rmn{S}$.
The bulge population is not considered for the source in the spiral arm, because
its contribution to the event rate can be neglected.}
\label{fig:origvel}
\end{figure}

With respect to the rest frame of the Galaxy, the Sun, located at 
a distance $R_0$ from the Galactic centre, shows a peculiar motion
${\vec V_{\sun}} = (V_{{\sun},X},V_{{\sun},Y},V_{{\sun},Z}) =
(9,12,7) \,\rmn{km}\,\rmn{s}^{-1}$
on top of the circular motion of the Galactic disk of
${\vec V_{{\sun},\rmn{circ}}} = (0, v_\rmn{circ}(R_0),0)$ with 
$v_\rmn{circ}(R_0) =  220\,\rmn{km}\,\rmn{s}^{-1}$.

One also might consider including the velocity of the Earth 
of $v_\oplus = 30~\rmn{km}\,\rmn{s}^{-1}$. While for event time-scales
$t_\rmn{E} \ll 1~\rmn{yr}$, this velocity is approximately constant 
(and roughly equivalent to the value at the closest angular approach between
lens and source), and
to next order, the acceleration of the Earth's motion can be included in
the model of the observed light curve by means of additional parameters, 
the full annual modulation affects the light curve for longer time-scales and
this parallax effect needs to be accounted for. In the last case, however, there is no
effective Earth's velocity that contributes to $v$. For the calculations in this paper,
the velocity of the Earth is neglected, so that with ${\cal T}(l,b)$ from
Eq.~(\ref{eq:galtrafo}), one obtains ${\vec v}^{0}_\rmn{O} = {\cal T}(l,b) 
(\vec V_{\sun,\rmn{circ}}+\vec V_{\sun})$ or
\begin{eqnarray}
& & \hspace*{-2.2em}
v^{0}_{\rmn{O},y} = - \sin l\,V_{\sun,X} + \cos l\,\left[V_{\sun,Y} 
+ v_\rmn{circ}(R_0)\right] \,,  \nonumber \\
& & \hspace*{-2.2em}
v^{0}_{\rmn{O},z} = - \cos l \sin b\,V_{\sun,X} \; - \nonumber \\
& & \hspace*{-1.7em} 
- \; \sin l \sin b \,\left[V_{\sun,Y} + v_\rmn{circ}(R_0)\right] + \cos b\,V_{\sun,Z}\,.
\end{eqnarray}

Fig.~\ref{fig:origvel} shows the distribution of the effective velocity for a source in the
Carina spiral arm or in the Galactic bulge for either bulge or disk lenses, where the same parameters
as for the distribution of the lens distance shown in Fig.~\ref{fig:origdist} have been adopted.
In consistence with the latter, 'typical' values of $x = 0.85$ for the source in the Galactic bulge or
$x = 0.35$ for the source in the spiral arm have been chosen.
The shift towards larger velocities for bulge lenses as compared to disk lenses for a bulge source reflects
the larger velocity dispersion of the bulge, whereas the smaller typical velocities for the source
in the spiral arm result from the smaller velocity dispersion for disk sources and lenses.

\section{Statistics of binary orbits}
\label{sec:orbitalproj}
In general, galactic microlensing light curves only depend on the components of
the orbital separation of a lens binary that are perpendicular to the line-of-sight.
Moreover, if the orbital period is sufficiently large as compared to the duration of the event,
only the actual projected orbital separation $\hat r$ rather than the semi-major axis
$a$ is relevant, where a best-fitting model parameter $d = \hat r/r_\rmn{E}$ can be determined
from the collected data. However, one is interested in statistical properties that refer
to $a$ rather than to $\hat r$. For a given orbital numerical eccentricity $e$, the 
absolute value of the orbital separation is given by
\begin{equation}
r(t) = \frac{a(1-e^2)}{1+e \cos \varphi}\,,
\end{equation}
where $r_\rmn{min} = a(1-e)$ and $r_\rmn{max} = a(1+e)$ are the minimal and
maximal separations corresponding to the phase angles $\varphi = 0$ or $\varphi = \upi$.
With $P$ denoting the orbital period and $v_\rmn{circ} = (2\upi/P)\,a$,
one moreover finds the maximal velocity
$v_\rmn{max} = v_\rmn{circ}\,\sqrt{(1+e)/(1-e)}$ occuring at the
minimal separation and the minimal velocity 
$v_\rmn{min} = v_\rmn{circ}\,\sqrt{(1-e)/(1+e)}$ occuring at the
maximal separation. The conservation of angular momentum then yields
\begin{equation}
[r(t)]^2\,\frac{\rmn{d}\varphi}{\rmn{d}t} = \frac{2\upi}{P}\,a^2\,\sqrt{1-e^2}\,.
\end{equation}
Therefore, the probability density of $\chi = r/a$, relating the semi-major axis $a$ and the
actual separation $r$ reads
\begin{eqnarray}
& & \hspace*{-2.2em}
\Phi_\chi^e(\rho;e) = \frac{2}{P} \int\limits_0^{P/2}
\delta\left(\chi - \frac{r(t)}{a}\right)\,\rmn{d}t \nonumber \\
& & \hspace*{-1.7em}
= \frac{1}{\upi\,a^2\sqrt{1-e^2}}\,
\int\limits_0^\upi [r(\varphi)]^2\,\delta\left(\chi - \frac{r(\varphi)}{a}\right)\,
\rmn{d}\varphi \nonumber \\
& & \hspace*{-1.7em}
= \frac{\chi}{\upi}\,
\frac{\Theta\left[
\chi-(1-e)\right]\,\Theta\left[
(1+e)-\chi\right]}
{\sqrt{
\left[\chi-(1-e)\right]\,
\left[(1+e)-\chi\right]}}\,,
\end{eqnarray}
which becomes $\Phi_\chi^0(\chi) = \delta(\chi -1)$ for a circular source, for which there is a
constant orbital radius $r = a$.

\begin{figure}
\includegraphics[width=84mm]{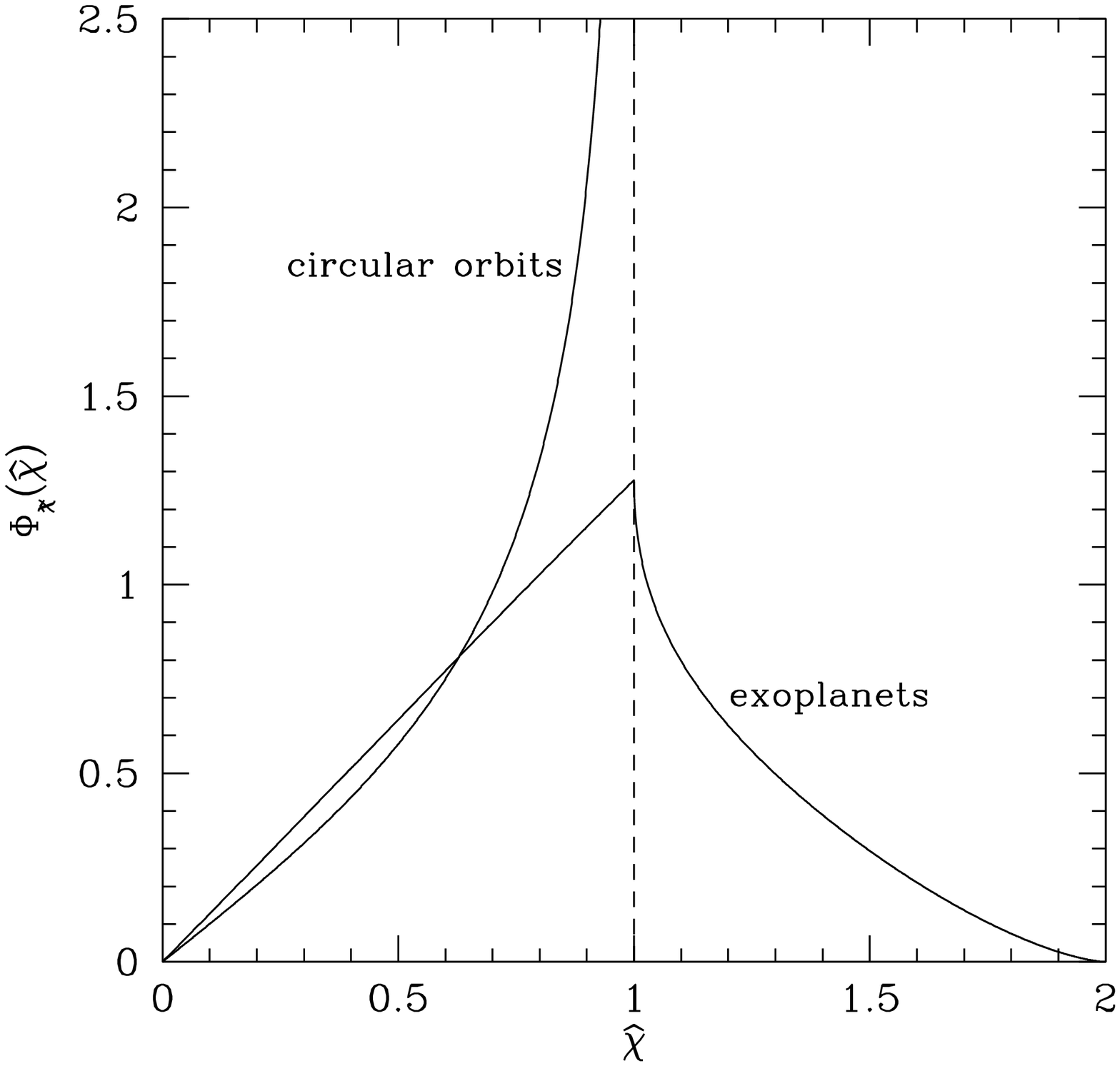}
\caption{Probability density $\Phi_{\hat \chi}(\chi)$ of the
projection factor $\hat \chi = {\hat r}/a$ between the 
semi-major axis $a$ of the orbit and the actual projected separation $\hat r$ perpendicular
to the line-of-sight. For exoplanets, a distribution $\Phi_e = (4/\upi)\,\sqrt{1-e^2}$ has
been assumed and the arising results are compared with the assumption of circular orbits.}
\label{fig:orbproj}
\end{figure}

An isotropic orientation of the orbit means that the position of the companion from the
primary at a given phase is uniformly distributed on a hemisphere with
radius $r$, so that a probability density of $\hat \chi = {\hat r}/a$ for a given $\chi = r/a$
reads
\begin{eqnarray}
& & \hspace*{-2.2em}
\Phi_{\hat \chi}^\chi(\hat \chi; \chi) = \int\limits_{0}^{\upi/2}
\delta(\hat \rho - \rho \cos \theta)\,\cos \theta \,\rmn{d}\theta \nonumber \\
& & \hspace*{-1.7em}
= \frac{\hat \chi}{\chi}\,\frac{\Theta(\chi-{\hat \chi})}{\sqrt{\chi^2-{\hat \chi}^2}}\,,
\end{eqnarray}
where the area of the hemisphere ($2\,\upi$) cancels out against the integral over
the azimuthal angle $\varphi$.

For a given orbital eccentricity, one obtains the probability density of $\hat \chi$ as
\begin{eqnarray}
& & \hspace*{-2.2em}
\Phi_{\hat \chi}^e(\hat \chi; e) = \int\limits_{0}^{\infty}
\Phi_{\hat \chi}^\chi(\hat \chi; \chi)
\,\Phi_\chi^e(\chi;e)\,\rmn{d}\chi \nonumber \\
& & \hspace*{-1.7em} 
= \frac{\hat{\chi}}{\upi}\,
\int\limits_{0}^{\infty} 
\frac{\Theta(\chi-\hat{\chi})\,\Theta\left[
\chi-(1-e)\right]\,\Theta\left[
(1+e)-\chi\right]}
{\sqrt{\left[\chi^2-\hat{\chi}^2\right]\,
\left[\chi-(1-e)\right]\,
\left[(1+e)-\chi\right]}}\,
\rmn{d}\chi \nonumber \\
& & \hspace*{-1.7em}  = \left\{
\begin{array}{lcl}
\frac{2\,\hat{\chi}}{\upi\,\sqrt{(1+e-\hat{\chi})\,
(1-e+\hat{\chi})}}\;
K\left(2\,\sqrt{\frac{e \hat{\chi}}{(1+e-\hat{\chi})\,
(1-e-\hat{\chi})}}\right) & & \\
\hfill \rmn{for} \quad \hat{\chi} \leq 1 - e & &  \\
\frac{1}{\upi}\,\sqrt{\frac{\hat{\chi}}{e}}\;
K\left(\frac{1}{2}\,\sqrt{\frac{(1+e-\hat{\chi})
(1-e+\hat{\chi})}{e \hat{\chi}}}\right) &  & \\
\hfill \rmn{for} \quad 1- e < \hat{\chi} < 1 + e & &  \\
 0 
 \hfill \rmn{for} \quad \hat{\chi} \geq 1 + e  & & 
\end{array}\right.\hspace*{-2em}\,,
\end{eqnarray}
so that for the orbital eccentricity $e$ being distributed following the
probability density $\Phi_e(e)$, the probability density of $\hat \chi$ results as
\begin{equation}
\Phi_{\hat \chi}(\hat \chi) = \int\limits_0^1 \Phi_e(e)\,
\Phi_{\hat \chi}^e(\hat \chi; e)\,\rmn{d}e\,,
\label{eq:phatchi}
\end{equation}
while for circular orbits, one obtains
\begin{equation}
\Phi_{\hat \chi}(\hat \chi) = \frac{\hat \chi}{\sqrt{1-{\hat \chi}^2}}\,\Theta(1-\hat \chi)\,.
\label{eq:phatchi0}
\end{equation}

Both for circular orbits and elliptical orbits that correspond to planetary systems,
$\Phi_{\hat \chi}(\hat \chi)$ is shown in Fig.~\ref{fig:orbproj}. For the latter case,
$\Phi_e(e) = (4/\upi) \sqrt{1-e^2}$ has been chosen in rough agreement 
with radial velocity searches, where $p_e(e)$ is approximately
constant for moderate $e$, but drops off to zero as $e \to 1$.

\end{document}